\documentclass[11pt]{article}
\usepackage{jheppub}
\usepackage{amsmath}
\usepackage{bm}
\usepackage{slashed}
\usepackage{graphicx}
\newcommand{\gf}{\mathfrak{g}}
\newcommand{\Gad}{G_\ad}
\newcommand{\ad}{\text{ad}}

\newcommand{\Tr}{\mathrm{Tr}\,}
\newlength{\dummysp}
\newcommand{\beq}{\begin{eqnarray}}
\newcommand{\eeq}{\end{eqnarray}}

\newcommand{\gappeq}{\mathrel{\rlap {\raise.5ex\hbox{$>$}}
{\lower.5ex\hbox{$\sim$}}}}
\newcommand{\lappeq}{\mathrel{\rlap{\raise.5ex\hbox{$<$}}
{\lower.5ex\hbox{$\sim$}}}}

\newcommand{\ben}{\begin{enumerate}}
\newcommand{\een}{\end{enumerate}}

\newcommand{\bit}{\begin{itemize}}
\newcommand{\eit}{\end{itemize}}

\def\[{\left [}
\def\]{\right ]}
\def\({\left (}
\def\){\right )}
\def\R{{\mathbb R}}
\def\S{{\mathbb S}}
\def\Z{{\mathbb Z}}
\pdfoutput=1

\title{Deconfinement and continuity between thermal and  (super) Yang-Mills theory for all gauge groups}

\author[]{Mohamed M. Anber,} \author[]{Erich Poppitz,} 
 \author[]{Brett Teeple}  
 
\affiliation[]{Department of Physics,   University of Toronto, 
Toronto, ON M5S 1A7, Canada}
\emailAdd{manber@physics.utoronto.ca}\emailAdd{poppitz@physics.utoronto.ca}\emailAdd{bteeple@physics.utoronto.ca}   
    
\abstract
{We study the phase structure of ${\cal{N}}$$=$$1$ supersymmetric Yang-Mills theory on $\R^3\times\S^1$, with massive gauginos, periodic around the $\S^1$,   with  $Sp(2N)$ ($N$$\ge$$2$), $Spin(N)$ ($N$$\ge$$5$), $G_2$, $F_4$, $E_6$, $E_7$, $E_8$ gauge groups. As the gaugino mass $m$ is increased, with    $\S^1$ size and strong coupling scale fixed, we find a first-order  phase transition  both for theories with and without a center. This semiclassically calculable transition is driven, as in  $SU(N)$ and $G_2$  \cite{Poppitz:2012sw,Poppitz:2012nz}, by a competition between   monopole-instantons and exotic topological ``molecules"---``neutral" or ``magnetic" bions. We compute the trace of the Polyakov loop  and its two-point correlator  near the transition. We find a behavior similar  to the one observed near the thermal deconfinement transition in the corresponding pure Yang-Mills (YM) theory in lattice studies (whenever available).  Our results  lend further support to the conjectured continuity, as a function of $m$,  between the quantum phase transition studied here and the thermal deconfinement transition in YM theory.   We also study the $\theta$-angle dependence of the transition, elaborate on the importance of the quantum-corrected moduli-space metric at large $N$, and offer comments for the future.}

\begin{document}
\maketitle
\flushbottom

\section{Introduction, summary, and outlook}
\label{Introduction}

The problems of confinement, deconfinement, and, more generally, the phases of  nonabelian four-dimensional Yang-Mills theories remain among the most difficult issues in quantum field theory. Theoretically-controlled analytical approaches to the problem are hard to come by. Remarkably, Seiberg-Witten theory and the AdS/CFT correspondence have provided some important hints and tools, but remain, in many aspects,  far from the nonsupersymmetric gauge theories of the real world or from  possible beyond-the-Standard-Model worlds.

In this paper, we study a small part of this large set of problems: the deconfinement transition in pure Yang-Mills theory. While our study is indirect, it offers the rare benefit of being under  theoretical control and we hope that it  provides relevant insights into the problem,  as we explain below.

 \subsection{Thermal Yang-Mills,  Polyakov loop,  GPY potential, and deconfinement}
 
\label{introone}
We begin with a reminder of some  well known  facts about thermal theories. The overview of this work begins in the next Section \ref{introtwo}.

The partition function of thermal quantum Yang-Mills theory is given by an Euclidean path integral on $\R^3 \times \S^1_\beta$; the size of the thermal circle is $\beta = {1 \over T}$, the inverse temperature. In thermal Yang-Mills theory, an important observable is the Polyakov loop around the thermal circle.
This is the   Wilson line  operator, $\Omega_{\cal R}(x^\mu)$, taken along a loop around the thermal $\S^1$:
\begin{equation}
\label{wilsonloop12}
\Omega_{\cal R}(x^\mu) \equiv   P e^{ i \int_{\S^1} A_3(x^\mu, x^3) d x^3}~.
\end{equation}
Here $P$ denotes  path ordering and the gauge field is taken in the representation  ${\cal R}$.\footnote{In the Introduction, it is convenient to think of $SU(N)$ pure Yang-Mills theory, with ${\cal R}$ taken to be in the fundamental representation. 
For future use, we label the $\S^1$ direction by $x^3$ and the rest of the Euclidean directions by $x^\mu$.}
The trace of the Polyakov loop, $\Tr \Omega_{\cal R}(x^\mu)$, is gauge invariant. Its insertion in the thermal partition function corresponds to the insertion, at $x^\mu$, of an infinitely heavy color-charged probe  (``quark") in the representation ${\cal R}$. The correlator of two Polyakov loops
\begin{equation}
\label{pol1}
\langle \Tr \Omega_{\cal R}(x^\mu)\Tr \Omega^\dagger_{\cal R}(0) \rangle~ = e^{- {V(r, T)\over T}},
\end{equation}
placed a distance $r= |x^\mu|$ apart plays an important role in thermal Yang-Mills theory:  when computed with the thermal partition function, it measures the potential of the two heavy quark sources,  as already indicated on the r.h.s. of (\ref{pol1})  \cite{Polyakov:1978vu, Susskind:1979up}.

At low temperature, in the confined phase, one expects  a linear confining potential between the quarks, i.e. $V(r,T) = \sigma(T) r$. Thus, as $r \rightarrow \infty$, we have $\langle \Tr \Omega_{\cal R}(x^\mu)\Tr \Omega^\dagger_{\cal R}(0) \rangle \rightarrow 0$ and thus $\langle \Tr \Omega_{\cal R}\rangle = 0$.
At high temperature, in a deconfined phase, one expects a screened Coulomb (i.e. Yukawa) potential between the heavy quark probes, i.e. $V(r,T) = v + b \;{e^{- m_e r} \over r}$, where $v$ and $b$ are constants and $m_e$ is the electric Debye screening mass.  Thus, as $r \rightarrow \infty$, $\langle \Tr \Omega_{\cal R}( x^\mu )\Tr \Omega^\dagger_{\cal R}(0) \rangle \ne 0$, hence  $\langle \Tr \Omega_{\cal R} \rangle \ne 0$. 
The conclusion   is that the Polyakov loop (say in the fundamental of $SU(N)$) has a qualitatively different behavior in the low- and high-temperature phases:  its two-point correlator behavior is qualitatively different and its expectation value changes from  $\langle \Tr \Omega_{\cal R} \rangle = 0$ in the confined phase to $\langle \Tr \Omega_{\cal R} \rangle \ne 0$ in the deconfined phase. This behavior of the Polyakov loop  (\ref{wilsonloop12}) expectation value and its correlator (\ref{pol1}) have been observed in lattice simulations of pure Yang-Mills theory for many years.

The high-temperature behavior of $\langle \Tr \Omega_{\cal R} \rangle$  and of its correlator  has  been understood also for many years. One expects that at $T \gg \Lambda$, where $\Lambda$ is the strong coupling scale of the theory, perturbation theory is a good guide to the dynamics, at least as far as the behavior of the Polyakov loop is concerned. Gross, Pisarski, and Yaffe (GPY) \cite{Gross:1980br} used this intuition and  computed  the one-loop Casimir potential for the eigenvalues of $\Omega_{\cal R}$.\footnote{For explicit expressions, see Appendix \ref{gauginoGPY}, Eqs.~(\ref{gpyheavy},\ref{gauginogpy}).}
As $\Omega_{\cal R}$ is a unitary operator, its eigenvalues   $e^{i \lambda}$ lie on the unit circle. GPY found that, at asymptotically high $T$, where the perturbative calculation is valid, the Casimir potential leads to clumping of the eigenvalues $\lambda$ and thus,  since all eigenvalues $\lambda$ are  the same, to  $\langle \Tr \Omega_{\cal R} \rangle  = \sum_\lambda e^{ i \lambda} \ne 0$. In gauge theories with a non trivial center of the gauge group, e.g. $\Z_{N}$ for $SU(N)$, the transition from the confining phase, $\langle \Tr \Omega_{\cal R} \rangle =0$, to the deconfined phase, $\langle \Tr \Omega_{\cal R} \rangle \ne 0$, is thus associated with center symmetry breaking. The trace of the Polyakov loop  in a representation $\cal{R}$ which transforms  under the center serves as an order parameter  \cite{Svetitsky:1982gs}.\footnote{Recall that in $SU(N)$, the $\Z_{N}$ center acts on the fundamental Polyakov loop $\Tr \Omega_{F} \rightarrow e^{i {2 \pi k\over N}}\Tr \Omega_{F}$. The centers of all Lie groups are listed in Table~\ref{tab:lie}. It is somewhat counterintuitive to have symmetry break at high temperature.  In particular, analogy with usual phase transitions leads one to expect domains and domain walls associated with $\Z_{N}$ breaking and the question what physical objects  at high-$T$ they correspond to has been discussed in the literature (e.g. \cite{Bhattacharya:1990hk,Smilga:1993vb,Aharony:1998qu,KorthalsAltes:1999xb,Armoni:2008yp}). We  only note  that the trace of the Polyakov loop is, indeed, seen to discontinuously  jump  in lattice simulations   as well as in analytical large-$N$ calculations in thermal pure YM on $\S^3 \times \S^1$ \cite{Aharony:2003sx}   and is thus a convenient order parameter (the limit of $\langle \Tr \Omega\rangle$ must be properly defined in finite volume, large-$N$, see  \cite{Svetitsky:1982gs,Aharony:2003sx}). In this  paper, we study an infinite volume quantum, rather than thermal, phase transition, where the usual intuition applies.   }

As  the temperature is lowered,  the behavior of the Polyakov loop  changes, as described above. In terms of its eigenvalues, the vanishing of the traces of  $\langle \Tr \Omega_{\cal R}^k \rangle$$=$$0$  $(k$$=$$1,...,$$N$$-$$1$ for $SU(N)$) in the confined phase means that as the temperature is lowered, the eigenvalues have to evenly spread around the unit circle, instead of clump,  to ensure that $\langle \sum_\lambda e^{ i \lambda} \rangle= 0$. The physics that causes this change of the behavior  of the Polyakov loop eigenvalues---from clumped at high-$T$ to evenly distributed around the circle at low-$T$ is the subject of this article. 

By dimensional analysis, it is  clear that in pure YM theory this change of behavior has to occur at $T$ of order $\Lambda$, i.e. at strong coupling, where perturbative methods are not applicable.
Theoretical studies of the behavior of  the Polyakov loop eigenvalues attempt to describe their behavior using an effective potential,  derived from field-theory model considerations or extracted from lattice data. Thus, studies of the Polyakov loop potential have employed lattice gauge theory, field theoretical models, and  functional renormalization group. An incomplete list for thermal $\R^3 \times \S^1$ is \cite{Pisarski:2001pe, Fukushima:2003fw, Ratti:2005jh,Braun:2010cy,Diakonov:2012dx, Greensite:2012dy,Dumitru:2012fw,Haas:2013qwp,Smith:2013msa};  in the rest of the paper, we will compare our findings to some of these works.

For such a complicated problem, it would be desirable to have an analytically calculable setup. Even if not completely realistic, it is likely to provide some insight into the relevant physics and inform model studies, an issue that we shall come back to later.

Analytic calculability of the deconfinement transition, as seen in the behavior of the Polyakov loop eigenvalues,  has so far been achieved in several circumstances. Upon compactification of thermal Yang-Mills theory on  $\S^3 \times \S^1_\beta$, using the size of $\S^3$  (smaller than $\Lambda^{-1}$) as a control parameter allowing perturbative calculations, the authors of \cite{Aharony:2003sx} showed that in the infinite-$N$ limit there is a sharp phase transition associated with deconfinement, and occurring in the weakly-coupled regime; as this is a small volume system, infinite-$N$ is necessary to have a phase transition. 

Another situation where analytic calculability is sometimes possible is to consider thermal YM, ``deformed" or with adjoint fermions, partially compactified on $\R^2 \times \S^1_L \times \S^1_\beta$, using $L \ll \Lambda^{-1}$ as a control parameter (further in this paper we will use $L=2\pi R$).\footnote{These studies lead to a description of deconfinement as  a genuine thermal deconfinement transition  in an electric-magnetic Coulomb gas in two dimensions, inheriting the symmetry properties of the original four-dimensional thermal theory 
\cite{Simic:2010sv,Anber:2011gn,Anber:2012ig,Anber:2013doa}.} While some of the ingredients of the latter setup are similar to the one of this paper, the general approach is  different. 

We now describe the main idea of  \cite{Poppitz:2012sw,Poppitz:2012nz}. The main purpose of this paper is to continue the study initiated in these references.

\begin{table}[t]
\begin{center}
\begin{tabular}{|c|c|c|c|} \hline
$\gf$   & \phantom{\framebox{$G$}}$G$\phantom{\framebox{$G$}}  
  &  $\Gad :=G/Z(G)$ &$\pi_1(\Gad)=Z(G)$ \\ \hline\hline
$A_{N-1}$ 	& $\text{\it SU}(N)$    	& $\text{\it PSU}(N)$
  & $\Z_N$\\
$B_N$	& $\text{\it Spin}(2N+1)$& $\text{\it SO}(2N+1)$    
  & $\Z_2$\\
$C_N$	&  $\text{\it Sp}(N)$ or $\text{\it USp}(2N)$ & $\text{\it PSp}(N)$ 
  & $\Z_2$\\
$D_{2N}$	& $\text{\it Spin}(4N)$	& $\text{\it PSO}(4N)$    
  & $\Z_2\times\Z_2$\\
$D_{2N+1}$& $\text{\it Spin}(4N+2)$& $\text{\it PSO}(4N+2)$
  & $\Z_4$\\
$E_6$	& $E_6$    	& $E_6^{-78}$  & $\Z_3$   \\ 
$E_7$	& $E_7$	        & $E_7^{-133}$ & $\Z_2$  \\ 
$E_8$	& $E_8$    	& $E_8$       & $1$   \\ 
$F_4$ 	& $F_4$    	& $F_4$       & $1$  \\ 
$G_2$ 	& $G_2$   	& $G_2$       &  $1$  \\ 
\hline
\end{tabular}
\caption{\label{tab:lie}
The simple Lie algebras $\gf$ together with  their associated 
compact simply-connected Lie groups $G$ and the compact adjoint Lie 
groups $\Gad$. The last column lists the center symmetry of $G$, 
which is isomorphic to the fundamental group of  $\Gad$.}
\end{center}
\end{table}
\subsection{Quantum phase transition vs. deconfinement: Continuity conjecture}
 \label{introtwo}
 
 To describe the main idea, let us schematically denote the perturbative one-loop GPY potential for the Polyakov loop eigenvalues by $g^2 V_{GPY}(\Omega)$, explicitly indicating its leading dependence on the coupling. If there exists a weakly coupled description of the transition,  changing the behavior of the $\Omega$ eigenvalues, there should be other, nonperturbative, contributions to the total potential, $V_{n.p.}(\Omega)$, which  lead to eigenvalue spreading, countering the eigenvalue-clumping effect of $V_{GPY}$. We imagine that these are  of order $e^{- {{\cal O}(1) \over g^2}}$ (possible pre-exponential $g^2$ dependence is neglected here)  as they can not be seen perturbatively to any finite order. Thus we expect the total potential to behave as: 
 \begin{equation}
 \label{vomega1}
 V(\Omega) = g^2 V_{GPY} (\Omega) + e^{- { 2 a \over g^2}} V_{n.p.}(\Omega) \;,
 \end{equation}
 with $2 a$ denoting a number of order unity.
It is clear that within the weak-coupling regime, the perturbative term will always dominate over the nonperturbative contributions, as $g^2 \gg e^{-  {{\cal O}(1) \over g^2}}$, unless there is some reason why the perturbative contribution is smaller  than indicated (or the nonperturbative one is larger\footnote{\label{five}This is what happens in the small-$\S^3\times \S^1_\beta$ studies, where the
spreading of eigenvalues at low-$T$ is a kinematic effect due to the Vandermonde determinant for the $\S^3$ zero-mode, an integral over which has to be included in the finite-volume system, see  \cite{Aharony:2003sx}. On the other hand, in ``deformed" YM or QCD(adj) on $\R^2\times S^1_L \times S^1_\beta$ at small-$L$ and weak coupling, a situation similar to the one in this paper is realized:  at small $L\over \beta$,  the perturbative $V_{GPY}$     is exponentially suppressed   by a $\sim e^{-{{\cal O}(1)}{ \beta\over L}}$ factor \cite{Simic:2010sv,Anber:2011gn,Anber:2012ig,Anber:2013doa}.}).

A setup where analytical calculability of the transition on $\R^3 \times \S^1$ is achieved because of a suppression of the perturbative contribution was proposed in \cite{Poppitz:2012sw,Poppitz:2012nz},  following related  remarks in \cite{Unsal:2010qh}. 
The idea is to consider an Euclidean theory on $\R^3 \times \S^1$, as in the thermal case. However, the theory is not exactly thermal pure Yang-Mills theory, but its ${\cal N}=1$ supersymmetric counterpart (SYM) with a supersymmetry breaking mass $m$ for the gaugino---the Weyl fermion in the adjoint representation which is the superpartner of the gluon in the massless limit. We shall call this theory SYM$^*$. The SYM$^*$ theory is connected to pure thermal Yang-Mills theory upon decoupling the gaugino, $m\rightarrow \infty$. Clearly, in the decoupling limit  the boundary conditions on the gaugino around the $\S^1$ are  irrelevant.\footnote{Unless a twist by an anomalous symmetry is involved, which is not the case here, however, see \cite{Poppitz:2008hr}.}

However, it is crucial for the calculability that the gaugino be periodic, i.e. non-thermal around the compact direction. This ensures the preservation of supersymmetry. In the $m \rightarrow 0$ limit, the theory is SYM on $\R^3 \times \S^1$ and supersymmetric nonrenormalization theorems guarantee that there is no potential for the Polyakov loop at any order of perturbation theory. Thus, when $m \ne 0$ is turned on, we have, instead of (\ref{vomega1}), a potential of the form:
 \begin{equation}
 \label{vomega2}
 V(\Omega) = g^2 m^2 \tilde{V}_{GPY} (\Omega) + e^{- {2 a \over g^2}} V^{SUSY}_{n.p.}(\Omega) + m  e^{- {a \over g^2}}  \tilde{V}_{n.p.}(\Omega)\;.
 \end{equation}
 A comparison of (\ref{vomega2}) with (\ref{vomega1}) shows that two things have changed in the potential upon the introduction of a small gaugino mass. First, the perturbative GPY contribution is now proportional to a small parameter, $m$, due to the fact that supersymmetry requires that it  vanish as $m\rightarrow 0$.
 Second, introducing a gaugino mass turns on  new nonperturtative contributions to the potential, indicated by the last term in (\ref{vomega2}). 
 
\begin{figure}[h]
\centering  
\includegraphics[width=.5 \textwidth]{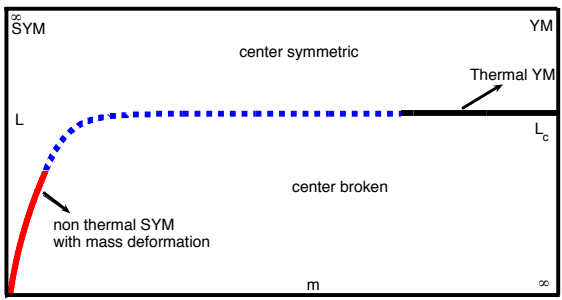}
\caption{\label{fig:phase}The conjectured phase diagram of SYM$^*$ in the $m$-$L$ plane. The calculable center-symmetry breaking quantum phase transition, occurring at small-$m,L$---the left-hand corner of the diagram, shown by a thick red line---is conjectured to be continuously connected, upon decoupling the gaugino, to the thermal deconfinement transition in pure YM theory, shown by the thick black line on the right. The dimensionless parameter which is varied, Eq.~(\ref{cmdefined}), is $c_m \sim {m \over L^2 \Lambda^3}$, with   $m \over \Lambda$ and $L \Lambda$ small. For all gauge groups,   the calculable quantum phase transition occurs for  $c_m$ of order unity. It should be possible to study this phase diagram on the lattice, see Section \ref{introfive}.}
\end{figure}

\begin{figure}[h]
\centering  
\includegraphics[width=.6 \textwidth]{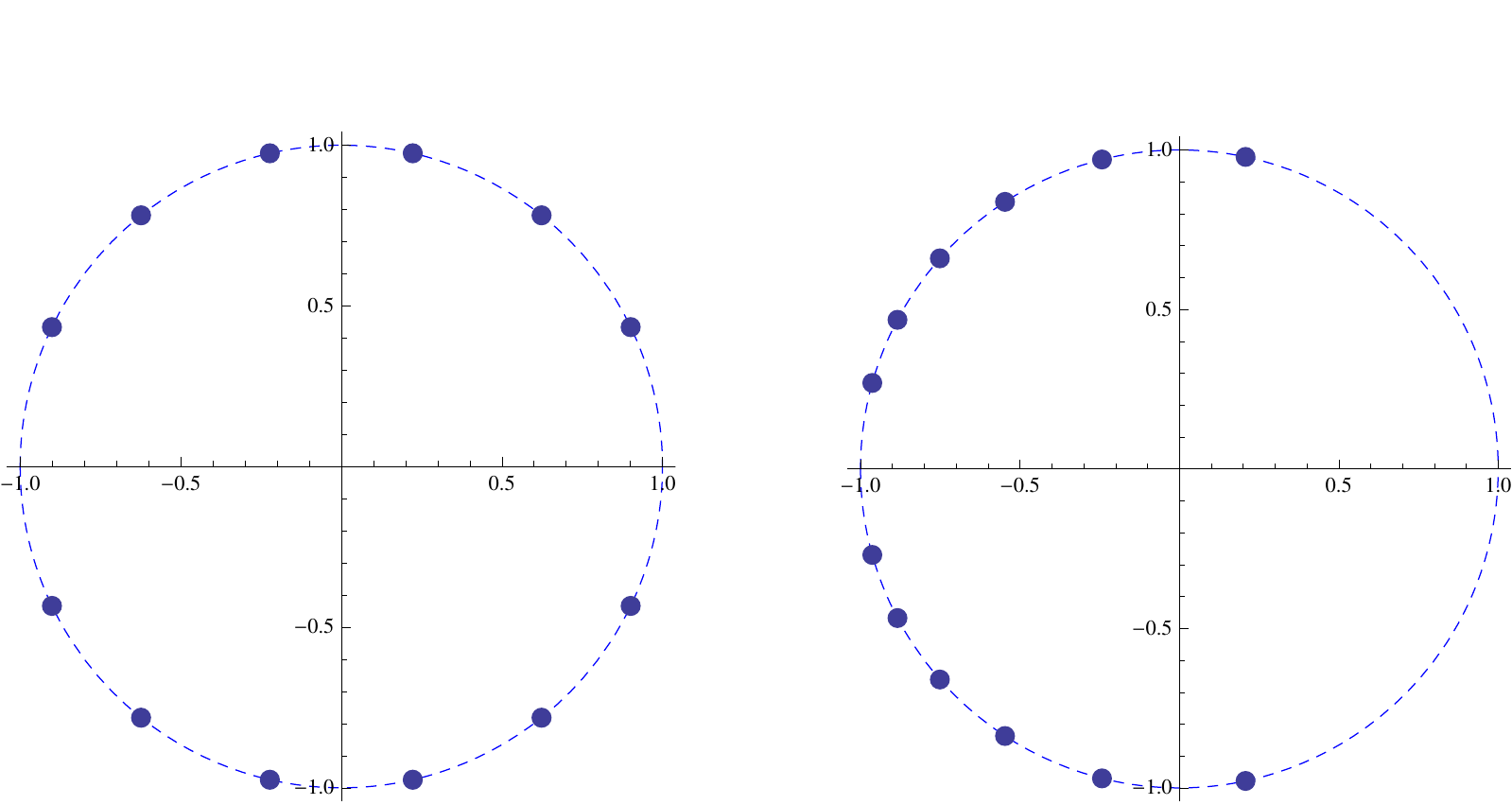}
\caption{\label{fig:speigenvalues}An example of the discontinuous change of the Polyakov loop eigenvalues: the $\Z_2$ center-symmetric  distribution of the eigenvalues of $\Omega$, in the fundamental representation of $Sp(12)$, for $c_m < c_{\rm cr} \sim 0.614$ (left panel) and the center-broken distribution for $c_m= c_{\rm cr}^+$ (right  panel). The eigenvalues on the right panel are plotted for ${g^2 \over 4\pi}=0.4$ to visually enhance the center-breaking effect. The discontinuous change of the eigenvalue distributions across $c_{\rm cr}$   looks similar for all gauge groups.}
\end{figure}

 As $m\rightarrow 0$, one is left with the supersymmetric nonperturbative contribution to the Polyakov loop potential, $V^{SUSY}_{n.p.}(\Omega)$, which prefers a uniform distribution of eigenvalues, as in the confined phase of thermal Yang-Mills theory.\footnote{Aspects of SYM on $\R^3 \times \S^1$ were studied in the 1990's \cite{Seiberg:1996nz,Aharony:1997bx,Davies:1999uw}, but center symmetry was not discussed.}
 On the other hand, as $m$ increases, one finds that  both $\tilde{V}_{GPY}$ and the new nonperturbative contribution $\tilde{V}_{n.p.}$ become important, and that both prefer the clumping of eigenvalues, as in the deconfined phase of YM theory. 
Thus, there is a phase transition, associated with the distribution of eigenvalues of $\Omega$, as a function of the gaugino mass $m$, at any fixed small size $L$ of the $\S^1$. This behavior is shown in  the left-hand corner of the phase diagram, in the $m$-$L$ plane, shown on Fig.~\ref{fig:phase} (the changing distribution of eigenvalues is shown on  Fig.~\ref{fig:speigenvalues} for one particular theory we study).
As will be explained in detail later in this paper, this transition is calculable at small $m$ and $L$; we also stress that the small-$m$ theory does not have a thermal interpretation and that this is a quantum phase transition.
 
Since at large $m$ the $\R^3 \times \S^1$  SYM$^*$ becomes thermal pure YM, it is natural to conjecture that the semiclassically calculable quantum phase transition is smoothly connected to the deconfinement transition in pure YM. Since \cite{Poppitz:2012sw}, considerable evidence, reviewed in the next Section, has been collected in support of this conjecture. 
 
\subsection{Review of previous evidence for the continuity conjecture}
 \label{introthree}
 
In \cite{Poppitz:2012sw}, the physics leading to the various terms in  (\ref{vomega2}) whose interplay determines the phase structure of the SYM$^*$ theory in the small-$L$, small-$m$ domain was discussed in detail, for an $SU(2)$ gauge theory. 
A  center-symmetry-breaking quantum phase transition was shown to occur  as the dimensionless parameter  $c_m \sim {m   \over L^2 \Lambda^3}$ is increased.\footnote{The precise definition of $c_m$ is in Eq.~(\ref{cmdefined}).} In $SU(2)$, the $\Z_2$-breaking transition is  second-order. This is also the known order of the deconfinement transition in nonsupersymmetric thermal $SU(2)$ YM theory, known from the lattice and also argued for by $\Z_2$ universality \cite{Svetitsky:1982gs}.

  Further evidence for the similarity of the small-$m$, small-$L$ center-breaking transition to the thermal deconfinement transition in YM theory with gauge group $SU(N)$ was given in \cite{Poppitz:2012nz}. 
For all $N>2$, a first-order transition was found, as  seen on the lattice in thermal pure YM theory, see the recent review of large-$N$ theories \cite{Lucini:2012gg}.

Since various topological objects play a crucial role in the calculable transition in SYM$^*$, it should not come as a surprise that, in all cases, the phase transition ``temperature"  ($c_m$) also acquires  topological $\theta$-angle dependence,   due to the ``topological interference" effect  \cite{Unsal:2012zj} (we note that  \cite{Parnachev:2008fy, Thomas:2011ee} gave earlier  discussions of $\theta$-dependence in the deconfinement transition). The $\theta$-dependence of the critical $c_m$ (or $L_{\rm cr}$ at fixed $m$) was  studied in \cite{Poppitz:2012nz,Anber:2013sga} and is  in  qualitative agreement with  recent lattice studies of $\theta$-dependence in thermal pure YM theory, see 
\cite{D'Elia:2012vv, D'Elia:2013eua} and references therein. 
In \cite{Anber:2013sga}, the  $\theta$-dependence of another quantity was also studied---the discontinuity of the trace of the Polyakov loop  at the transition, and found a dependence later confirmed by  the  lattice \cite{D'Elia:2013eua}. 
 
A discontinuous transition in the small-$m,L$ regime was also found to occur in SYM$^*$ theories without a center. The case of $G_2$ SYM$^*$ was studied in 
   \cite{Poppitz:2012nz}. This theory is similar to real QCD in that fundamental  quarks  can be screened (in $G_2$, by three gluons). Proceeding along the lines described above for $SU(2)$,     a discontinuous transition of the Polyakov loop eigenvalues from an almost uniformly distribution on the unit circle to  a more clumped one, upon increasing  $c_m$ was found. This is also the behavior    seen on the lattice for  thermal   $G_2$ YM theory
\cite{Pepe:2006er,Cossu:2007dk}.\footnote{We revisit this theory in Section \ref{g2revisit}, correcting small omissions in  \cite{Poppitz:2012nz} and giving more discussion.}
  
 \subsection{Outline and  summary}
 \label{introfour}
 
 The main purpose of this paper is to continue the study initiated by \cite{Poppitz:2012sw} and complete it for all gauge groups. We ask whether  the mechanism behind the discontinuous transition in the behavior of the eigenvalues of $\Omega$ is indeed universal, as anticipated in \cite{Poppitz:2012nz}, and whether there is qualitative agreement of the features of the transition with all available lattice studies of thermal pure-YM theories. 
 
 We find that the answer   is ``yes" on both accounts. For all simple Lie groups, except for $SU(2)$, we find a discontinuous transition which occurs for  $c_m$ of Eq.~(\ref{cmdefined}) of order unity. 
  The physical mechanism behind this transition is, in all cases, the same, and the potential for the Polyakov loop $\Omega$  has the schematic form  (\ref{vomega2}). For all groups, it turns out that for  $c_m$ below and near the critical value, $c_{\rm cr}$, the perturbative GPY contribution is suppressed relative to the two nonperturbative terms in (\ref{vomega2}). The transition is thus driven by a competition between the second and third terms. The second term is also present in SYM theory on $\R^3 \times \S^1$, and is due to novel topological excitations: magnetic bions and neutral (or center stabilizing) bions, discussed earlier in the literature  \cite{Unsal:2007jx,Poppitz:2011wy,Argyres:2012ka}. The third term is due to monopole-instantons and   has a center-destabilizing effect, increasing with the gaugino mass.
  
  The explicit form of the three terms comprising $V(\Omega)$ of Eq.~(\ref{vomega2}), for arbitrary gauge group,  is given later in this paper: the ${\cal O}(g^2 m^2)$  GPY term in Eq.~(\ref{gauginogpy2}),  the supersymmetric nonperturbative ${\cal{O}}(e^{- {2 a \over g^2}})$ term due to magnetic and neutral bions in Eq.~(\ref{bionpotential2}), and  the ${\cal{O}}(m e^{- {a\over g^2}})$ term due to monopole-instantons in Eq.~(\ref{monopolepotential2}). Further, it turns out that $a = {8 \pi^2 \over c_2 g^2}$, with $c_2$ the dual Coxeter number of the gauge group. The factor of two difference in the exponent in the two nonperturbative terms in (\ref{vomega2}) is due to the composite nature of the magnetic and neutral bions \cite{Unsal:2007jx,Poppitz:2011wy,Argyres:2012ka}.   
 
 \subsubsection{The Polyakov loop potential  $\mathbf{V(\Omega)}$  and the moduli space metric}
 
 The first part of the paper contains the derivation of the various  contributions to $V(\Omega)$. Most of this part  reviews known material, but we strive to fill in all detail left out of the study of $SU(N)$ and $G_2$  \cite{Poppitz:2012nz}. 
 The only new contribution  is the calculation of the one-loop determinants around the monopole-instantons for a general gauge group and the related  loop correction to the moduli space metric (or K\" ahler potential in the dual picture). The corrections to the moduli space metric are  important in two circumstances: in taking the abelian large-$N$ limit, see Section \ref{largencoupling}, and in the study of Polyakov loops without center symmetry of Section \ref{results44}. To describe them properly, we also include a detailed description of the chiral-linear duality.

In Section \ref{abelianization}, we describe  the abelianization of SYM with arbitrary gauge group   on $\R^3 \times \S^1$ at small $L$. We include this material for completeness and to introduce notation appropriate to our goals.\footnote{The 
Lie-algebraic notation used throughout is reviewed in Appendix \ref{groupsummary} and is used to calculate the one-loop GPY contribution of the massive gaugino  in Appendix \ref{gauginoGPY}, for arbitrary gauge groups.} 
 Section \ref{treeduality} introduces in detail some relevant supersymmetric technology---the linear-chiral supermultiplet duality, the supersymmetrization of  the three-dimensional photon-scalar (dual photon) duality. We use a basis conveniently related to the conventional four-dimensional superfield and Weyl-spinor notation.\footnote{All our notation  is the one of Wess and Bagger \cite{Wess:1992cp}.} Section \ref{fundamentaldomain} is devoted to describing the physically inequivalent range  of the fields that enter the low energy theory: the holonomies $\pmb \phi$, which determine the eigenvalues of $\Omega$ in any representation, in Section \ref{holonomyperiod}, and the dual photons $\pmb \sigma$, in Section \ref{dualphotonperiod}. Throughout this paper, the gauge group is taken to be the universal cover, i.e. all representation of (probe) matter fields are permitted.
  
In Section \ref{symnonp}, we discuss the perturbative and nonperturbative physics of SYM, including a small gaugino mass. 

In Section \ref{moninst}, we begin by listing the semiclassical objects---the fundamental and ``twisted" \cite{Lee:1997vp,Lee:1998bb,Kraan:1998pm} (or ``Kaluza-Klein") monopole-instantons---which contribute to the superpotential  \cite{Seiberg:1996nz,Aharony:1997bx,Davies:1999uw}. In Section \ref{quantumkahlervertex}, we study the quantum corrected monopole-instanton vertex. Details of the calculation are presented in   Appendix \ref{monopoledets}. We then generalize the tree-level linear-chiral supermultiplet duality of Section \ref{treeduality} to include loop corrections and use it to calculate the one-loop contributions to the moduli space metric.

 In Section \ref{bionpotential0}, with all ingredients in place, we present the final form of the neutral and magnetic bion induced potential, Eq.~(\ref{bionpotential2}), valid for all gauge groups and find the supersymmetric ground state, which, in all cases,  preserves center symmetry. 
 The physics of the loop-corrected moduli space metric at the center symmetric point and the subtleties of the large-$N$ limit are discussed in Section \ref{largencoupling}. A nonzero gaugino mass is introduced in Section \ref{softly}, where the monopole-instanton contribution to the potential, Eq.~(\ref{monopolepotential2}), is obtained.  The perturbative contribution to $V_{GPY}$ due to the massive gaugino is calculated in Appendix \ref{gauginoGPY}. It is argued, in Section \ref{softly}, to be suppressed in the semiclassically calculable regime with respect to the monopole- and bion-induced potentials.
 
   This completes the part of the paper devoted to deriving $V(\Omega)$.
  
   \subsubsection{The phase structure and $\mathbf{\theta}$-dependence for general gauge groups }
   
   The second part of the paper  studies the phase structure, as well as the behavior of the Polyakov loop and its two-point function, using the potential $V(\Omega)$ found in the first part of the paper. 
   We study $Sp(2N)$, $Spin(N)$, $G_2$, $F_4$, and $E_{6,7,8}$, thus completing the list of simple Lie groups left out from  \cite{Poppitz:2012sw,Poppitz:2012nz}.
 
 We begin in Section \ref{results41}, where we give the expressions for the Polyakov loop and its two-point function, and use it to define the string tension  in the center symmetric phase (some details are relegated to Appendix \ref{polyakovloop}).
   
   The symplectic $Sp(2N)$ and   $Spin(N)$  groups are studied in Section \ref{results42}. The $Sp(2 N)$ and $Spin(N)$, $N$-odd, groups have $\Z_2$ centers, where universality arguments would suggest that the deconfinement transition, if continuous, is in the same universality class as the one in $SU(2)$ (the 3d Ising model). We find that the quantum phase transition in softly-broken SYM is  first order for all groups besides $SU(2)$. This is in accord with the available lattice studies, only performed for $Sp(4)$ \cite{Holland:2003kg}, and  functional renormalization group arguments, made for $Sp(4)$ and $E_7$ in \cite{Braun:2010cy,Haas:2013qwp}. The general parameterizations of the Polyakov loop potential of \cite{Dumitru:2012fw} have also been argued to be able to account for the discontinuous transition. The small-$\S^3 \times \S^1_\beta$ studies of $SU(N)$  \cite{Aharony:2003sx}  were also performed for general groups in \cite{HoyosBadajoz:2007ds} and a  first-order large-$N$ transition was  found.
   
   The authors of \cite{Holland:2003kg} conjecture that the transition is first order for all gauge groups apart from $SU(2)$, arguing that this is due to the large difference between the number of the thermally excited   degrees of freedom below and above the deconfinement transition for groups of higher rank. Our findings are consistent with this observation. While the quantum center-breaking transition in SYM$^*$ occurs in the abelian regime, without any change in the number of degrees of freedom, the high rank of the $\Z_2$-center gauge groups  exhibiting a discontinuous transition  is reflected in the complexity of $V(\Omega)$. For $SU(2)$, there is only  single field $b$ (describing $\Omega$ of rank unity, see Section \ref{abelianization}) but in general groups there are $r$=rank($G$) fields $(b^1, ... b^r)$. As we shall see later, cubic terms in the potential $V(\Omega)$  due to neutral and magnetic bions naturally appear for higher rank groups, indicating that a first order transition is likely.
   
\begin{figure}[h]
\centering  
\includegraphics[width=.4 \textwidth]{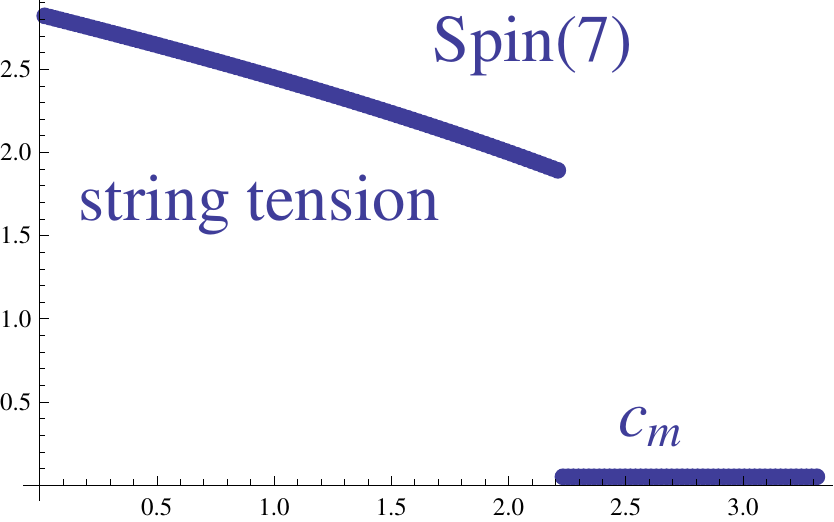}
\caption{\label{fig:string}The discontinuous change of the string tension (in units of $m_0 R^{-1}$, see Section \ref{bionpotential0} for a definition of the scales and Section \ref{results41} for more details) for probes in the spinor representation of $Spin(7)$  as a function of $c_m \sim {m\over L^2 \Lambda^3}$ .}
\end{figure}

   We also study  the  string tension  as a function of $c_m$. For one representative case, $Spin(7)$, its behavior is shown on Fig.~\ref{fig:string}: a discontinuous jump  to zero at $c_{\rm cr}$, as  in lattice simulations of thermal pure YM, is seen. This is the behavior found in all cases with center symmetry.

The Polyakov loop discontinuity and string tensions for high-rank groups, up to $Sp(12)$ and $Spin(16)$, are also calculated. We use these results to study the abelian large-$N$ limit. We show that the  discontinuity of $\langle\Tr \Omega\rangle$ for $Sp(2N)$, normalized to unity, appears to have a large-$N$ limit similar to the one found for $SU(N)$ \cite{Poppitz:2012nz}. This is consistent with the large-$N$ orbifold equivalence \cite{Lovelace:1982hz,Unsal:2006pj}, see also \cite{Bershadsky:1998cb} for a discussion in perturbation theory (we warn against numerical comparisons for such small ranks, see Section \ref{results42}). On Fig.~\ref{fig:speigenvalues}, we showed an example of the change of the eigenvalue distribution for the Polyakov loop in the fundamental of $Sp(12)$.

   The exceptional groups with center symmetry, $E_6$ and $E_7$, are studied in Section \ref{results43}. We find, respectively, a discontinuous transition associated with $\Z_3$ and $\Z_2$ breaking at some $c_{\rm cr}$.
   The exceptional groups without center symmetry, $G_2, F_4$ and $E_8$ are studied in Section \ref{results44}.
   The results for $F_4$ and $E_8$ are similar to the $G_2$ case already studied in \cite{Poppitz:2012nz}. In each case, there is a discontinuous behavior of the eigenvalues of $\Omega$,  as already mentioned, seen on the lattice for  thermal   $G_2$ YM theory
\cite{Pepe:2006er,Cossu:2007dk}.

\begin{figure}[h]
\centering 
\includegraphics[width=.37 \textwidth]{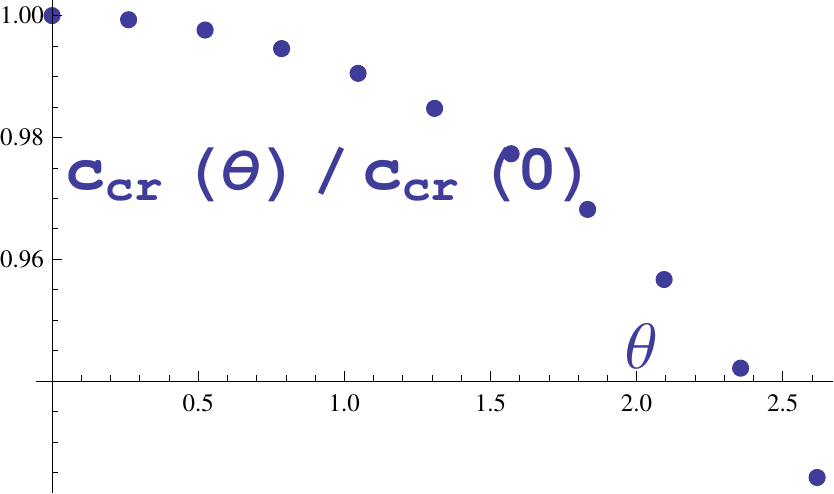}
\hfill
\includegraphics[width=.37 \textwidth]{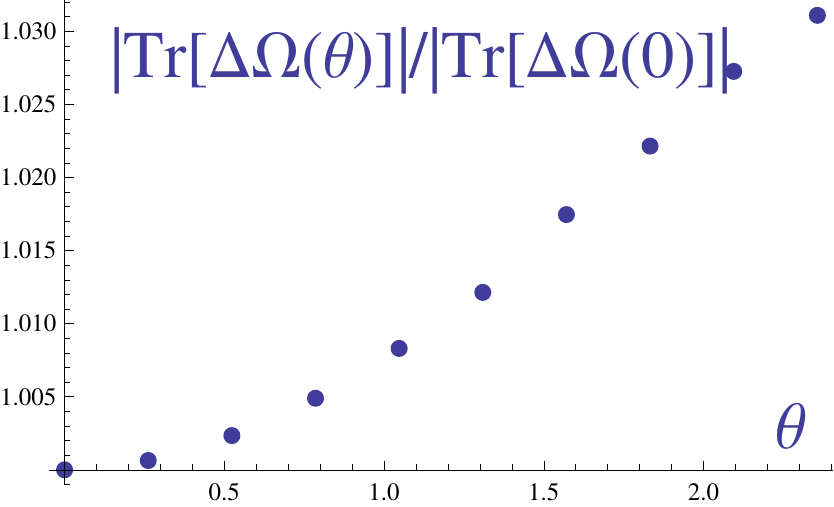}
\caption{\label{fig:SO6 c and omega}  An illustration, for  $Spin(6)$ gauge group,  of the $\theta$-dependence of the normalized critical transition mass, $ c_m$, $c_{\rm \small cr}(\theta) \over c_{\rm \small cr}(\theta=0)$, and the normalized Polyakov loop discontinuity $|{\rm \small Tr}\langle \Delta\Omega(\theta)\rangle|\over |{\rm \small Tr}\langle \Delta\Omega(0)\rangle|$ for $0 < \theta < {10 \over 12}\pi $. The right panel is for the spinor-representation Polyakov loop. The behavior is qualitatively similar for all gauge groups. It has been recently observed in lattice simulations for $SU(N)$ \cite{D'Elia:2012vv, D'Elia:2013eua}.}
\end{figure}

Finally, the $\theta$-angle dependence of  $c_{\rm cr}$ and of the discontinuity of the Polyakov loop, $|\Tr \Delta \Omega|$, is studied in a few representative cases for each gauge group.\footnote{Except for $E_7$ and $E_8$, where extracting the $\theta$-dependence is numerically challenging due to the high rank.} In each case it is found that, for $0< \theta <\pi$, $c_{\rm cr}(\theta) < c_{\rm cr}(0)$ and that $|\Tr \Delta \Omega(\theta)|>|\Tr \Delta \Omega(0)|$. The  $\theta$-dependence of $c_{\rm cr}$ and the discontinuity of the Polyakov loop is illustrated on Fig.~\ref{fig:SO6 c and omega} for the $Spin(6)$ case. 

We note that  \cite{D'Elia:2012vv} gave theoretical arguments for the expected behavior of $T_c$ as a function of $\theta$ using large-$N$ arguments, which was confirmed by their results. 
 A somewhat different-flavor (based on topological excitations and topological interference) argument for the behavior of $T_c$ as a function of $\theta$ for thermal transitions in deformed YM theory on $\R^2\times \S^1_L \times \S^1_\beta$ was given in \cite{Unsal:2012zj}; the qualitative behavior also agrees with what we find here and with lattice observations. 
 
Finally, we mention another quantity whose $\theta$-dependence can be studied on the lattice: the string tension inferred from the Polyakov loop two-point correlator, like the one plotted on Fig.~\ref{fig:string}. From the results of \cite{Poppitz:2012nz}, it is easy to see that the string tension for $SU(2)$ decreases upon increase of $\theta$, but the dependence on the topological angle is weak, at least in the semiclassical regime,  where it is suppressed by additional powers of $g^2$. The $SU(N )$-results of \cite{Anber:2013sga} imply that $\theta$ dependence of the string tension is similarly suppressed. We have also checked that this is the case for $Sp(2N)$, but leave a detailed study for the future. Lattice studies of the string tension's topological angle dependence exist, see  \cite{DelDebbio:2006df} for a study in $SU(N)$, $N=3,4,6$ pure-YM theory. They have shown that it decreases upon increase  of $\theta$ and that the dependence is rather weak in all cases studies (and getting weaker upon increase of  $N$)---a behavior qualitatively similar to what we find. 

  \subsection{Outlook: similarity between the SYM$^\mathbf*$ transition and deconfinement---is it just a curiosity?}
 \label{introfive}
 
A possible answer to the question is: ``It may be."
However, we believe, but can not prove, that there is more. 
 The finding of this and earlier papers that the behavior of various quantities (the Polyakov loop trace and its correlator,  as illustrated on Figs.~\ref{fig:speigenvalues},\ref{fig:string},\ref{fig:SO6 c and omega}) is quite similar to their behavior in the thermal deconfinement transition studied on the lattice is only part of the story. The most interesting aspect of the calculable transition in SYM$^*$ on $\R^3 \times \S^1$ is that it clearly elucidates the nature and  role of the topological excitations contributing to $V(\Omega)$: the neutral bions, which stabilize center symmetry, and  the monopole-instantons, leading to its destabilization.
  The setup used in this paper has the advantage that these excitations are  unambiguously identified, due to the weakly-coupled nature of the small-$m,L$ dynamics (in particular, it requires no gauge fixing procedures  or model assumptions).  The price to pay is  the fact that this is a quantum and not a thermal transition and that there is an extra parameter, the gaugino mass, not present in thermal YM. The relation between  the SYM$^*$ and  thermal YM  transitions, while perhaps continuous, as on Fig.~\ref{fig:phase}, involves decoupling the gaugino, which entails a loss of theoretical control.

Nonetheless, we think that the description of the $\R^3 \times \S^1$ calculable transition in SYM$^*$ comes closer, in many aspects, to the  thermal YM transition---compared to previous analytic weak-coupling studies on $\S^3 \times \S^1_\beta$ or $\R^2 \times \S^1_L \times \S^1_\beta$. The small-$\S^3 \times \S^1_\beta$ transition \cite{Aharony:2003sx} is a 
finite-volume large-$N$ transition. The physics of confinement (the uniform spreading of eigenvalues of the Polyakov loop  at low $T$) is purely kinematical in origin, see Footnote~\ref{five}, while the center-symmetry destabilization at high $T$ is due to the perturbative GPY potential. In contrast, the $c_m < c_{\rm cr}$ phase of SYM$^*$ on $\R^3 \times \S^1$ is a genuine (albeit abelian) confining phase, where confinement is due to magnetic Debye screening \cite{Polyakov:1976fu} in the magnetic bion plasma \cite{Unsal:2007jx}. The mechanism of the center-breaking transition involves a rich dynamics due to the competition between  the bions responsible for center symmetry and confinement, magnetic monopole-instantons, and (at larger    gaugino masses) perturbative GPY contributions---all objects that can in principle be identified in pure thermal-YM.
The other calculable example, the small-$L$ transition in ``deformed" YM \cite{Simic:2010sv} theory  on   $\R^2 \times \S^1_L \times \S^1_\beta$, on the other hand, is a genuine infinite volume thermal phase transition, albeit in two dimensions. The physics of the two-dimensional electric-magnetic Coulomb gases that describe  this transition is also quite rich  \cite{Anber:2011gn,Anber:2012ig,Anber:2013doa}.
However, the transition in ``deformed" YM $SU(N)$ theory  is  continuous, in the two-dimensional $\Z_{N}$ universality class \cite{Lecheminant:2006hj}, in contrast with the discontinuous transition observed on the lattice in the large-$L$ limit.  
    
Coming back to the SYM$^*$ studies of this paper, one clear and important lesson learned here and in \cite{Poppitz:2012sw,Poppitz:2012nz} is that ``topological" excitations  (whether carrying topological charge or ``molecules" of total topological charge zero) play a crucial role in the transition.\footnote{The role of topological excitations and $\theta$ dependence in deconfinement has been emphasized in \cite{Parnachev:2008fy}, however, the $\theta$ dependence of $T_c$ \cite{Unsal:2012zj,Poppitz:2012nz} or the discontinuity of the Polyakov loop \cite{Anber:2013sga} was not discussed there.} This is also evident from the fact that many quantities characterizing the deconfinement transition have $\theta$-angle dependence, now seen on the lattice \cite{D'Elia:2012vv,D'Elia:2013eua}. It is not clear to us how approaches which do not appear to invoke topology, i.e. the functional renormalization group \cite{Haas:2013qwp}, capture this dependence.

An interesting question to ask is whether the nature and role of various topological excitations\footnote{Some of the excitations considered here---``calorons" but not neutral or magnetic bions---have been used to model deconfinement  \cite{Diakonov:2007nv}. We will not describe these models here, see Section 4 of \cite{Poppitz:2012sw} for  discussion.}  in the  SYM$^*$ transition on $\R^3 \times \S^1$ can be used to gain some theoretical  understanding of deconfinement in pure\footnote{We do not even mention the realistic case of QCD with light quarks in this paper. The interplay between deconfinement and continuous chiral symmetry breaking is  complex and ill-understood, see \cite{Shuryak:2014gja} for a recent review. We have nothing to say about this outstanding problem here. One can certainly include massive quarks  in the setup studied in this paper. This was done in \cite{Poppitz:2013zqa} for   $SU(2)$ massive SQCD; when quarks are taken lighter than the dual photon mass, however, the semiclassical description breaks down.} YM theory. 
A step in this direction was recently made in \cite{Shuryak:2013tka}, where an instanton-monopole liquid model (see \cite{Schafer:1996wv} for a review of instanton-liquid models of the QCD vacuum) of the deconfinement transition in pure YM, incorporating the center-stabilizing neutral-bions first identified   in SYM${^*}$ was given.\footnote{We stress that all topological excitations discussed here, under the assumption of a constant expectation value for the holonomy, can be identified in pure YM theory as well, but a consistent semiclassical approximation is hard to argue for.} After fixing the monopole-instanton density (a model parameter) from lattice data on caloron densities, a comparison of the model's predictions  on electric and magnetic masses  with lattice data  yielded order-of-magnitude agreement. Thus these models appear to capture some of the physics of deconfinement, but we note that the quality of the lattice data used in the comparison can be improved. In general, the question to what extent one can use an abelian description to study deconfinement in pure YM theory deserves further study (it has been the subject of some discussion, see e.g.~\cite{Poppitz:2012sw} and \cite{Faccioli:2013ja,Shuryak:2014gja} for a  purist and  phenomenological   perspective, respectively).
 
Another interesting recent avenue of research may also be relevant. We will not dwell on the structure of neutral bions (which are so crucial for the center-stabilizing dynamics) in this paper. However, in earlier work   \cite{Poppitz:2011wy,Argyres:2012ka,Poppitz:2012sw}  it was argued that the existence of these topological ``molecules" is rather nontrivial to show  and requires the use of either supersymmetry or the ``Bogomolnyi-Zinn-Justin prescription." It was further argued  
 that,  ultimately, the necessity of including neutral bion contributions in the path integral can be traced to the divergence of the perturbative series and  that the center-stabilizing (or neutral) bions may be the semiclassical realization of 't Hooft's infrared renormalons \cite{Argyres:2012vv,Dunne:2012ae}.\footnote{There are many recent studies of ``resurgent series" in field theory, mostly in lower dimensional theories reduced to quantum mechanics by compactification. We recommend \cite{Cherman:2014ofa} for a recent review and references.} Thus, one   view on the continuity shown on Fig.~\ref{fig:phase} is that it is an example of ``resurgence" in action. In the most optimistic scenario,\footnote{See M. \" Unsal's talk at XQCD-2013  
at  \url{http://www.xqcd13.unibe.ch/downloads/Wed07_Unsal.pdf}.}  after a ``Borel-\' Ecalle resummation" of the semiclassical series at small $L$ and $m$ (not a small feat, as it involves all orders in perturbation theory,  in the semiclassical expansion, and in $m$) one is  led to results which can be extended over the entire phase diagram, including the large-$m$ strongly-coupled limit near the pure-YM deconfinement transition. While admittedly bold, we think this is a fascinating dream worthy of pursuit.

We finally note that  lattice simulations of SYM$^*$ at their current state of the art  (see \cite{Giedt:2008xm,Demmouche:2010sf,Bergner:2014saa} and  Section \ref{future}) might  be able to test whether the dashed line on the phase diagram shown on Fig.~\ref{fig:phase} can be replaced by a solid one.

\section{SYM on $\mathbf{\R^3 \times \S^1}$: Abelianization and tree-level duality }
\label{abelianization}

We consider ${\cal N}=1$ supersymmetric gluodynamics  (SYM) on $\R^3 \times \S^1$ with a compact gauge group $G$. 
The boundary conditions on the gauginos are supersymmetric. As usual in supersymmetric theories, the presence of flat directions, along with holomorphy and symmetries, greatly enables the study of the dynamics. While SYM theory on $\R^4$ does not have flat directions, once the theory is compactified on $\R^3 \times \S^1$, a perturbatively exact\footnote{As long as supersymmetry is manifest, no perturbative potential  generating a holonomy potential (i.e. lifting the flat direction) is generated. Upon adding a small gaugino mass, such a potential appears, but its strength is controlled by the smallness of the gaugino mass, see Appendix \ref{gpypotential}.} flat direction appears, corresponding to the freedom to turn on a nontrivial holonomy of the gauge field along the $\S^1$.
We perform our compactification along the third direction such that $x_3\sim x_3+2\pi R$.\footnote{Greek letters run over the three-dimensional space $\mu, \nu,... = 0,1,2$, while the Roman letters run over $m, n,... = 0,1,2,3$.  We also use the mostly positive signature $(-,+,+,+)$. } In gauge invariant terms, the flat direction can be described in terms of the expectation value of the eigenvalues of the Wilson loop operator $\Omega_{\cal R}(x^\mu)$ around the $\S^1$:
\begin{equation}
\label{wilsonloop1}
\Omega_{\cal R}(x^\mu) \equiv   P e^{ i \int_{\S^1} A_3(x^\mu, x^3) d x^3}~.
\end{equation}
 In thermal theories, the Wilson loop around the compact time direction  is also called the Polyakov loop and we shall adopt this name as well. 
 
 Along a generic flat direction all eigenvalues of $\langle \Omega_{\cal R}\rangle$  are distinct and the only gauge transformations commuting with $\langle \Omega_{\cal R}\rangle$ correspond to the maximal torus of the gauge group. From a three-dimensional perspective, the operator (\ref{wilsonloop1}) is a (compact) adjoint scalar field and one can think of the breaking of $G \rightarrow U(1)^r$  by $\langle \Omega_{\cal R}\rangle$ as due to  an adjoint Higgs field. 
Thus, for a generic expectation value for the holonomy, the gauge group is broken to $U(1)^r$, where $r$ is the rank of the group.
The long-distance degrees of freedom are those of a three dimensional $U(1)^r$ supersymmetric gauge theory without matter fields (since  the original theory is SYM, with only adjoint fields, no light fields charged under the unbroken $U(1)^r$ are present).

We shall now assume that the scale of the breaking  $G \rightarrow U(1)^r$ is large enough, i.e.   larger than the strong coupling scale, so that the gauge coupling is frozen at a small value and weak-coupling methods are applicable. Naturally, this assumption has to be justified a posteriori  in a self-consistent manner: as we shall see later in the paper, the mass of the lightest $W$-boson is self-consistently determined to be of order $1\over {c_2 R}$, where $c_2$ is the dual Coxeter number of the gauge group. Thus, for sufficiently small $c_2 R$, the assumed separation of scales holds.

We now proceed, assuming abelianization, to describe the dynamics of the theory at distances larger than $c_2 R$. As already explained, at large distances the only degrees of freedom are the $U(1)^r$ gauge fields and superpartners. 
The dimensionally reduced classical action of the theory can be obtained starting from the four dimensional vector multiplet $\pmb{\cal V}$ written in the Wess-Zumino gauge, where $\pmb v_m = (v_m^1,... v_m^r)$, $\pmb \zeta_\alpha = (\zeta_\alpha^1, ..., \zeta_\alpha^r)$  denote  the Cartan components of the gauge field  and gaugino, respectively%
\begin{eqnarray}
\pmb{\cal V}=-\theta \sigma^m \bar \theta \pmb v_m+i \theta^2\bar \theta\bar{\pmb\zeta}-i\bar\theta^2\theta \pmb \zeta+\frac{1}{2}\theta^2\bar \theta^2 \pmb D\,.
\label{V}
\end{eqnarray}
We stress again that the bold symbols used in this paper   denote $r$-dimensional quantities (in the example above, $\pmb{\cal V} = ({\cal V}_1,... {\cal V}_r)$ is used to denote the $r$ components of the massless $U(1)^r$ superfield).

Before proceeding to study the action of the unbroken $U(1)^r$ theory, we pause and mention the global symmetries of SYM on $\R^3 \times \S^1$. Their realization as a function of the size of $\S^1$ and soft mass parameter are our main interest here. SYM on $\R^4$ has a classical $U(1)$ chiral symmetry acting on the Weyl-fermion gaugino field. It is broken by the anomaly to a $\Z_{2 c_2}$ (this is most easily seen by noticing that the 't Hooft vertex in an instanton background involves $2 c_2$ zero modes). Notice that a $\Z_2$ subgroup of this discrete chiral symmetry acts on the gauginos as fermion number, hence only $\Z_{c_2}$ is a genuine chiral symmetry. In the softly broken theory the chiral symmetry is explicitly broken by the gaugino mass parameter.

In gauge theories with nontrivial center, there is an additional global symmetry when the theory is considered on 
$\R^3 \times \S^1$. In this paper, we only consider gauge theories where the gauge group is the covering group, i.e. all matter representations are allowed.
Among the groups we consider, the ones with nontrivial center, see Table~\ref{tab:lie},  are:  $Sp(2N)$ with a $\Z_2$ center,  $Spin(2 r+1)$, where the center is $\Z_2$, $Spin(2 r)$ where the center is $\Z_4$ for odd $r$ and $\Z_2 \times \Z_2$ for even $r$, as well as $E_6$ and $E_7$ which have a $\Z_3$ and $\Z_2$ center, respectively. In each of these cases, the observable that transforms under the center symmetry is a nontrivial line operator---the Polyakov loop (\ref{wilsonloop1}) in the defining representation (this is the fundamental representation of $Sp(2N)$, $E_6$ and $E_7$, and the spinor representation in the $Spin(N)$ case).

\subsection{Tree-level linear-chiral duality}
\label{treeduality}

In three dimensions, the gauge field dynamics of a supersymmetric theory with four supercharges is described by a real linear multiplet (instead of a chiral spinor superfield).
We define the three-dimensional real linear multiplet superfield $\pmb W$ (once again we stress that $\pmb W = (W^1, ..., W^r)$ denotes an $r$-vector comprised of the components along the Cartan subalgebra) in terms of (\ref{V}) as follows\footnote{The real linear multiplet and the subsequent linear-chiral duality transformation is described  in detail in, e.g. \cite{deBoer:1997kr,Intriligator:2013lca}, albeit in a basis intended for a three dimensional theory.  We give the details of the duality in a different basis, keeping  the connection to four dimensions  as manifest as possible: already our definition of the linear multiplet (\ref{W}), indicates that $x^3$ is the compact direction.}
\begin{eqnarray}
\pmb W=-\frac{1}{2}\bar \sigma^{3\dot\alpha \alpha}\bar D_{\dot\alpha}D_\alpha \pmb {\cal V}\,.
\label{W}
\end{eqnarray}
$\pmb W$ obeys the real ($\pmb W = \pmb W^\dagger$) and linear  ($D^2 \pmb W = \bar{D}^2 \pmb W = 0$) multiplet constraints and is invariant under three-dimensional gauge transformations, $\pmb {\cal{V}} \rightarrow \pmb {\cal{V}} + \pmb\Lambda + \pmb \Lambda^\dagger$, where $\pmb \Lambda$ is chiral and $x^3$-independent.
  From (\ref{W}) and (\ref{V}) we obtain
\begin{eqnarray}
\pmb W&=&\; \pmb v^3+\theta\pmb\lambda+\bar\theta\bar{\pmb\lambda}+\frac{1}{2}\theta\sigma_\kappa\bar\theta \epsilon^{\kappa\mu\nu}\pmb v_{\mu\nu} \nonumber  \\
&& + \theta\sigma^3 \bar\theta \pmb D+\frac{i}{2}\theta^2 \partial_\mu \pmb\lambda\sigma^\mu \bar \theta-\frac{i}{2}\bar\theta^2\theta \sigma^\mu \partial_\mu \bar{\pmb \lambda} -\frac{1}{4}\theta^2\bar\theta^2\Box\pmb v^3,
\label{wlinear}
\end{eqnarray}
where we have defined $\pmb \lambda^\alpha \equiv -i \bar{\pmb \zeta}_{\dot \alpha}\bar \sigma^{3 \dot\alpha \alpha}$ and $\bar{\pmb \lambda}^{\dot\alpha} = i \bar\sigma^{3 \; \dot\alpha \alpha} \pmb \zeta_\alpha$ (also note that $\epsilon^{012} = +1$, $\epsilon^{0123}=+1$, $\Box \equiv \eta^{\mu\nu} \partial_\mu \partial_\nu$). The dimensionally reduced action for the $U(1)^r$ components of the gauge supermultiplet  is given by
\begin{eqnarray}
\nonumber
S&=&\frac{2\pi R}{g^2}\int d^3x d^4 \theta \left[-2\pmb W^2\right]\\
&=&\frac{2\pi R}{g^2}\int d^3 x\left\{ -\partial_\mu \pmb v^3\cdot\partial^\mu \pmb v^3-\frac{1}{2}\pmb v_{\mu\nu}^2-i2\pmb {\bar\lambda}\cdot \bar\sigma^{\mu}\partial_\mu \pmb \lambda +  {\pmb D}^2 \right\}\,,
\label{the action as integral over WS}
\end{eqnarray}
where $\pmb v_{\mu\nu}=\partial_\mu \pmb v_\nu-\partial_\nu \pmb v_\mu$ and $\pmb W^2 \equiv  \pmb W \cdot \pmb W =  \sum_{i=1}^r W_i^2$.\footnote{Notice that the normalization in (\ref{the action as integral over WS})   differs by a factor of $1/2$ compared to the standard normalization. However, this normalization is arbitrary since any prefactor can be absorbed in the normalization of $g$. } In this Section, we do not yet precisely specify the scale the coupling $g^2$ is taken at and only note that it should be taken larger than the strong coupling scale in order for the abelian weakly-coupled description to hold. 
Next, we define the three dimensional scalar field $\pmb \phi$
 \begin{equation}
 \pmb v^{3}\equiv\frac{\pmb \phi}{2\pi R},
 \label{phidefinition}
\end{equation}
in terms of which,  we have, from now on neglecting the vanishing $\pmb D$ term,\begin{eqnarray}
S=\frac{2\pi R}{g^2}\int d^3 x\left\{ -\frac{\partial_\mu \pmb \phi\cdot\partial^\mu \pmb \phi}{4\pi^2R^2}-\frac{1}{2}\pmb v_{\mu\nu}^2-i2\pmb {\bar\lambda}\cdot \bar\sigma^{\mu}\partial_\mu \pmb \lambda\right\} .
\label{main Lagrangian with chern term}
\end{eqnarray} 

Let us first  remind the reader how one can describe the abelian dynamics via  a dual description of the three dimensional photon in non-supersymmetric terms. One 
 introduces an auxiliary Lagrangian
\begin{eqnarray}
S_{\mbox{\scriptsize aux}}=\frac{1}{4\pi}\int d^3x \epsilon_{\mu\nu\rho}\partial^\mu \pmb \sigma\cdot \pmb {v^{\nu\rho}}\,,
\label{auxiliary S}
\end{eqnarray}
and regards  $\pmb v^{\nu\rho}$ as an arbitrary two-form field. 
Varying $S_{\mbox{\scriptsize aux}}$ with respect to $\pmb \sigma$ enforces  the Bianchi identity $\epsilon_{\mu\nu\rho} \pmb v^{\nu\rho}=0$ on $\pmb v^{\nu\rho}$ and leads to the original gauge theory. On the other hand, by varying  $S+S_{\mbox{\scriptsize aux}}$ with respect to $\pmb v^{\alpha\beta}$, one   finds
\begin{eqnarray}
\pmb v^{\nu\rho}=\frac{g^2}{8\pi^2R} \;\partial_\mu \pmb \sigma \;\epsilon^{\mu\nu\rho}~.
\label{v in terms of sigma}
\end{eqnarray}
Substituting (\ref{v in terms of sigma}) into  $S+S_{\mbox{\scriptsize aux}}$ one obtains the dual lagrangian in terms of the dual photon field $\pmb \sigma$:  
\begin{eqnarray}
S\rightarrow S+S_{\mbox{\scriptsize aux}}=\frac{1}{2\pi R}\int d^3 x \left\{-\frac{1}{g^2}\left(\partial_\mu \pmb \phi\right)^2-\frac{g^2}{16\pi^2}\left(\partial_\mu \pmb\sigma \right)^2-\frac{i2(2\pi R)^2}{g^2}\pmb {\bar\lambda}\cdot \bar\sigma^{\mu}\partial_\mu \pmb \lambda  \right\}\,.
\label{total action of the system}
\end{eqnarray}

Now, we shall perform the duality transformation in superspace. To this end, 
we introduce a chiral multiplet $\pmb X$, with component expansion
\begin{eqnarray}
\pmb X=\pmb z+\sqrt{2}\theta\pmb\psi +\theta^2 \pmb F + i\theta \sigma^\mu\bar \theta \partial_\mu \pmb z-\frac{i}{\sqrt{2}}\theta^2\partial_\mu \pmb \psi\sigma^\mu \bar \theta+\frac{1}{4}\theta^2\bar\theta^2\Box \pmb z,
\label{def chiral field X}
\end{eqnarray}
and introduce the superspace generalization of (\ref{auxiliary S}):
\begin{eqnarray}
S_{\mbox{\scriptsize aux SUSY}}= - \frac{1}{2\pi}\int d^3x d^4 \theta \; \pmb W \cdot \left( \pmb X + \pmb X^\dagger \right),
\label{auxiliary S SUSY}
\end{eqnarray}
which is added to   $S$  of Eqn.~(\ref{the action as integral over WS}). As in the non-supersymmetric case,   now $\pmb W$ is regarded as an unconstrained real multiplet, i.e. $\pmb W^\dagger = \pmb W$,  but without the linearity constraint. The superspace expansion of the unconstrained real superfield $\pmb W$  can be chosen as (notice that the choice of the higher components made below is different from the one in \cite{Wess:1992cp}, for obvious reasons)\footnote{Strictly speaking, $\pmb W$ here should be denoted by a different symbol. In order to not clutter notation, we shall not do so here (but   we use $\pmb Y$ in Section \ref{quantumkahlervertex}, when we study the duality with loop corrections to the moduli space metric included), hoping this will not cause confusion.} 
\begin{eqnarray}
\label{w1}
\pmb W&=&\pmb v^3+ \theta \pmb \lambda + \bar\theta \bar{\pmb \lambda}+\frac{i}{2}\theta^2 \left(\pmb M+i\pmb N\right)-\frac{i}{2}\bar\theta^2 \left(\pmb M-i\pmb N\right)-\theta \sigma^m  \bar\theta \pmb E_m\\
&& \nonumber
+\frac{i}{2}\theta^2( \bar{\pmb \eta} \bar\theta + \partial_\mu \pmb\lambda\sigma^\mu \bar \theta)-\frac{i}{2}\bar\theta^2(\theta {\pmb \eta}+ \theta \sigma^\mu \partial_\mu \bar{\pmb \lambda})
+\frac{1}{2}\theta^2\bar\theta^2\left[\pmb P-\frac{1}{2}\Box\pmb v^3 \right]\,.
\end{eqnarray}
Varying (\ref{auxiliary S SUSY})
with respect to the chiral superfields $\pmb X$ and $\pmb X^\dagger$ imposes the linear multiplet constraints  $D^2 \pmb W = \bar{D}^2 \pmb W = 0$. They are  solved by (\ref{W}) and  lead back via (\ref{wlinear}--\ref{phidefinition}) to  (\ref{main Lagrangian with chern term}); in fact the component expansion   (\ref{w1}) of the unconstrained $\pmb W$ is chosen such that upon imposing linearity constraints it reduces to (\ref{wlinear}).

 Instead, we are interested in varying  $S+ S_{\mbox{\scriptsize aux SUSY}}$, the sum of Eqns.~(\ref{main Lagrangian with chern term},\ref{auxiliary S SUSY}), with respect to $\pmb W$, which gives:
\begin{equation}
\label{XviaW}
\pmb W = - {g^2 \over 16 \pi^2 R} \left( \pmb X + \pmb X^\dagger \right)~.
\end{equation}
In components, this relation reads
\begin{eqnarray}
\pmb v^3&=&- {g^2 \over 16 \pi^2 R} (\pmb z+\pmb z^\dagger),~ \pmb E_\mu= i   {g^2 \over 16 \pi^2 R} \partial_\mu\left( \pmb z-\pmb z^\dagger\right), ~\pmb \lambda=   {g^2\over 8  \sqrt{2}\pi^2 R} \pmb \psi\,, \nonumber \\
 \pmb M+i\pmb N&=&  i {g^2 \over 8 \pi^2 R} \pmb F,~\pmb P=\pmb \eta= \pmb E_3 = 0\,.  \label{equality between X and Y}
\end{eqnarray}
Now,  recalling that the lowest component, denoted by   $\pmb W\big\vert$, of $\pmb W\big\vert=\pmb v^3$ from (\ref{wlinear}, \ref{w1}) and the redefinition (\ref{phidefinition}), we find that the (tree-level) relation between the lowest component of the chiral superfield $\pmb X$ and the gauge-field holonomy around $\S^1$ is:
\begin{eqnarray}
\label{zphirelation}
\pmb z = - {4 \pi \over g^2} \pmb \phi + i \pmb \sigma \equiv i \tau \pmb \phi + i \pmb \sigma~.
\end{eqnarray}
In the second equality above,  we introduced the usual form of the tau-paramter (with  the vacuum $\theta$-angle to zero; in the softly broken theory it is  re-introduced through the phase of the gaugino mass parameter):
\begin{eqnarray}
\tau \equiv i {4 \pi^2 \over g^2} + {\theta \over 2 \pi} \bigg\vert_{\theta =0}~.
\end{eqnarray}
We also denoted by $\pmb \sigma$ the imaginary part of the lowest component of $\pmb X\big\vert=\pmb z$---it is easily seen that the non-supersymmetric duality relation  (\ref{v in terms of sigma}) is contained in (\ref{equality between X and Y}).
Finally, substituting (\ref{XviaW}) into  $S+ S_{\mbox{\scriptsize aux SUSY}}$, we find the tree-level action of the dual chiral superfield:
\begin{eqnarray}
S_{dual}=\frac{g^2}{64 \pi^3 R}\int d^3x d^4\theta \left(\pmb X+\pmb X^\dagger \right)^2
\label{susy dual}
\end{eqnarray}
Both (\ref{the action as integral over WS}) and  the dual action (\ref{susy dual}) are subject to  perturbative and nonperturbative quantum corrections. These will be important and considered in subsequent sections.

\subsection{The fundamental domain  of the moduli fields}
\label{fundamentaldomain}

Before we begin to 
 study the ground states and realization of symmetries of SYM on $\R^3 \times \S^1$, it is important to understand the range of the physically inequivalent values of the fields in the long-distance theory whose tree-level action is (\ref{susy dual}). Both $\pmb \phi$ and $\pmb \sigma$ are compact variables,   whose expectation values determine the possible ground state(s) of the theory. Their periodicities, and thus the possible inequivalent grounds states, are determined by the global properties of the gauge group as we now describe.

\subsubsection{The holonomy and the Weyl chamber}
\label{holonomyperiod}

We begin with the holonomy.
Recall the relation of the modulus field $\pmb \phi$ and the holonomy (\ref{phidefinition})  which we  rewrite as $\pmb \phi = 2 \pi R \pmb v^3$. Gauge transformations act on the gauge field $v_m$ as $v_m \rightarrow U_{\cal R}^\dagger v_m U_{\cal R} + i \partial_m U_{\cal R}^\dagger U_{\cal R}$. Here $U_{\cal R}$ is a gauge group element  in a representation ${\cal R}$. Of special interest for determining the periodicity of the holonomy are the large gauge transformations:
\begin{equation}
U_{\cal R}(x_3, \pmb a) = e^{ i {x_3 \over R} \pmb a \cdot H_{\cal R}}, ~~ U_{\cal R}(x_3 + 2 \pi R, \pmb a) = e^{ i 2 \pi \pmb a \cdot  H_{\cal R}} U(x_3, \pmb a)_{\cal R}~,
\label{largegauge}
\end{equation}
where $H_{\cal R}$ are the Cartan generators in the representation $\cal R$ and $\pmb a \cdot H_{\cal R}$ denotes a sum over the $r$ Cartan generators.
The transformations  (\ref{largegauge}) 
act on $v_3$ by a shift, $v_3 \rightarrow v_3 + {  \pmb a \cdot   H \over R}$, and, clearly,  generate  shifts on $\pmb \phi$:
\begin{equation}
\label{holonomyshift}
  \pmb \phi \rightarrow \pmb \phi + 2 \pi \pmb a .
 \end{equation}
 Thus, values of the modulus field $\pmb \phi$ differing by $2 \pi \pmb a$ are physically (gauge) equivalent.  
The possible values of $\pmb a$ depend  on the global properties of the gauge group. Notice that these have to be treated carefully \cite{Aharony:2013hda}. In this paper, we shall only consider the case where the gauge group is the covering group. For example, for orthogonal groups, this means that the gauge group is $Spin(N)$ rather than $SO(N)$, so that  spinor representations are allowed. When all representations of the covering group are allowed,  the eigenvalues of $H_{\cal R}$ are in the weight lattice, i.e. they are integer linear combinations of the fundamental weights $\pmb w_a$, $a=1,...r$. 
If a representation $\cal R$ is permitted, for example as a massive probe,  gauge transformations in that representation have to be periodic around the $\S^1$ to avoid introducing discontinuities (in the form of multivalued fields, see e.g. the discussion in \cite{Argyres:2012ka}).

Periodicity of the  large gauge transformations (\ref{largegauge})  thus requires that the eigenvalues of $e^{ i 2 \pi \pmb a \cdot  H_{\cal R}}$ should equal unity for all $\cal R$. Since all $\cal R$ are allowed, it must be that $e^{ i 2 \pi \pmb a \cdot \pmb w_a} = 1$ for any weight $\pmb w_a$. Thus, since $\pmb w_a \cdot \pmb \alpha^*_b = \delta_{ab}$, the allowed $\pmb a$ are spanned by the dual root lattice, $\pmb a = \sum\limits_{a=1}^r n_a \pmb \alpha^*_a$. We conclude that when the gauge group is the covering group, we have that
\begin{equation}
\label{holonomyperiodicity}
  \pmb \phi \equiv  \pmb \phi + 2 \pi \pmb \alpha^*_a, \; a = 1, ... r,
 \end{equation}
i.e. the holonomy modulus is periodic, with periodicity in the dual root lattice (notice that  the periodicity (\ref{holonomyperiodicity})  is different from  \cite{Davies:1999uw}, which argued for periodicity in the dual weight lattice instead, in effect assuming that only electric charges   in the root lattice are permitted). 

To end the discussion of the $\pmb \phi$ fundamental cell, we note that there are further restrictions on the $\pmb \phi$ moduli, in addition to the large gauge identifications (\ref{holonomyperiodicity}). These come from further discrete identifications (Weyl reflections) acting on the fundamental cell of the  dual root lattice. The resulting fundamental region of the moduli space  $\pmb \phi$ on $\R^3 \times \S^1$ is called the ``Weyl chamber". It is described, for all gauge groups,  in Appendix B of \cite{Argyres:2012ka}.\footnote{One can also  think of the Weyl chamber as the smallest region in $\pmb \phi$-space  such that no massless $W$-bosons, including  any Kaluza-Klein modes (apart from the ones corresponding to the zero roots) appear for any values of $\pmb \phi$ away from the region's boundary. This follows by carefully studying the $W$-boson spectrum, which is given, see Section \ref{determinant}, by $|{p\over R} + { \pmb\beta \cdot \pmb \phi  \over 2 \pi R}|$, where $p$ is any  integer and $\pmb \beta$ is any root; showing that this leads to (\ref{weylchamber}) is left as an exercise.}
The upshot is that the Weyl chamber, or the region of physically inequivalent values of $\pmb \phi$,  is given by the $\pmb \phi$ obeying the inequalities \begin{equation}
\label{weylchamber}
\pmb \alpha_a \cdot \pmb \phi > 0, a = 1,...,r, \; {\rm and } \; - \pmb \alpha_0 \cdot \pmb \phi < 2 \pi~.
\end{equation}
Here $\pmb \alpha_0$ is the affine (lowest) root. The Weyl chamber of $\pmb \phi$ can be geometrically  described as the region in an $r$-dimensional space, which is the inside of the volume whose  boundary is given by the  $r+1$ hyperplanes  ($\pmb \alpha_a \cdot \pmb \phi=0, a=1,...r$, and $\pmb \alpha_0 \cdot \pmb \phi =- 2 \pi$)---a triangle for $r=2$, a tetrahedron for $r=3$, etc.
 As already mentioned, whenever one of the  $r+1$ inequalities (\ref{weylchamber}) 
 becomes an equality, massless nonabelian gauge bosons appear, rendering the abelian description of the theory invalid.
 
 \subsubsection{Periodicity of the dual photon fields}
\label{dualphotonperiod}
Let us now discuss the periodicity of the dual photon fields $\pmb \sigma$.
We shall determine the fundamental period of the dual photon field $\pmb \sigma$ by further  compactifying the large spatial directions of $\R^3 \times \S^1$, denoted by $x,y$, over a two-torus $\mathbb T_2$.  The periodicity of the dual photon fields then follows from magnetic flux quantization on $\mathbb T_2$. Similar to the periodicity of the holonomy, the global properties of the gauge group affect the result by restricting the sets of allowed electric probes. When the gauge group is the covering group, the dual photons $\pmb \sigma$ have periodicity in the  weight lattice $\pmb w_a$:
\begin{eqnarray}
\label{sigmaperiod}
\pmb \sigma \equiv \pmb \sigma + 2 \pi \pmb w_a~, \; a = 1, ...,r.
\end{eqnarray}
The rest of this Section presents a pedestrian derivation of this fact and the familiar reader can immediately proceed further. 
 
 Consider first the Wilson loop for a single abelian gauge field, given by:
\begin{eqnarray}
{\cal W}_{C}=\exp\left[i \oint_C  dl A\right]\,,
\label{Wilson loop in 3D}
\end{eqnarray} 
where the contour $C$ lies in the $x-y$ plane, or in other words on the two-torus surface. Using Stokes' theorem, the line integral above can be written as
\begin{eqnarray}
{\cal W}_{C}=\exp\left[i \oint_C  dl_\mu A^\mu\right]=\exp\left[i\int_{\Sigma \subset \mathbb T_2} ds B^3\right]=\exp\left[-i\int_{\Sigma_o} ds B^3\right]\,,
\label{wilson1}
\end{eqnarray} 
where $B^3=v^{12}$ is the magnetic field in $2+1$ D, $\Sigma$ is the interior surface enclosed by $C$, while $\Sigma_o$ is the exterior or complementary surface, i.e. $\Sigma_o=\mathbb T_2-\Sigma$. The last equality results from the fact that the line integral is equivalent, by Stokes' theorem, to the integral over the internal and external areas enclosed by the loop. From the last equality in (\ref{wilson1}) we infer the Dirac (flux) quantization condition
\begin{eqnarray}
\exp\left[i\int_{\mathbb T_2} ds B^3\right]=1\,,\mbox{or}\quad \int_{\mathbb T_2} ds B^3=2\pi n\,, \quad n \in \Z\,.
\label{quantization condition}
\end{eqnarray}

In our case of multiple abelian fields, the Wilson loop (\ref{Wilson loop in 3D}) measures the magnetic field probed by an electric charge that belongs to a representation ${\cal R}$. Thus,   we promote $B^3$ (as well as $A^\mu$) to a matrix in the Cartan subalgebra: 
\begin{eqnarray}
B^3= \pmb B^3 \cdot  H_{{\cal R}}\,,
\end{eqnarray}
where $H_{{\cal R} }$ are the Cartan generators of the group in the representation ${\cal R}$,  and note that every eigenvalue should obey the quantization condition (\ref{quantization condition}).  Using (\ref{v in terms of sigma}) the magnetic field $\pmb B$ on the torus can be expressed in terms of the dual photon  field  $\pmb \sigma$:
\begin{eqnarray}
\pmb B^3=\pmb v^{12}=\frac{g^2}{8\pi^2 R} \; \dot{\pmb \sigma} ,
\label{B sigma and phi}
\end{eqnarray}
where $\dot{\pmb \sigma}\equiv \partial_t {\pmb \sigma}$. In addition, since we have a compact space, $\mathbb T_2$, we can ignore all higher modes of the fields in (\ref{total action of the system}) and  keep only the zero modes. In what follows, we shall only consider the  $\pmb\sigma$ field, whose zero mode is denoted by $\pmb\sigma_0$. Its action takes the form
($A_{\mathbb T_2}$ is the area of the torus)
\begin{eqnarray}
S=\frac{A_{\mathbb T_2}}{2\pi R}\int dt  \frac{g^2}{16\pi^2} \dot{\pmb \sigma_0}^2\,,
\label{zero mode action}
\end{eqnarray}  
and the equation  of motion is solved by
\begin{eqnarray}
\nonumber
\dot{\pmb \sigma_0} &=&\pmb U\,,
\end{eqnarray}
where $\pmb U$ is a constant  of motion. From (\ref{B sigma and phi}) we find
\begin{eqnarray}
\pmb B^3=\frac{g^2}{8\pi^2 R}\pmb U\,.
\end{eqnarray} 
 The allowed values of $\pmb U$ are determined using the Dirac quantization condition (\ref{quantization condition}):
\begin{eqnarray}
\frac{g^2 A_{\mathbb T_2}}{8\pi^2 R}\;\pmb U\cdot  H_{{\cal R}}=2\pi (n_1, ... ,n_{{\rm dim}{\cal R} }) \,.
\label{3fr}
\end{eqnarray}

As already stated, we   restrict our attention to the case when the gauge group is the covering group, i.e. all   representations  are allowed. Thus, as before, the eigenvalues of $H_{{\cal R}}$ lie in the weight lattice spanned by the fundamental weights $\pmb w_a$. Since for any weight and any root vector  $\pmb\beta$, see Appendix \ref{groupsummary},
\begin{eqnarray}
\frac{\pmb w_a \cdot \pmb\beta}{\pmb \beta^2}=\frac{n}{2}\,,\quad n \in \Z\,,
\end{eqnarray}
the solution of (\ref{3fr}) for $U$ can be found in terms of root vectors. 
 Since any root $\pmb \beta$ can be written as a superposition of the simple roots $\pmb \alpha$, we find that any $\pmb U$ obeying (\ref{3fr})  can be written as a integer linear combination of simple co-roots (recall that $\pmb \alpha^* = {2 \pmb \alpha \over  \pmb\alpha^2}$, see (\ref{coroots})): 
\begin{eqnarray}
\pmb U  = \frac{16\pi^3 R}{g^2A_{\mathbb T_2}}\sum_{i=1}^{r}n_a \pmb\alpha^*_a\,,\; \mbox{thus} ~~
\pmb B^3 = \frac{2\pi}{A_{\mathbb T_2}}\sum_{a=1}^{r}n_a \pmb\alpha^*_a\,,
\label{quantized B}
\end{eqnarray}
where $\pmb\alpha^*$ are the co-roots, and  $n_a$ are integers.

Now we go back to the action  (\ref{total action of the system}) and 
  find the  momenta conjugated to the fields $\pmb\sigma$ 
\begin{eqnarray}
\nonumber
\pmb \Pi_\sigma = \frac{\delta S}{\delta \dot{\pmb \sigma}}=\frac{g^2 }{16 \pi^3R} \; \dot{\pmb \sigma}  . \label{momenta}
\end{eqnarray}
When the $x$-$y$ plane is compactified over the two-torus we use the flux quantization condition (\ref{quantized B}) and the relation of $\pmb B$ and $\dot{\pmb \sigma}$ (\ref{B sigma and phi}) to find that the momentum of   $\pmb \sigma_0$ is quantized:
\begin{eqnarray}
\label{momentum}
\pmb \Pi_{\sigma_0}&=&\frac{g^2  A_{\mathbb T_2} }{16 \pi^3R} \dot{\pmb \sigma}_0 =\frac{g^2 A_{\mathbb T_2}}{16 \pi^3R}\pmb U=\frac{A_{\mathbb T_2}}{2\pi}\pmb B^3=\sum_{a=1}^{r}n_a\pmb\alpha^*_a\;, \end{eqnarray}
which already shows that $\pmb \sigma_0$ is  a compact variable. To find the period, note that 
the Hamiltonian of the $\pmb \sigma_0$ field is 
\begin{eqnarray}
H=\frac{8\pi^3 R}{g^2 A_{\mathbb T_2}}\pmb{\Pi}_{\sigma_0}^2  
\end{eqnarray}
and that, using (\ref{momentum}),   its energy   is
\begin{eqnarray}
H_{\sigma_0}=\frac{8\pi^3 R }{g^2A_{\mathbb T_2}}\pmb \Pi_{\sigma_0}\cdot\pmb \Pi_{\sigma_0}=\frac{8\pi^3 R }{g^2A_{\mathbb T_2}}\left[ \sum_{a=1}^r n_a\pmb\alpha_a^*\right]^2\,.
\label{energy for sigma}
\end{eqnarray}
This energy can also be obtained by promoting the field $\pmb\sigma_0$ to a quantum field (operator) $\pmb\sigma_0\rightarrow \hat{\pmb\sigma}_0$ and $\pmb\Pi_{\sigma_0}\rightarrow \pmb{\hat\Pi}_{\sigma_0}=-i\pmb{\partial}_{\sigma_0}$. Thus, the quantum mechanical Hamiltonian reads
\begin{eqnarray}
\hat H_{\sigma_0}=-\frac{8\pi^3 R }{g^2A_{\mathbb T_2}}\pmb{\partial}_{\sigma_0}\cdot \pmb{\partial}_{\sigma_0}\,.
\end{eqnarray}
The wave function of the state of  energy (\ref{energy for sigma}) is 
\begin{eqnarray}
\psi=\exp\left[i\pmb\sigma_0\cdot \sum_{i=1}^r n_a \pmb \alpha^*_a \right]\,.
\end{eqnarray}
The periodicity of $\pmb \sigma_0$ is determined by the shifts $\pmb\sigma_0\rightarrow \pmb\sigma_0+\pmb\Lambda$ under which $\psi$ remains single valued. Thus, we have $\pmb\Lambda =2\pi \pmb w_a$, $a=1,...,r$, since $\pmb\alpha_a^*\cdot \pmb w_b=\delta_{ab}$. The period can also be obtained by applying the Bohr-Sommerfeld quantization condition: $2\pi n=\int d\pmb \sigma \cdot \pmb\Pi_{\sigma_0}=\pmb \Lambda \cdot \pmb\Pi_{\sigma_0}=\sum_{a=1}^r n_a\pmb \Lambda \cdot \pmb\alpha_{a}^*$, which can be solved by demanding $\pmb \Lambda=2\pi \pmb w_a$, $a=1,...,r$. 
To summarize, when the gauge group is the covering group, the dual photons $\pmb \sigma$ have periodicity in the  weight lattice $\pmb w_a$ as already declared in the beginning of this Section, see (\ref{sigmaperiod}).
 
\section{SYM on $\mathbf{\R^3 \times \S^1}$: Monopole-instantons, loops, and moduli-space metric}
\label{symnonp}

At small $\S^1$ size $R$ the physics in SYM is weakly coupled, as the gauge coupling is frozen at a small value. Thus, semiclassical calculations become possible. In order to contribute to the superpotential, instanton solutions should have two gaugino zero modes. We begin with   a summary of the monopole-instanton solutions relevant for the generation of a superpotential.

\subsection{The relevant monopole-instantons and their contributions to the superpotential}
\label{moninst}

 It turns out that, for all gauge groups, the monopole-instantons contributing to the superpotential are labeled by the simple co-roots $\pmb\alpha_a^*$, $a = 1,...r$, and the affine (lowest) co-root $\pmb\alpha_0^*$. 
 The monopole-instanton solutions associated with the simple  co-roots $\pmb \alpha_a^*$, $a=1,...,r$, have long range magnetic field   given by
\begin{eqnarray}
\label{magnchargealpha}
B_\mu^{\pmb \alpha_a }=-\frac{x_\mu}{|x|^3}\frac{\pmb\alpha^*_a \cdot   H}{2}\,, ~a=1,...,r,
\end{eqnarray}
and  action and instanton number:
\begin{eqnarray}
\label{actionalpha}
S^{\pmb \alpha_a }=\frac{4\pi }{g^2}\pmb \alpha^*_a \cdot\pmb \phi\,,\quad {\cal K}^{\pmb \alpha_a}=\frac{1 }{2\pi}\pmb \alpha^*_a \cdot\pmb \phi\,, ~ a=1,...,r.
\end{eqnarray}
The $\pmb \alpha_0^*$-solution is also self-dual, has long-range magnetic field
\begin{eqnarray}
\label{magnchargealpha3}
B_\mu^{\pmb \alpha_0 }=-\frac{x_\mu}{|x|^3}\frac{\pmb \alpha _0^* \cdot   H}{2}\,,
\end{eqnarray}
and   action and  instanton number
\begin{eqnarray}
\label{actionalpha3}
S^{\pmb \alpha_0 }=\frac{4\pi }{g^2}(2 \pi + \pmb \alpha^*_0 \cdot\pmb \phi)\,,\quad {\cal K}^{\pmb \alpha_0 }=  \frac{2\pi + \pmb \alpha^*_0 \cdot\pmb \phi }{2\pi}\,.
\end{eqnarray}
These $r+1$ monopole-instantons carry two gaugino zero modes each, hence they contribute to the superpotential.

The form of the superpotential ${\cal W}$, which, in the semiclassical regime, is due to the above monopole-instantons, can be obtained by demanding: (1) the holomorphy of ${\cal W}$ as a function of the chiral field $\pmb X$, (2) single-valuedness of ${\cal W}$   under  shifts of $\pmb X$ corresponding to periodic shifts\footnote{Notice that periodicity of $\pmb \sigma$  is a property of the long-distance theory, as explained in the previous Section, as opposed to shifts of $\pmb \phi$ (\ref{holonomyperiodicity}), which are due to large gauge transformations
 (\ref{largegauge}) about which the long-distance theory is ignorant.} of $\pmb \sigma$ (\ref{sigmaperiod}), i.e. under $\pmb X \rightarrow \pmb X+ 2\pi i \;\pmb\omega_a$, and (3)   that ${\cal W}$ should respect the discrete $R$-symmetry:\footnote{The discrete chiral $R$-symmetry acts on the dual photon fields $\pmb \sigma$ by a shift, due to the usual intertwining of the topological shift symmetries of the dual photons with chiral symmetries (acting on the gauginos) on $\R^3 \times \S^1$. This can be understood by considering the 't Hooft vertex of monopole-instantons or as due to a one-loop effect via induced Chern-Simons terms, see \cite{Aharony:1997bx}.} $\pmb X\rightarrow \pmb X+i\frac{2\pi}{c_2}\pmb \rho$, where $\pmb \rho$ is the Weyl vector, $c_2=\sum_{a=0}^r k_a^*$ is the dual Coxeter number and $k_a^*$ are the dual Kac labels. These conditions imply that the superpotential is:
\begin{eqnarray}
{\cal W}=\kappa\frac{2\pi R}{g^2}\mu^3\left(\sum_{a=1}^r\frac{2}{\pmb \alpha_a^2}e^{\pmb \alpha_a^*\cdot \pmb X}+\frac{2}{\pmb \alpha_0^2}e^{\pmb \alpha_0^*\cdot \pmb X+2\pi i\tau} \right)\,,
\label{superpotential}
\end{eqnarray}
where $\pmb\alpha_0$ is the affine root, and $\kappa$ is a numerical coefficient whose precise value is not important for our purposes. The discrete $R$ symmetry amounts to multiplying the superpotential by an overall factor $e^{i{2\pi \over c_2}}$.
Naturally, the symmetries and holomorphy requirements alone do not determine the coefficients   given in (\ref{superpotential}). These were first calculated in \cite{Davies:1999uw}. The one-loop determinants for $SU(2)$ SYM were first calculated in \cite{Poppitz:2012sw}  and for general gauge groups---in this paper. These will play some role in our further discussion, as they fix the scale of the coupling appearing in (\ref{superpotential}) and also determine the one-loop corrections to the K\" ahler potential (\ref{susy dual}); see also Section \ref{largencoupling}.
 
\subsection{The quantum corrected monopole-instanton vertex, its supersymmetric completion, and the K\"ahler potential}

\label{quantumkahlervertex}

The computation of the quantum corrections to the monopole background is similar to the  one for $SU(2)$ in \cite{Poppitz:2012sw} and is presented in  Appendix~\ref{monopoledets}.   The calculation  uses the index theorem on $\mathbb R^3 \times \mathbb S^1$. Here, we shall present the result and discuss its consequences. We assume that the monopole corresponds to the simple root $\pmb\alpha_a$, $a=1,...,r$; the result for the $\pmb \alpha_0$ monopole-instanton background follows along similar lines.

We begin with the bare vertex corresponding to the $\pmb\alpha$-th monopole instanton. As any semiclassical vertex, it is accompanied by  a 't Hooft suppression factor, $e^{-S}$, where $S$ is the classical action of the solution. 
In the case of an $\R^3 \times \S^1$ monopole instanton, there is an additional complication---monopole instantons have long-range magnetic Coulomb interactions and their 't Hooft vertices come with an additional factor $\sim e^{i \sigma}$, where $\sigma$ is the dual photon, as first explained in \cite{Polyakov:1976fu}.
By the duality (\ref{v in terms of sigma}), an insertion of $e^{i \sigma}$ in the partition function corresponds to the insertion of a magnetic charge, thus this factor in the 't Hooft vertex accounts for the long-range magnetic Coulomb interactions of the monopole instantons. 

The bare 't Hooft vertex for the $\pmb \alpha$-th monopole instanton is thus accompanied by a factor
 \begin{eqnarray}
 \label{barethooft}
 e^{-S_0^{\pmb \alpha} +i\pmb \alpha^*\cdot \pmb \sigma}  = e^{-\frac{4\pi}{g^2(\Lambda_{PV})}\pmb \alpha^*\cdot \pmb \phi +i\pmb \alpha^*\cdot \pmb \sigma}  =e^{i\tau\pmb \alpha^*\cdot \pmb \phi+i\pmb \alpha^*\cdot \pmb \sigma}=e^{\pmb \alpha^*\cdot \pmb z} ,
\end{eqnarray}
where, in the last equality we used (\ref{zphirelation}), with the  coupling taken to be the bare one, $g(\Lambda_{PV})$, with $\Lambda_{PV}$ the Pauli-Villars scale.
The bare 't Hooft vertex (\ref{barethooft}) is subject to perturbative quantum corrections: despite the supersymmetry of the monopole-instanton background, the one-loop determinants in this background do not cancel. 
  From the calculations in Appendix \ref{monopoledets}, see Eqn.~(\ref{quantum corrections to monopole}), we find that the exponent in the bare 't Hooft vertex  (\ref{barethooft}) is quantum-modified as follows: 
  \begin{eqnarray}
  \label{oneloopthooft}
e^{\pmb \alpha^*\cdot \pmb z}~ \rightarrow ~ e^{\pmb\alpha^*\cdot\left[\pmb z -\frac{3}{2}\sum\limits_{\pmb \beta_+}\left(\pmb \beta\log \frac{\Gamma\left(\frac{\pmb \beta\cdot \pmb \phi}{2\pi}\right)}{\Gamma\left(1-\frac{\pmb \beta\cdot \pmb \phi}{2\pi}\right)} \right)\right]}\,. 
\end{eqnarray}
The sum in the exponent is over all positive roots, denoted by $\pmb \beta_+$.
In addition to the  field-dependent shift in (\ref{oneloopthooft}),  the quantum corrections change the  scale of the coupling appearing in $\pmb z$, (\ref{zphirelation}). It  is  not $\Lambda_{PV} $, but the appropriate low-energy scale $2\over R$, similarly, the scale $\mu$ in the superpotential ${\cal W}$ is also replaced by $2\over R$. The superpotential is similar to the one  already shown in (\ref{superpotential}):
\begin{eqnarray}
{\cal W}=\kappa\frac{2\pi R}{g^2}\left(\frac{2}{R}\right)^3\left(\sum_{a=1}^r\frac{2}{\pmb \alpha_a^2}e^{\pmb \alpha_a^*\cdot \pmb X}+\frac{2}{\pmb \alpha_0^2}e^{\pmb \alpha_0^*\cdot \pmb X+2\pi i\tau} \right)\,,
\label{superpotential expression}
\end{eqnarray}
except for  $\mu ={ 2\over R}$ and the fact that the lowest component of $\pmb X$ is, instead of $\pmb z$, given by:
\begin{eqnarray}
\pmb X\big\vert \equiv \pmb z-\frac{3}{2}\sum_{\pmb\beta_+}\left(\pmb \beta\log \frac{\Gamma\left(\frac{\pmb \beta\cdot \pmb \phi}{2\pi}\right)}{\Gamma\left(1-\frac{\pmb \beta\cdot \pmb \phi}{2\pi}\right)} \right)\,.
\label{lowest components of X}
\end{eqnarray}
As already mentioned, all $g$ factors are taken at $\mu={2\over R}$. In all subsequent equations we use $g^2$ to denote $g^2({2\over R})$.

Next, as previously done in $SU(2)$ SYM \cite{Poppitz:2012sw} and massive SQCD \cite{Poppitz:2013zqa}, we shall use the one-loop correction to the 't Hooft vertex, along with supersymmetry, to 
calculate the loop corrections to the moduli-space  metric. 
To do this,  
we generalize the tree-level chiral-linear duality of Section \ref{treeduality} to include one-loop effects. We begin by replacing the sum of (\ref{the action as integral over WS}) and (\ref{auxiliary S SUSY}) with 
\begin{eqnarray}
S = \int d^3x d^4\theta \left[ - \frac{4 \pi R}{g^2} \;  \pmb Y \cdot \pmb Y  + 2 \pi R f(\pmb Y) - \frac{1}{2 \pi} \pmb Y\cdot \left(\pmb X+
\pmb X^\dagger \right) \right]\,.
\label{general susy duality}
\end{eqnarray}
With respect to the duality transformation in Section \ref{treeduality}, the two novel elements are  the use of $\pmb Y$ to denote the unconstrained real superfield (instead of $\pmb W$) and the inclusion of the term $f(\pmb Y)$, which represents the one-loop correction  to the moduli space metric (the kinetic terms of $\pmb v_{\mu\nu}$ and $\pmb v_3$). 
Naturally, these one-loop corrections can be calculated separately; however, we shall  use  supersymmetry to relate them to the already calculated one-loop correction to the 't Hooft vertex by requiring consistency of the duality transformation with (\ref{lowest components of X}). 

As before, varying (\ref{general susy duality}) with respect to the chiral superfields $\pmb X$, $\pmb X^\dagger$ imposes the linear multiplet constraint on $\pmb Y$ and leads back to the one-loop corrected action including the yet unknown function $f(\pmb Y)$. 
To go in the other direction, we now define the superfield Legendre transform:
\begin{eqnarray}
{\cal K}(\pmb X+\pmb X^\dagger)=- \frac{4 \pi R}{g^2} \; \pmb Y \cdot \pmb Y  + 2 \pi R f(\pmb Y) - \frac{1}{2 \pi} \pmb Y\cdot \left(\pmb X+
\pmb X^\dagger \right) \bigg\vert_{\pmb Y = \pmb Y(\pmb X+ \pmb X^\dagger)}~, 
\label{leg1}
\end{eqnarray}
such that
\begin{eqnarray}
S=\int d^3x d^4\theta \;{\cal K}(\pmb X+ \pmb X^\dagger)\,
\label{khaler action}
\end{eqnarray}
is the $D$-term part of the action of the dual chiral superfield $\pmb X$, whose total action reads
\begin{eqnarray}
S_T=\int d^3x d^4\theta {\cal K}(\pmb X+ \pmb X^\dagger)+\int d^3x d^2\theta\left[ {\cal W}(\pmb X)+\bar{\cal W}(\pmb X^\dagger)\right]\,.
\label{total supersymmetric action}
\end{eqnarray}
From the properties of the Legendre transform, we have from (\ref{leg1})
\begin{eqnarray}
\frac{\partial {\cal K}}{\partial ( X^i+ X^{\dagger \;i})}= - \frac{1}{2 \pi}Y^i\,.
\label{D for khaler}
\end{eqnarray}
Also, varying  (\ref{leg1}) with respect to $Y_i$ we find
\begin{eqnarray}
X^i+X^{i\;\dagger}= - {16 \pi^2 R \over g^2} \; Y^i+4 \pi^2 R \; \frac{\partial {f}}{\partial Y_i}\,.
\label{one of the main khaler equations}
\end{eqnarray}

We now recall the form of the lowest component of $\pmb X$ in terms of the holonomy, given by   (\ref{lowest components of X}),   and remembering that  $\pmb z=- {4 \pi \over g^2} \pmb\phi+ i \pmb\sigma$ and $Y^i \big\vert =  \phi^i/2 \pi R$, we use (\ref{one of the main khaler equations}) to find for the lowest component of the derivative of the one-loop correction $f(\pmb Y)$:\footnote{Upper and lower indices of $Y$ and $X$ are lifted and lowered with $\delta_{ij}$.}
\begin{eqnarray}
4 \pi^2 R \; \frac{\partial {f}}{\partial Y_i}\; = \;- 3 \sum_{\pmb \beta_+}\left( \beta_i\log \frac{\Gamma\left(\frac{\pmb \beta\cdot \pmb \phi}{2\pi}\right)}{\Gamma\left(1-\frac{\pmb \beta\cdot \pmb \phi}{2\pi}\right)} \right)\,.
\label{the main equation for the relation between F and phi}
\end{eqnarray}
This expression is the main new result of this Section. It implies that the one-loop correction to the moduli-space metric of $\pmb Y$, denoted by  $2 \pi R f(\pmb Y)$ in (\ref{general susy duality}), can be written as 
\begin{eqnarray}
\label{efofY} 
2 \pi R f(\pmb Y)= - {3 \over 2 \pi R} \sum\limits_{\pmb \beta_+} \left( \psi^{(-2)}\left(R \;\pmb\beta \cdot \pmb Y \right) +  \psi^{(-2)}\left(1-R \; \pmb\beta \cdot \pmb Y  \right) \right) ~,
\end{eqnarray}
where $\psi^{-2}(z)$  is the polygammma function of negative order ($\psi^{(-2)}(z)' = \log \Gamma(z)$). As a check, note that  in the $R\rightarrow 0$ limit, using $\psi^{(-2)}(z) \approx - z \log z$, this yields the  three-dimensional result $2 \pi R f(\pmb Y) \approx {3 \over 2 \pi} \sum\limits_{\pmb \beta_+} \pmb\beta \cdot \pmb Y \log(  \pmb\beta \cdot \pmb Y R)$, see  \cite{Smilga:2004zr} where this result is obtained for $SU(2)$.

Next, in order to determine the K\" ahler potential for $\pmb X$, we take the derivative of (\ref{D for khaler}) with respect to $X_j+X_j^\dagger$  and the derivative of (\ref{one of the main khaler equations}) with respect to $Y_k$, to obtain
\begin{eqnarray}
\nonumber
{\cal K}_{i\bar j}\equiv \frac{\partial^2 {\cal K}}{\partial ( X_i+ X^\dagger_i)\partial ( X_j+ X^\dagger_j)}&=& - \frac{1}{2 \pi}\frac{\partial Y_i}{\partial ( X_j+ X^\dagger_j)}\,,\\
- {16 \pi^2 R \over g^2}\; \delta_{jk}+ 4 \pi^2 R\; \frac{\partial^2 {f}}{\partial Y_j\partial Y_k}&=&\frac{\partial(X_j+X_j^\dagger)}{\partial Y_k}\,,
\end{eqnarray} 
respectively, where ${\cal K}_{i\bar j}$ is the K\"ahler metric. From these two equations, we find
\begin{eqnarray}
 {\cal K}_{i\bar j}\left(- {16 \pi^2 R \over g^2}\; \delta_{jk}+ 4 \pi^2 R\; \frac{\partial^2 {f}}{\partial Y_j\partial Y_k}\right)= - \frac{1}{2 \pi }\delta_{ik}\, , 
\end{eqnarray}
thus,  the inverse K\"ahler metric is given by
\begin{eqnarray}
{\cal K}^{i\bar j}={ 32  \pi^3 R \over g^2}  \left[\delta_{ij} - {g^2 \over 4} \frac{\partial^2 f}{\partial Y_i\partial Y_j}\right]\,.
\label{kinv1}
\end{eqnarray}
Then, from  (\ref{the main equation for the relation between F and phi}) we find (remembering that $Y_i \big\vert = \phi_i/2 \pi R $)
\begin{eqnarray}
{\cal K}^{i\bar j}
=\frac{32\pi^3 R}{g^2}\left[\delta_{ij}+\frac{3g^2}{16\pi^2}\sum_{\pmb \beta_+}\beta_i\beta_j\left[\psi\left(\frac{\pmb \phi\cdot\pmb \beta}{2\pi}\right)+\psi\left(1-\frac{\pmb \phi\cdot\pmb \beta}{2\pi}\right)\right]\right]\,,
\label{final expression of Kij}
\end{eqnarray}
where $\psi(x)=\Gamma'(x)/\Gamma(x)$,
and the K\" ahler metric is, to one-loop order:
\begin{eqnarray}
{\cal K}_{i\bar j}
=\frac{g^2}{32\pi^3 R}\left[\delta_{ij}-\frac{3g^2}{16\pi^2}\sum_{\pmb \beta_+}\beta_i\beta_j\left[\psi\left(\frac{\pmb \phi\cdot\pmb \beta}{2\pi}\right)+\psi\left(1-\frac{\pmb \phi\cdot\pmb \beta}{2\pi}\right)\right]\right]\,.
\label{kij down}
\end{eqnarray}
We conclude that, at one-loop level around the monopole-instanton background, the  bosonic part of the Lagrangian (\ref{total supersymmetric action}) is given by ($X^j$ below denotes the lowest component of the chiral superfield):
\begin{eqnarray}
S=\int d^3 x\left[K_{i\bar j}\partial_\mu X^i \partial^\mu X^{\dagger \bar j}+{\cal K}^{i\bar j}\frac{\partial{\cal W}}{\partial X^i}\frac{\partial\bar{\cal W}}{\partial X^{j^\dagger}}\right]\,. 
\label{action1}
\end{eqnarray}
 As already stated earlier, the one-loop correction to the monopole-instanton background is relevant for calculating the expectation value of $\Tr \Omega$ as well as its correlation function, because of the induced ${\cal O}(g^2)$ shift in the expectation value for $\pmb \phi$,  see  Section \ref{bionpotential0}. 
The corrections to $K_{i\bar{j}}$, or equivalently, the moduli space metric of $\pmb Y$ in (\ref{efofY}), become relevant   for large ranks of the gauge group (large-$c_2$, or large-$N$), where they are important for setting the scale of the coupling and taking the abelian large-$N$ limit.
 We discuss this in detail in Section \ref{largencoupling}. But, first, we study the supersymmetric ground state of (\ref{action1}).

\subsection{The supersymmetric vacuum, the neutral- and magnetic-bion potentials}
\label{bionpotential0}

The supersymmetric vacuum of (\ref{action1}) is the solution of
\begin{eqnarray}
\label{aaa}
\frac{\partial{\cal W}}{\partial X^i}=0\,,
\end{eqnarray}
where ${\cal W}$ is given by (\ref{superpotential expression}). The solution of this equation was given in \cite{Davies:2000nw} (as one can also check upon substitution in (\ref{aaa})):
\begin{eqnarray}
\pmb X_{0}=\sum_{a=1}^r\pmb \omega_a \log\left(\frac{k_a^* \pmb\alpha_a^*}{2}\right)+\left( \frac{2\pi i(\tau+u)}{c_2}-\log|\upsilon| \right)\pmb \rho,~{\rm with}~
|\upsilon|=\left[\prod_{a=0}^r\left(\frac{k_a^*\pmb \alpha_a^2}{2}\right)^{k_a^*}\right]^{\frac{1}{c_2}},
\end{eqnarray}
and $u=1,2,...c_2$ labels the $c_2$ different supersymmetric vacua. Using (\ref{lowest components of X}) we find
\begin{eqnarray}
\nonumber
\pmb \phi_0&=&-\frac{3g^2}{8\pi}\sum_{\pmb\beta_+}\left(\pmb \beta\log \frac{\Gamma\left(\frac{\pmb \beta\cdot \pmb \phi_0}{2\pi}\right)}{\Gamma\left(1-\frac{\pmb \beta\cdot \pmb \phi_0}{2\pi}\right)} \right)+\left( \frac{2\pi }{c_2}+\frac{g^2}{4\pi}\log|\upsilon| \right)\pmb \rho-\frac{g^2}{4\pi} \sum_{a=1}^r\log\left(\frac{k_a^*\pmb\alpha_a^2}{2}\right)\pmb\omega_a\,,\\
\pmb\sigma_0&=& \frac{2\pi u}{c_2}\pmb\rho\,,
\label{explicit phi0}
\end{eqnarray}
where $\pmb \phi_0$ can be found self consistently. We do so by substituting the approximation 
\begin{eqnarray}
\label{phivev}
\pmb \phi_0 \cong\pmb \phi_0^{(0)}+\frac{g^2}{4\pi}\pmb \phi_0^{(1)}+{\cal O}(g^4)
\end{eqnarray}
into (\ref{explicit phi0}) to find:
\begin{eqnarray}
\nonumber
\pmb \phi_0^{(0)}&=&\frac{2\pi}{c_2}\pmb \rho \\
\pmb \phi_0^{(1)}&=&-\frac{3}{2}\sum_{\pmb \beta_+}\pmb \beta \log \frac{\Gamma\left(\frac{\pmb \beta \cdot \pmb \rho}{c_2}\right)}{\Gamma\left(1-\frac{\pmb \beta \cdot \pmb \rho}{c_2}\right)}+\pmb \rho \log|v|-\sum_{a=1}^r\log\left(\frac{k_a^*\pmb\alpha_a^2}{2}\right)\pmb\omega_a\,.
\label{approximate phi0}
\end{eqnarray}
The above expressions for $\pmb \phi_0^{(0)}$ and $\pmb \phi_0^{(1)}$, which determine (\ref{phivev}), will be important for the calculation of the Polyakov loop expectation values. 
 
The potential can be calculated, as per (\ref{action1}), from  (\ref{superpotential expression}), (\ref{kinv1}), omitting (for the moment) the one-loop correction to    the K\" ahler metric
\begin{eqnarray}
\nonumber
V_{\mbox{\scriptsize bion}}&=&{\cal K}^{i\bar j}\frac{\partial{\cal W}}{\partial X_i}\frac{\partial\bar{\cal W}}{\partial X_j^\dagger}=\frac{32\pi^3 R}{g^2} \delta^{ij}\frac{\partial{\cal W}}{\partial X_i}\frac{\partial\bar{\cal W}}{\partial X_i^\dagger}\\
\label{bionpotential1}
&=&64 \pi^2 \kappa^2\left(\frac{2\pi R}{g^2}\right)^3\left(\frac{2}{R}\right)^6\left[\sum_{a,b=1}^r\frac{\pmb \alpha_a^*\cdot \pmb \alpha_b^*}{\pmb \alpha_a^2\pmb \alpha_b^2} e^{\pmb \alpha_a^*\cdot \pmb X+\pmb \alpha_b^*\cdot \pmb X^\dagger}+\frac{\pmb \alpha_0^{*2}}{\pmb \alpha_0^{4}}e^{\pmb \alpha_0^*\cdot \left(\pmb X+\pmb X^\dagger\right)+2\pi i(\tau-\tau^*) }\right.\\
\nonumber
&&\left.+\sum_{a=1}^r \frac{\pmb \alpha_a^*\cdot \pmb \alpha_0^*}{\pmb \alpha_a^2\pmb \alpha_0^2} \left(e^{\pmb \alpha_a^*\cdot \pmb X+\pmb \alpha_0^*\cdot \pmb X^\dagger-2\pi i\tau^*}+e^{\pmb \alpha_a^*\cdot \pmb X^\dagger+\pmb \alpha_0^*\cdot \pmb X+2\pi i\tau}\right)\right]_{\pmb X=\pmb i(\tau\pmb \phi+\pmb\sigma)-\frac{3}{2}\sum_{\pmb\beta_+}\left(\pmb \beta\log \frac{\Gamma\left(\frac{\pmb \beta\cdot \pmb \phi}{2\pi}\right)}{\Gamma\left(1-\frac{\pmb \beta\cdot \pmb \phi}{2\pi}\right)} \right)}\,.\nonumber
\end{eqnarray}
The subscript ``bion" in $V_{\mbox{\scriptsize bion}}$ refers to the nature of the topological excitations giving rise to the various terms; these will be explained after (\ref{vbion0}).
We are interested in examining this potential near the supersymmetric vacuum (\ref{explicit phi0}, \ref{phivev}, \ref{approximate phi0}). To this end, we define
\begin{eqnarray} \label{phi11}
\pmb \phi=\pmb \phi_0+\frac{g^2}{4\pi}\pmb b\,,\quad\pmb \sigma=\pmb \sigma_0+\pmb \sigma'\,.
\end{eqnarray} 
and find, recalling (\ref{lowest components of X}), that, in terms of the fluctuations around the supersymmetric vacuum, $\pmb \sigma'$ and $\pmb b$, $\pmb X$ is expressed  as
\begin{eqnarray}
\pmb X&=&-\frac{8\pi^2}{g^2 c_2}\pmb\rho +\left(i\frac{2\pi u}{c_2} -\log |v|\right)\pmb \rho+\sum_{a=1}^r \log\left(\frac{k_a^*\pmb\alpha_a^2}{2}\right)\omega_a+g^2 {\pmb x_0}-\pmb b +i\pmb \sigma'\,,
\label{expansion in X}
\end{eqnarray}
where
\begin{eqnarray}
g^2 {\pmb x_0}&=&-\frac{3 g^2}{16 \pi^2}\sum_{\pmb \beta_+}\pmb \beta \;  (\pmb\beta\cdot \pmb b+\pmb\beta\cdot\pmb \phi_0^{(1)})  \left(\psi\left(\frac{\pmb \beta \cdot \pmb \rho}{c_2}\right)+\psi\left(1-\left(\frac{\pmb \beta \cdot \pmb \rho}{c_2}\right) \right)\right)\,.
\label{Og2 corrections of X}
\end{eqnarray}
In fact, to leading order in $g^2$ in $V_{\mbox{\scriptsize bion}}$, we can neglect $g^2 {\pmb x_0}$ in (\ref{expansion in X}). Collecting everything and using
\begin{eqnarray}
\nonumber
\pmb \rho\cdot \pmb \alpha_0^*&=&1-c_2\,,\\
\nonumber
\pmb \alpha_a^*\cdot \pmb X&=&-\frac{8\pi^2}{g^2 c_2}+i\frac{2\pi u}{c_2}-\log |v|+\log\left(\frac{k_a^* \pmb \alpha_a^2}{2}\right)-\pmb\alpha_a^*\cdot \pmb b+i\pmb \alpha_a^* \cdot \pmb \sigma'\,,\\
\nonumber
\pmb \alpha_0^*\cdot \pmb X&=&-\frac{8\pi^2}{g^2 c_2}(1-c_2)+ \left(i\frac{2\pi u}{c_2} -\log |v|\right)(1-c_2) -\sum_{a=1}^rk_a^*\log\left(\frac{k_a^* \pmb \alpha_a^2}{2}\right) \\
&&-\pmb\alpha_0^*\cdot \pmb b+i\pmb \alpha_0^* \cdot \pmb \sigma'~,
\end{eqnarray}
 we find, after some algebra, a compact expression for the bion-induced potential:\footnote{As a consistency check on (\ref{bionpotential2}) notice that at the supersymmetric minimum $\pmb\sigma'= \pmb b=0$, we have
\begin{eqnarray}
\nonumber
\frac{V_{\mbox{\scriptsize bion}}(\pmb\sigma'=0,\pmb b=0)}{V_{\mbox{\scriptsize bion}}^0}&=&\sum_{a=0,b=0}^r k_a^*k_b^*\pmb \alpha_a^*\cdot \pmb \alpha_b^*=\left(\sum_{a=0}^r k_a^*\pmb \alpha_a^*\right)^2
=\left(\pmb\alpha_0^*+\sum_{k=1}^rk_a^*\pmb\alpha_a^*\right)^2=0\,,
\end{eqnarray}
since by definition $\pmb\alpha_0^*=-\sum_{k=1}^rk_a^*\pmb\alpha_a^*$. If we also keep the $g^2{\pmb x_0}$ part of $X$ in (\ref{Og2 corrections of X}) and  the ${\cal O}(g^2)$ correction of the K\" ahler potential (\ref{final expression of Kij}) we find the ${\cal O}(g^2)$ correction of the bion potential:
\begin{eqnarray}
\nonumber
\frac{V_{\mbox{\scriptsize bion}}^{g^2}(\pmb\sigma',\pmb b)}{V_{\mbox{\scriptsize bion}}^0}=&&\frac{3g^2}{16\pi^2}\sum_{\pmb\beta_+}\left(\psi\left(\frac{\pmb \beta \cdot \pmb \rho}{c_2}\right)+\psi\left(1-\left(\frac{\pmb \beta \cdot \pmb \rho}{c_2}\right)\right)\right)\\
\nonumber
&&\times \sum_{a,b=0}^r k_a^*k_b^* \left[\pmb \beta\cdot \pmb \alpha_a^*\pmb \beta\cdot \pmb \alpha_b^* -\frac{(\pmb b+\pmb \phi_0^{(1)})\cdot \pmb \beta}{\pi}\left(\pmb\alpha_a^*\cdot \pmb \beta+\pmb\alpha_b^*\cdot \pmb\beta\right)\pmb\alpha_a^*\cdot \pmb\alpha_b^*\right]\\
&&\times e^{-\left(\pmb \alpha_a^*+\pmb \alpha_b^*\right)\cdot \pmb b}\cos\left(\left(\pmb \alpha_a^*-\pmb \alpha_b^*\right)\cdot \pmb \sigma'\right)\,, 
\end{eqnarray}
and, 
for another consistency check, notice   that $V_{\mbox{\scriptsize bion}}^{g^2}(\pmb\sigma',\pmb b)$ also vanishes at the supersymmetric vacuum $\pmb\sigma'=\pmb b=0$.
}

\begin{eqnarray}
 V_{\mbox{\scriptsize bion}} = {V_{\mbox{\scriptsize bion}}^0}  \sum_{a=0,b=0}^r k_a^*k_b^*\pmb \alpha_a^*\cdot \pmb \alpha_b^*e^{-\left(\pmb \alpha_a^*+\pmb \alpha_b^*\right)\cdot \pmb b}\cos \left(\pmb \alpha_a^*-\pmb \alpha_b^*\right)\cdot \pmb \sigma' \,,
 \label{bionpotential2}
\end{eqnarray}
where 
\begin{eqnarray}
V_{\mbox{\scriptsize bion}}^0=16 \pi^2 \kappa^2\left(\frac{512\pi^3 }{g^6 R^3 |v|^2}\right)e^{-\frac{16 \pi^2}{g^2 c_2}}\,.
\label{vbion0}
\end{eqnarray}

The compact form of the potential (\ref{bionpotential2}) is very useful for numerical implementation for any gauge group, but does not do justice to the physics. As this has been discussed in previous papers, we only note that 
terms in the potential with $a=b$ correspond to neutral, or center-stabilizing bions \cite{Poppitz:2011wy,Argyres:2012ka}. On the other hand, terms with  $a \ne b$ terms represent magnetic bions \cite{Unsal:2007jx}, this is also clear from the already-mentioned fact that their 't Hooft vertex includes a factor $e^{i\left(\pmb \alpha_a^*-\pmb \alpha_b^*\right)\cdot \pmb \sigma'}$, responsible for their long-range magnetic interactions. The former are responsible for the stabilization of center symmetry and the latter---for the generation of a mass gap for the dual photon and confinement. The physics of the various contributions to the potential (\ref{bionpotential2}) has been extensively discussed in \cite{Poppitz:2012sw,Poppitz:2012nz}.

 \subsection{Moduli-space metric at the center-symmetric point and  large-$\mathbf{N}$ limit(s)}
\label{largencoupling}

In this Section, we study the role of the  loop corrections to the moduli-space metric near the center-symmetric point (the supersymmetric vacuum  (\ref{phivev})) and their importance for taking the large-$N$ limit. 

We stress that this interlude is needed here, because, when we combine the bion induced potential from Section \ref{bionpotential0} with the soft-breaking contributions of the following Section \ref{softly}, we shall use the results from this Section to write all quantities in terms of renormalization group invariants that have a proper large-$N$ limit.
To this end, let us go back to the ``electric variables" in terms of the real superfield $\pmb Y$ (this should be now viewed as a linear superfield, i.e. should really be labeled by $\pmb W$), whose one-loop effective action is given by (\ref{general susy duality}, \ref{efofY}):
\begin{eqnarray}
S = \int d^3x d^4\theta \left[ - \frac{4 \pi R}{g^2} \;  \pmb Y \cdot \pmb Y   - {3 \over 2 \pi R} \sum\limits_{\pmb \beta_+} \left( \psi^{(-2)}\left(R \;\pmb\beta \cdot \pmb Y \right) +  \psi^{(-2)}\left(1-R \; \pmb\beta \cdot \pmb Y  \right) \right)\right]\,.
\label{modulispace1}
\end{eqnarray}
 We are interested in  the moduli-space metric near the center-symmetric point, i.e. we take  $\pmb Y=   {1 \over c_2 R} \; \pmb \rho + \pmb y$. This corresponds to expanding around the supersymmetric vacuum (\ref{phivev}). The kinetic term for the fluctuation of the lowest component of $\pmb Y$, i.e. the holonomies' fluctuations around the supersymmetric vacuum, $\pmb y\vert = (v^{3 \; 1}, ...,v^{3 \; i},... v^{3 \; r})$, is then given by:
 \begin{eqnarray}
S &=&  - {3  R c_2\over4 \pi }   \int d^3x d^4\theta  \sum\limits_{i,j=1}^r y^i y^j \left[{16 \pi^2 \over 3 c_2 g^2} \; \delta_{ij}  + 
\sum\limits_{\pmb \beta_+} {\beta_i \beta_j \over c_2} \left( \psi({\pmb \beta \cdot \pmb \rho\over c_2}) + \psi(1-{\pmb \beta \cdot \pmb \rho\over c_2})\right)
\right]  \nonumber \\
&\equiv&  - {3  R c_2\over 8 \pi }\; g_{ij} \; \partial_\mu  v^{3\; i}  \partial^\mu  v^{3\; j}\label{modulispace2}~.
\end{eqnarray}
Above, we introduced the moduli space metric $g_{ij}$ at the center symmetric point:
\begin{equation}
\label{modulispace3}
g_{ij} = 
 {16 \pi^2 \over 3 c_2 g^2} \; \delta_{ij}  + 
\sum\limits_{\pmb \beta_+} {\beta_i \beta_j \over c_2} \left( \psi({\pmb \beta \cdot \pmb \rho\over c_2}) + \psi(1-{\pmb \beta \cdot \pmb \rho\over c_2}) \right) \equiv {16 \pi^2 \over 3 c_2 g^2} \delta_{ij}  + k_{ij}~,\end{equation}
where the second equation defines the matrix $k_{ij}$.

Let us now study the properties of $g_{ij}$ for the case of $SU(N)$ (the $Spin(N)$ and $Sp(2N)$ cases result in identical conclusions). Using the standard $N$-dimensional basis\footnote{The positive roots are $\pmb \beta_{ij} = \pmb e_i - \pmb e_j$ for $1 \le i < j \le N$, with $\pmb e_i$---an orthonormal basis in $\R^{N}$, and the Weyl vector is $\pmb \rho = \sum_{a = 1}^{N} {N+1 - 2a \over 2}\;\pmb e_a$.} for the  roots of $SU(N)$,   the one-loop contribution to the moduli space metric becomes ($c_2 = N$)
\begin{equation}
\label{modulispace4}
k_{ij} = \delta_{ij} {2 \over N} \sum_{k=1}^{N -1} \psi({k \over N})  - {(1 - \delta_{ij})} { \psi({|i-j|\over N}) + \psi(1 - {|i-j| \over N} )\over N}~.
\end{equation}
After an orthogonal transformation to  a basis where $k_{ij} = \delta_{ij} k^{diag}_i$, we have
for the moduli space metric:
\begin{eqnarray}
\label{modulispace5}
g_{ij}^{diag} &=& \delta_{ij} \left( {16 \pi^2 \over 3 N g^2({2 \over R})}  + k_i^{diag} \right) \\
&=& \delta_{ij} \left( {16 \pi^2 \over 3 N g^2({2 \over  N R})}  +2 \log N +  k_i^{diag} \right) \label{modulispace6}
\end{eqnarray}
Above, we have written two expressions for the moduli space metric. These will be used in what follows   to discuss the nature of two possible large-$N$ limits: one where the radius $R$ is kept fixed  and another one,  where, instead the scale of the lightest W-bosons, $1 \over N R$, is kept fixed (requiring $R \rightarrow 0$ as well as weak coupling, see below).

 To proceed further, we admit that we have not been able to analytically find the eigenvalues of $k_{ij}$ of (\ref{modulispace4}). This is a drawback of our analysis, as we can not rigorously establish an asymptotic large-$N$ limit on the eigenvalues $k_i^{diag}$ and we have to resort to numerics.
 Numerically, however, the   explicit form of the moduli space metric (\ref{modulispace4}) allows for studies for rather large $N$. We find that the eigenvalues $k_i^{diag}$ are all negative and the largest (by absolute value) negative eigenvalue logarithmically increases with $N$.
 
\begin{figure}[h]
\centering
\includegraphics[width=.6 \textwidth]{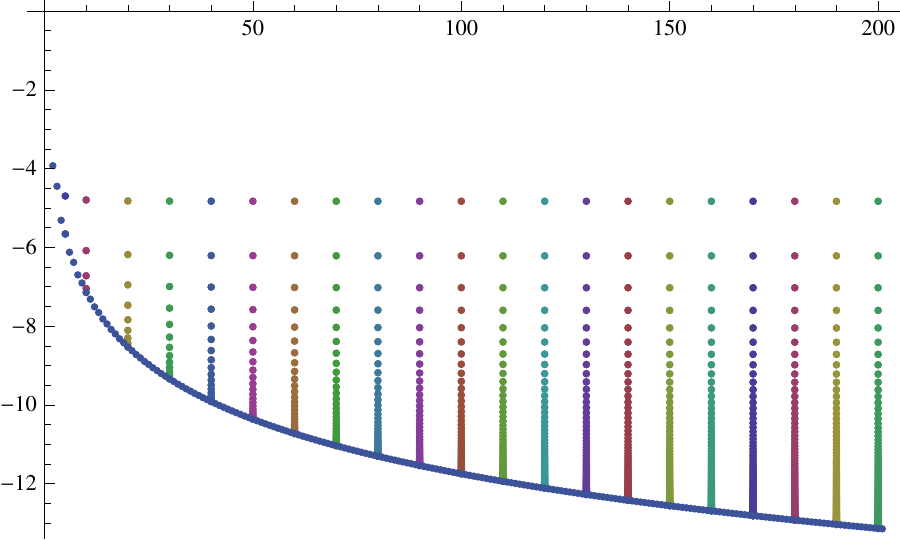}
\caption{\label{fig:kdiag} The continuous line on the bottom shows the most negative eigenvalue of $k_{ij}$ for all  $2\le N \le 200$. All eigenvalues $k_i^{diag}$ are shown for  $N=10,20,...,190,200$. The $N$ dependence of the most negative eigenvalue  is well-fitted by  ${\rm min}_i k_i^{diag} \sim - 2.009 \log N$. If the $\S^1$ radius $R$ is kept fixed at large $N$, one finds, from (\ref{modulispace5}), a singularity in the moduli space metric at fixed $N g^2({2\over R})$, owing to the onset of strong coupling.  
On the other hand, when the mass of the lightest $W$-boson ${1 \over N R} = m_W$ is kept fixed, the moduli space metric is smooth, as follows from (\ref{modulispace6}) and the $\simeq-2\log N$ asymptotics of the minimal value of $k_i^{diag}$. See discussion in text.}
\end{figure}

  On Figure~\ref{fig:kdiag}, we show (the thick continuous line) the behavior of the minimal  $k_i^{diag}$ for all $N$ up to 200. We also show all eigenvalues $k_i^{diag}$ for a subset of $N$ as indicated on the Figure.
   The behavior of the most negative eigenvalue shown on the Figure can be fitted by an analytic expression of the form $k_{min} \sim - 2.009 \log N$. Even more telling, however, is to numerically compute the maximum of $|2 \log N + k_i^{diag}|$ (which still corresponds to a negative value of $2 \log N + k_i^{diag}$) for the same values of $N$. One finds that for $N$-even, this quantity monotonically approaches a constant, $N$-independent, value min($2 \log N + k_i^{diag}) \sim -2.540...$ for $N$ up to 200; for $N$-even the same value is approached with oscillations in the fifth digit.
   
 From the above discussion we can draw conclusions about the two kinds of large-$N$ limit one can take. 
To begin, recall that the two-loop RG invariant scale is (notice that $\Lambda$ is fixed in the $c_2 g^2 = const$ large-$N$ limit):
\begin{equation}
\label{lambdascale}
\Lambda^3 = { 16 \pi^2 \mu^3 \over 3 c_2 g^2(\mu) } e^{- {8 \pi^2 \over c_2 g^2(\mu)}}~.
\end{equation}

 If we  take $N \rightarrow \infty$, keeping $R$ fixed and the 't Hooft coupling $N g^2({2\over R})$ fixed (or equivalently, keeping $\log {1\over \Lambda R}$ fixed as follows from (\ref{lambdascale})), we find, from the logarithmic behavior of  $k^{diag}_{i_{min}} \sim - 2.009 \log{N}$   discussed above, that the moduli space metric at the center symmetric point becomes singular for sufficiently large $N$, for arbitrarily small but fixed 't Hooft coupling. This is due to the onset of strong coupling in this limit (clearly, this is unavoidable, as in the large-$N$, fixed-$R$ limit the masses of the lightest $W$-bosons vanish and nonabelian physics sets in). 
  
 On the other hand, we can take $N$ large by keeping $N R$, or equivalently $m_W = {1 \over N R}$, fixed, while also ensuring that $\Lambda$ (\ref{lambdascale}) is fixed and $\Lambda R N \ll 1$, so that the weak coupling approximation that led to (\ref{modulispace6}) is valid. Then, we find that for arbitrarily small fixed 't Hooft coupling, $N g^2 (2 m_W)$, the moduli space is  smooth owing to the finite $N \rightarrow \infty$ limit of the sum $2 \log N + k_i^{diag}$ 
  from (\ref{modulispace6}), as discussed above. We notice that the one-loop correction to the moduli space metric, $k_{ij}$ was important in order to be able to reach this conclusion.\footnote{\label{abc}While it is clear that there is no singularity of the moduli space metric in the abelian large-$N$ limit, we also note  that if a truly large-$N$ limit is taken, one may have  to incorporate  the fact that  the $N -1$ fields have different couplings. While most of the eigenvalues of $k_{ij}$ 
 are clustered near the most negative one, upon a glance on  Figure~\ref{fig:kdiag} it is clear that there are eigenvalues which are smaller by absolute value. The corresponding fields can have couplings much lower than $c_2 g^2(2 m_W)$: an extreme example is the eigenvalue closest to zero, which is $N$-independent and whose coupling is  essentially $c_2 g^2({2 \over R})$, which vanishes at infinite-$N$ with $N R$-fixed. Thus, to properly take the limit, one has to study in more detail the density distribution  of eigenvalues as $N \rightarrow \infty$. As our ``large-$N$" studies do not go beyond $16$, in this paper  we shall not further dwell on this issue.} 
 
The upshot of this discussion is that the sum over the large number of Kaluza-Klein modes of the off-diagonal (non-Cartan) W-bosons that appears in (\ref{modulispace3}) leads, at large-$N$, to a running of the coupling (of the most negative eigenvalue $k_i^{diag}$, see Footnote~\ref{abc}) according to its four-dimensional beta function. Notice that this happens despite the fact that $R \rightarrow 0$. This four dimensional running,  changing the scale of the coupling from $2\over R$ in (\ref{modulispace5}) to to the much smaller $2 \over N R$ scale in (\ref{modulispace6}), is what makes the abelian large-$N$ limit possible. The fact that we are working in the center-symmetric point in moduli space is important for this (for a recent discussion of volume independence and the role of center symmetry, see  \cite{Unsal:2010qh}).    
   
To conclude, we note that in the following  Sections, when we study the abelian large-$N$ limit, we shall take the coupling that appears in the kinetic term to be  $g^2(2 m_W)$ for all fields and shall also ignore the small remainder ($2 \log N + k_i^{diag}$) in (\ref{modulispace6}). 
The coupling from the K\"ahler potential  was assumed to be at the $m_W$ scale in  \cite{Poppitz:2012nz}, essentially for consistency of the abelian large-$N$ limit, but the role of the corrections to the moduli space metric to establishing this scale in the K\" ahler potential was not appreciated in Ref.~\cite{Poppitz:2012nz}, at least by one of its authors (EP).

\subsection{Softly broken supersymmetry, the total potential, and large-$\mathbf{N}$}
\label{softly}

Now, we continue the work of Section \ref{bionpotential0} and explicitly break supersymmetry by adding a soft mass term. The gaugino mass term lifts the monopole-instanton zero mode, with the resulting monopole vertex giving a contribution to the potential, for the $\pmb \alpha$-th monopole:
\begin{eqnarray}
V_m^{\pmb \alpha} =  \kappa' \; \frac{R m}{g^2 }\;\frac{2}{\pmb \alpha_a^2}\left(\frac{2}{R}\right)^3 \left(- {8 \pi^2 \over g^2} \delta_{a0}-\pmb \alpha_a^*\cdot \mbox{Re} \; \pmb z\right)  e^{2\pi i\tau\delta_{a0}+\pmb \alpha_a^*\cdot \pmb X}\,.
\end{eqnarray}
The derivation follows the lines of Appendix A of \cite{Poppitz:2012sw}. The origin of the various factors is as follows: $m$ is the gaugino mass, the $- {8 \pi^2 \over g^2} \delta_{a0}- \pmb\alpha_a^*\cdot \mbox{Re}  \pmb z$ factor is from the integral over the monopole-instanton fermion zero modes, saturated by  the fermion-bilinear insertion ${1\over g^2}\int_{\R^3 \times \S^1} \Tr \lambda \lambda$ from the gaugino lagrangian in (\ref{lambda1}), and the rest is from the superpotential vertex (\ref{superpotential}); the constant $\kappa'$ differs from $\kappa$ in the superpotential (\ref{superpotential}) by a numerical factor whose value is inessential for us. 

The total monopole potential is given by a sum over $a=0,1,...,r$ as well as the antimonopole contributions. After using (\ref{zphirelation}), we find:
\begin{eqnarray}
V_m&&= \label{monopole1} \\
-&&\kappa'  \frac{4 \pi R m}{g^4} \left(\frac{2}{R}\right)^3\left[\sum_{a=1}^r\frac{2 \pmb \alpha_a^*\cdot \pmb \phi_0^{(0)}}{\pmb \alpha_a^2}\; \;e^{\pmb \alpha_a^*\cdot \pmb X}+\frac{2\left(2\pi+\pmb \alpha_0^*\cdot \pmb \phi_0^{(0)}\right)}{\pmb\alpha_0^2} e^{2\pi i\tau+\pmb \alpha_0^*\cdot \pmb X} + {h.c.}\right]_{\pmb X={\rm Eqn.}(\ref{expansion in X})}.
\nonumber 
\end{eqnarray}
where we kept only the ${\cal O}(1)$ part of the expectation value of $\pmb \phi$ (\ref{phivev}) in the pre-exponential factor.
Using manipulations similar to the ones carried over in the previous Section for $V_{\rm bion}$ (\ref{bionpotential2}), we find a compact expression for the monopole potential:
\begin{eqnarray}
 {V_m}=-{V_m^0} \; \sum_{a=0}^rk_a^*e^{-\pmb \alpha_a^*\cdot \pmb b}\cos\left(\pmb \alpha_a^*\cdot\pmb \sigma'+\frac{\theta+2\pi u}{c_2}\right)\,,
 \label{monopolepotential2}
\end{eqnarray}
where 
\begin{eqnarray}
\label{vmzero}
V_m^0={m 
\kappa' \over R^2} \frac{128 \pi^2}{c_2 g^4  |v|}\;e^{-\frac{8\pi^2}{c_2g^2}}\,.
\end{eqnarray}
For future  use, we   restored the appropriate $\theta$ dependence, which re-appears in (\ref{monopolepotential2})  along with the fermion mass insertion.

Collecting everything, the total potential we shall study is then given by
\begin{eqnarray}
\label{vtotal}
 &&{V_T\over {V_{\mbox{\scriptsize bion}}^0}} =  \\
&&  \sum_{a=0,b=0}^r k_a^*k_b^*\pmb \alpha_a^*\cdot \pmb \alpha_b^*e^{-\left(\pmb \alpha_a^*+\pmb \alpha_b^*\right)\cdot \pmb b}\cos\left(\left(\pmb \alpha_a^*-\pmb \alpha_b^*\right)\cdot \pmb \sigma'\right) 
 -c_m\sum_{a=0}^rk_a^*e^{-\pmb \alpha_a^*\cdot \pmb b}\cos\left(\pmb \alpha_a^*\cdot\pmb \sigma'+\frac{\theta+2\pi u}{c_2}\right) \,, \nonumber 
\end{eqnarray}
where $c_m=V_m^0/V_{\mbox{\scriptsize bion}}^0$. Using $\pmb \phi=\pmb\phi_0+\frac{g^2}{4\pi}\pmb b$, and $\pmb \sigma=\pmb\sigma_0+\pmb\sigma'$, we find the total bosonic action:
\begin{eqnarray}
{\cal L}_{\mbox{\scriptsize bosonic}}= 
  { g^2_{  {\rm eff}} \over 32  \pi^3   R} \left[ (\partial_\mu   \pmb b )^2+ (\partial_\mu \pmb  \sigma' )^2  \right]+V_T + V_{GPY}\,~,
\label{bosonic potential}
\end{eqnarray}
where $g^2_{ {\rm eff}} \simeq g^2(m_W)$ is the diagonal coupling from (\ref{modulispace2}),(\ref{modulispace6}), which as we argued above, is essentially the four-dimensional coupling at the lowest W-boson mass scale. 

In (\ref{bosonic potential}), $V_T$ is the total nonperturbative  potential from (\ref{vtotal}), while 
the last $V_{GPY}$ term is the one-loop perturbative contribution to the potential for the holonomy (the GPY potential). It is calculated in Appendix \ref{gauginoGPY} and, to leading order in $m^2$, is given by Eq.~(\ref{gauginogpy}):
\begin{eqnarray}
\label{gauginogpy2}
V_{GPY}^{{\cal O}(m^2)}=   - \frac{m^2}{2 \pi^3 R}\sum_{p=1}^\infty \sum_{\pmb \beta_+} {\cos(p \pmb \beta\cdot \pmb \phi)\over p^2} ~.\end{eqnarray} 
We shall argue shortly, as in \cite{Poppitz:2012sw}, that the contribution of the perturbative fluctuations can be neglected compared with the monopole-instanton and bion potentials in the semiclassical regime.

We now rewrite the bosonic Lagrangian (\ref{bosonic potential}) using the renormalization group invariant scale $\Lambda$   (\ref{lambdascale}), with an eye towards also exploring the abelian ``large-$N$" (i.e. large-$c_2$) limit. 
From  (\ref{bionpotential1}), we have  (with the factor of $g^2$ coming from the inverse K\" ahler metric replaced by the diagonal  $g^2(m_W)$; recall that the rest of the $g^2$ factors are at $\mu = {2\over R}$):
 \begin{equation}
 \label{vbionzero}
V_{\mbox{\scriptsize bion}}^0= {16 \pi^2 \kappa^2   \over g^2_{\rm eff}}  \left(\frac{512\pi^3 }{g^4 R^3 |v|^2}\right)e^{-\frac{16 \pi^2}{g^2 c_2}}\, = {108 \kappa^2 \over \pi |v|^2} \left({ 8 \pi^2 \over {3 c_2 g^2 ({2 \over c_2 R}) }} \right) \; \Lambda^6 \;  (c_2 R)^3  \simeq {108
 \kappa^2\over \pi |v|^2}  {\Lambda^6 \over m_W^3} \log {m_W  \over \Lambda }~.
 \end{equation}
 
We recall that the large-$N$ limit is taken in the abelian sense. This was already discussed in Section \ref{largencoupling}, so we shall be brief:  as $N$ (or $c_2$) becomes large, both $\Lambda$ and $m_W = {1\over c_2 R}$ are kept fixed. At the same time, we require $m_W \gg \Lambda$ for the validity of the weak-coupling semiclassical approximation. 
 As already discussed in \cite{Poppitz:2012nz}, keeping the lightest $W$-boson mass fixed means that $R$, the $\S^1$ radius, has to become smaller. Note that general QCD-like theories exhibiting abelian confinement   admit a (non-'t Hooftian) large-$N$ limit, see \cite{Unsal:2008ch,Poppitz:2011wy}. In Seiberg-Witten  theory this was discussed in \cite{Douglas:1995nw}, where, instead of taking $R$ to zero, one has to take the ${\cal N}=2$ mass deformation to zero as $N\rightarrow \infty$. This condition arises for the same reason as in the present case:   in all cases, the abelian large-$N$ limit requires that there be a hierarchy between the mass of the heaviest dual photon and the lightest W-boson.

Next, we consider the monopole-instanton induced potential. Its strength is given by (\ref{vmzero}). We recall that one of the $g^2$ factors in $V_m$ comes from the   gaugino mass insertion, $\sim {m\over g^2} \lambda \lambda$ (\ref{lambda1}). We also recall that in softly-broken four-dimensional SYM the ratio ${m\over c_2 g^2}$ is a one-loop RG invariant (the quantity  $\alpha m \over \beta(\alpha)$,  with $\alpha = {g^2\over 4\pi}$, is an invariant to all loop orders in $\alpha$ and leading order in $m$ \cite{Hisano:1997ua}).
We shall keep the value of $\hat{m} = m/(c_2 g^2)$ in $V_m^0$ as our parameter to be adjustable in what follows:
\begin{equation}
 \label{vmonzero}
V_{m}^0= {24 m \kappa' R \over  g^2  |v|} \frac{16 \pi^2}{3 c_2 g^2 R^3}\;e^{-\frac{8\pi^2}{c_2g^2}} ={3  \kappa' \over \  |v|}\; {\hat{m}} \; {\Lambda^3 \over m_W} ~, ~~ \hat{m} = {m \over c_2 g^2}~.
 \end{equation}
 Finally, we find that the dimensionless parameter $c_m$ has a finite large-$N$ limit and, up to inessential constants, with $\log {m_W\over \Lambda}$ accuracy is given by: 
 \begin{equation}
 \label{cmdefined}
 c_m = {V_m^0 \over V_{\rm bion}^0} \sim  {\hat{m} \; m_W^2 \over \Lambda^3 } ~.
  \end{equation}
  We note that the two potentials in (\ref{bosonic potential}) are of similar order of magnitude for   $c_m$ of order unity, which is when the transitions we study occur.
  
Finally, we come to the GPY potential (\ref{gauginogpy}), which we expand around the supersymmetric value (\ref{phivev}), i.e. substitute (\ref{phi11}) $\pmb \phi = \pmb \phi_0 + {g^2 \over 4 \pi} \pmb b $, to find:
\begin{equation}
\label{Vgpy2} 
V_{GPY}^{{\cal O}(m^2)} = - \frac{m^2}{2 \pi^3 R}\sum_{p=1}^\infty \sum_{\pmb \beta_+} {\cos({ 2 \pi p\over c_2} \pmb \beta\cdot \pmb \rho  + p {g^2 \over 4 \pi} \pmb \beta \cdot (\pmb\phi_0^{(1)} + \pmb b))\over p^2} ~.
\end{equation}
We see that the $\pmb b$ dependence of $V_{GPY}$ arises at order at least---if linear terms in $\pmb b$ actually arise from (\ref{Vgpy2}) (for $SU(2)$, they do not)---$V_{GPY}^0 \le {m^2 g^2 \over R} = \left({m  \over c_2 g^2}\right)^2 {(c_2  g^2)^3 \over c_2 R} = \hat{m}^2 m_W (c_2  g^2)^3$. Thus, comparing the GPY potential to the bion and monopole contributions, we have, for values of $\hat{m} \sim \Lambda^3 m_W^{-2}$ such that $c_m \sim 1$, i.e.  near the transition:\footnote{Accounting for the logarithmic precision of (\ref{cmdefined}) does not change this conclusion.}
\begin{equation}
\label{cpert}
c_{GPY} \equiv {V^{0}_{GPY}\over V_{\rm bion}^0} \le (c_2  g^2)^4 \sim {1 \over (\log  {m_W \over \Lambda})^4 } ~.
\end{equation}
Remembering that within the region of validity of semiclassics, we have $m_W \gg \Lambda$, we find that the GPY potential is suppressed in the semiclassical regime by at least the factor on the r.h.s. of (\ref{cpert}).

\section{The phase structure of  SYM$\mathbf{^*}$  on $\mathbf{\R^{  3}\times  \S^{  1}}$, center symmetry, and deconfinement}
\label{results}

In this Section, we study the discontinuous change of the $\Omega$ eigenvalue distribution as $c_m$ is increased, as outlined in the Introduction.
For every gauge group, we study the realization of center symmetry, whenever present, or the shift of the vacuum away from the supersymmetric one ($\pmb b = \pmb \sigma' = 0)$. We do this by minimizing the total potential (\ref{vtotal}) with respect to $\pmb b$ and $\pmb \sigma'$ for increasing values of the parameter $c_m$. In each case, we find that as $c_m$ is increased, there is a discontinuous transition to a new vacuum, where  $\pmb \sigma' =0$  but $\pmb b \ne 0$ ( for $\theta=0$ .) 
In each theory, we study the trace of the Polyakov loop as well as its two-point correlation function. We now give more details and describe the interpretation of the results.

\subsection{Polyakov loop, correlator, and string tension}
\label{results41}

To quantify the discontinuous nature of the transition and to study the breaking of the center symmetry (where appropriate) we calculate the expectation value of the Polyakov loop in the relevant representation. For general groups, we compute the trace to leading order in $g^2$:
\begin{eqnarray}
\label{traceomega}
\mbox{Tr}\langle\Omega\rangle=\sum_{\pmb \nu}e^{i\pmb \nu \cdot \langle \pmb\phi \rangle}\cong\sum_{\pmb \nu}e^{i\pmb \nu \cdot {\pmb\phi_0}^{(0)}}\left(1+i\frac{g^2}{4\pi}\pmb \nu \cdot\pmb \phi_0^{(1)}+i\frac{g^2}{4\pi}\pmb \nu \cdot \langle \pmb b \rangle \right)\,.
\end{eqnarray} 
Here $\pmb \nu$ denotes the weights of the corresponding representation and $\langle \pmb b \rangle$ is the expectation value above the transition. We define the magnitude of the discontinuity in the Polyakov loop as:
\begin{eqnarray}
\frac{4\pi}{g^2c_2}|\mbox{Tr}\langle\Delta\Omega\rangle|=\frac{1}{c_2}\sum_{\pmb \nu}e^{i\pmb \nu \cdot {\pmb\phi_0}^{(0)}}\pmb \nu\cdot \langle \pmb b \rangle\,.
\end{eqnarray} 
For the $Sp(2N)$ theories we use the fundamental representation as a probe of center symmetry breaking. For $Spin(2N+1)$ we use the spinor representation and for $Spin(2N)$ we use the two chiral spinor representations to probe center symmetry breaking. For $E_6$  ($E_7$) we have not calculated traces of $\Omega$ in any representation; instead, we have verified, using the action of the $\Z_3$ ($\Z_2$) center on $\pmb b$ (given in the appropriate Section below) that there is center-symmetry breaking with the required number of degenerate vacua appearing above $c_{\rm cr}$, see the relevant Sections below.

For theories with center symmetry\footnote{For $G_2, F_4$, and $E_8$ theories without center symmetry, the only difference is that an $r$-independent constant contribution can  also appear at $c_m=0$ and $c_m = c_{{\rm cr}_-}$. See Eq.~(\ref{polyakov loop correlator coefficients G2}) for $\langle \Omega(x)\Omega^\dagger(y)\rangle$ in the $G_2$ case.}  we parametrize the Polyakov loop correlator  as follows
\begin{eqnarray}
\langle\Tr \Omega(x) \Tr \Omega^\dagger(y)\rangle=\left\{ \begin{array}{cc}\frac{g^2 R}{4r} \; A_0 \; e^{-\hat\sigma_0  m_0 r}\,,& c_m=0\\
\frac{g^2 R}{4r} \; A_{\rm cr} \; e^{-\hat\sigma_{\rm cr} m_0 r}\,,&~\;\; c_m=c_{\mbox{\scriptsize cr}_-} \\ 
\left(\frac{g^2}{4\pi^2}\right)^2 D_{\rm cr_+}  +
\frac{g^2 R}{4r} \;C \;e^{-\lambda_C  m_0 r} \,,&~~ c_m=c_{\mbox{\scriptsize cr}_+}\,,
\end{array}  \right.
\label{polyakov loop correlator coefficients}
\end{eqnarray}
 where $r = |x-y|$. We compute $\langle\Tr \Omega(x)\Tr\Omega(y)^\dagger\rangle$ in the supersymmetric vacuum  at zero gaugino mass (at $c_m=0$), just below the transition ($c_m = c_{{\rm cr}_-}$) and just above the transition ($c_m = c_{{\rm cr}_+}$). See Appendix \ref{polyakovloop} for a derivation of the formulae we use to obtain (\ref{polyakov loop correlator coefficients}).
 The numerical values of $A_0,\hat\sigma_0$,  $A_{\rm cr},\hat\sigma_{\rm cr}$, $D_{\rm cr}, C$, and $\lambda_C$ are displayed in tables for  each gauge group. A  discussion of their significance is given below.
 
 The scale $m_0$ appearing in the exponents in (\ref{polyakov loop correlator coefficients}) is given by 
 $m_0^2 = {16 \pi^3 R \over g^2(m_W)} V_{\rm bion}^0$. Thus, $m_0$  can  be written, via (\ref{vbionzero}), as:
 \begin{equation}
 \label{m0}
 m_0  = 18 \sqrt{2} \; {\kappa\; \Lambda^3 \over |v| \;m_W^2} \; \log {m_W \over \Lambda}~,
\end{equation}
a quantity which is ${\cal O}(N^0)$ in the large-$N$ limit (recall $\Lambda$ and $m_W = (c_2 R)^{-1}$ are fixed in the large-$c_2$ limit). When computing (\ref{polyakov loop correlator coefficients}), in each case we kept only the leading decaying exponential.

It is instructive to discuss the interpretation of the correlator (\ref{polyakov loop correlator coefficients}) in the simplest case of $Sp(2)$ ($=SU(2)$), where it can be easily computed analytically, as this is the only $r=1$ group. While $SU(2)$ SYM was already discussed in \cite{Poppitz:2012sw}, the Polyakov loop correlator was not studied there (see \cite{Poppitz:2013zqa} for a related discussion for $SU(2)$ massive SQCD). The trace of the fundamental representation Polyakov loop operator (\ref{traceomega}) in $Sp(2)$, where $\pmb \phi_0^{(1)} =0$, is $\Tr  \Omega(x) = i  (e^{i {g^2 \over 4 \pi} b(x)} - e^{- i{g^2 \over 4 \pi} b(x)} )$. Its vanishing in the supersymmetric vacuum $\langle b \rangle =0$ is evident. The two point correlator, computed in the $\langle b \rangle = 0$ vacuum, i.e. below the transition, follows immediately from the expressions in Appendix \ref{polyakovloop} and yields\footnote{Recall that in $SU(2)$ the transition is second order and  $\hat\sigma$, a number of order unity at $c_m=0$, smoothly goes to zero as $c_m$ approaches the critical value. This is inessential for our present discussion.}
\begin{equation}
\label{su2correlator}
\langle \Tr  \Omega(x) \Tr \Omega^\dagger(0)\rangle = 2 \sinh {g^2 R \over 4 r} e^{ -\hat\sigma m_0 r}\bigg\vert_{ r \gg {g^2 R \over 4}} \sim {g^2 R  \over 2 r} e^{ - \hat\sigma m_0 r}~.
\end{equation}
As already indicated on the r.h.s., at distances $r > {\cal O}(g^2 R)$, the correlator can always be approximated by the leading term in the expansion of the $\sinh$ and is thus within the parametrization of (\ref{polyakov loop correlator coefficients}) with $A=2$. Also recall that our effective three-dimensional description of the dynamics is valid at scales $r>{\cal O}(R)$, the inverse lowest W-boson mass. Hence, in the entire region of validity of the three dimensional description, the parametrization given in  (\ref{polyakov loop correlator coefficients}) is expected to hold. 
 
Next, we recall that the scale $m_0^{-1}$ is exponentially large compared to the compactification radius $R$,  as $m_0^{-1} \sim R e^{ {8 \pi^2 \over c_2 g^2}}$ (the Debye screening length in the magnetic bion plasma, responsible for the dual photon mass, is of order $(Rm_0)^{-{1\over 3}}$ in units of $R$). At scales between $R$ and $m_0^{-1}$, the exponential is negligible and the behavior of the correlator is determined by the $1/r$ term. On the other hand, at scales  larger than the screening length, the Polyakov loop correlator is dominated by the exponential term. Thus, as in the confined phase of thermal Yang-Mills theory, we have  \begin{equation}
\label{su2correlator1}
\langle \Tr  \Omega(x) \Tr \Omega^\dagger(0)\rangle\bigg\vert_{ r \gg  m_0^{-1}} \simeq  e^{ -  {\hat\sigma m_0 \over R} r R} \equiv e^{ - \sigma \;  r R}~.
\end{equation}
The above equation defines the ``string tension"   in the relevant representation, $\sigma = \hat\sigma m_0 R^{-1}$. The numbers $\hat\sigma$ shown for the various groups will correspondingly be called ``dimensionless string tensions".

Above the transition, the field $\pmb b$ (for all groups we study) acquires an expectation value and the correlator (\ref{polyakov loop correlator coefficients}) acquires the constant term $D_{\rm cr_+}$. This term dominates the behavior of 
$\langle \Tr  \Omega(x) \Tr \Omega^\dagger(0)\rangle$ at large distances and leads to a vanishing string tension at $c>c_{\rm cr}$. In all theories we study in this paper, this vanishing is discontinuous. This behavior of the string tension as a function of the dimensionless $c_m$, for one representative case, $SO(7)$, is displayed on Fig. \ref{fig:string}, already shown in the Introduction.

\subsection{The symplectic and orthogonal groups}
\label{results42}

The $Sp(4)$ theory is one of the few non-$SU(N)$ pure Yang-Mills theories in $\R^4$ whose thermal physics has been studied on the lattice. Despite the fact that the $\Z_2$ center symmetry allows for a continuous transition in the 3d Ising universality class, it was found that the transition is first order \cite{Holland:2003kg,Holland:2003mc}.  The authors  conjectured that the large change of the number of relevant degrees of freedom below and above the transition in an $Sp(2N)$ (as well as $SO(N)$) theory may be responsible for a first order nature of the transition (there are order $N^0$ confined degrees of freedom while the number of gluons liberated above $T_c$ scales as $N^2).$\footnote{At large $N$ a $\Z_N \rightarrow U(1)$ center symmetry emerges in $SO(N)$ and $Sp(2N)$ theories and one expects  a discontinuous transition similar to the one in $SU(N)$ theories, see  \cite{Armoni:2007kd}.} 
The authors also conjectured that  the transition is first order for all gauge groups but $SU(2)$. Indeed, we shall see that our findings---if, indeed, our quantum phase transition continuously connects to the corresponding deconfinement transition in pure YM---confirm this. 

We note also that the deconfinement phase transition in  $Sp(4)$ and $E_7$, both with a $\Z_2$ center, was studied in the framework of the functional renormalization group, see Ref.~\cite{Braun:2010cy}, whose results also indicated a first order transition. The small-$\S^3 \times \S^1_\beta$ studies mentioned in the Introduction were also performed for general groups in \cite{HoyosBadajoz:2007ds}, where, similarly, a  first-order large-$N$ transition was found.

\subsubsection{$\mathbf{Sp(2N)}$}

This is the symplectic group which is defined as the set of $2N \times 2N$ unitary matrices ${\cal M}$ which preserve the anti-symmetric scalar product
\begin{eqnarray}
{\cal M}^T J{\cal M}=J\,,\quad\mbox{where}\,\, J=\left(\begin{array}{cc}0& \mathbb{I}_{N}\\-\mathbb{I}_{N}&0\end{array}\right)\,.
\end{eqnarray}
The roots can be expressed in terms of the standard orthonormal basis $\{
\pmb e_i\}\,,i=1,2,...N$ ($\pmb e_i \cdot \pmb e_j = \delta_{ij}$), of an $N$-dimensional space. The rank of the group is $N$ and the simple roots are given by 
\begin{eqnarray}
\{\pmb\alpha\}=\left\{(\pmb e_i-\pmb e_{i+1})/\sqrt{2}\,, 1\leq i\leq N-1,\mbox{and}\,\, \pmb \alpha_{N}=\sqrt{2}\pmb e_N\right\} \,,
\end{eqnarray}
and the affine root is $\pmb \alpha_0=-\sqrt{2}\pmb e_{1}$. The set of the positive roots is
\begin{eqnarray}
\{\pmb\beta_+\} =\left\{(\pmb e_i\pm\pmb e_{j})/\sqrt{2}, 1\leq i< j\leq N\,, \mbox{and}\, \sqrt{2}\pmb e_{i}, 1\leq i\leq N\right\}\,.
\end{eqnarray} 
The dual Kac labels are (we always start with $k_0^*=1$)  $\{k_a^*\}=\{1,1,...,1\}$, and the dual Coxeter number $c_2=N+1$. The weights of the fundamental representation are given by
\begin{eqnarray}
\{\pmb \nu\}=\left\{\begin{array}{c} \frac{\pmb e_i}{\sqrt{2}}\,,\quad i=1,...,N\\ -\frac{\pmb e_i}{\sqrt{2}}\,,\quad i=N,...,2N\,.  \end{array}  \right.
\end{eqnarray}
The symplectic group has a $\mathbb Z_2$ center symmetry which acts on the modulus field $\pmb \phi$ as
\begin{eqnarray}
\phi_i\rightarrow \pi -\phi_{N+1-i}\,,\quad i=1,2,...N\,.
\label{center action sp}
\end{eqnarray}
In fact, it is trivial to check that the supersymmetric minimum is invariant under the $\mathbb Z_2$ center.

By minimizing the full potential $V_T = V_{\mbox{\scriptsize bion}}+c_m V_m$, we find that the broken minima correspond to a first order phase transition for all $N\ge2$. For each case, we find that there are two broken minima which are mapped to each other via (\ref{center action sp}). From Table~\ref{SP group values}, we see that the critical value of $c_m$ decreases as we increase $N$.  In Table  \ref{SP group values}, we give the numerical value of the discontinuity of the Polyakov loop $\frac{4\pi}{g^2c_2}\times \Delta |\mbox{Tr}\langle \Omega\rangle |$ for various values of $N$. The discontinuity of the trace of the Polyakov loop, normalized  to unity, can be fitted as follows
\begin{eqnarray}
\label{fit}
{1 \over 2N} |\mbox{Tr}\langle\Delta\Omega\rangle|\cong \left(0.217 - {0.178 \over N} - {0.005\over N^2}\right) \frac{g^2c_2}{4\pi}
\end{eqnarray} 
as appropriate to the large $N$ limit. However, we caution against taking the fit seriously, and especially against a comparison with the numbers  with the similar $SU(N)$ fit from \cite{Poppitz:2012nz} (motivated by  large-$N$ (orbifold) equivalences \cite{Lovelace:1982hz,Unsal:2006pj}) as the values of $N = 2,...,7$ are rather small; nonetheless, the behavior in (\ref{fit}) is qualitatively similar to the one found for $SU(N)$.

\begin{table}
\centerline{
\begin{tabular}{|c|c|c|c|c|c|c|c|c|c|}
\hline
    & $c_{\mbox{\scriptsize cr}}$& $\frac{4\pi}{g^2c_2}\times  |\mbox{Tr}\langle \Delta\Omega\rangle |$&   $A_0$ &  $\hat\sigma_0$& $A_{\rm cr}$&  $\hat\sigma_{\rm cr}$ & $C$ & $\lambda_C$ &$D_{\rm cr_+}$  \\ \hline
    $sp(4)$ &  $3.8$ & $0.501$ & $3$ & $2.828$ & $3$ & $0.593$ & $2.756$  & $0.878$ & $2.26$\\ \hline  
    $sp(6)$ &  $2.08$ & $0.963$ & $4$ & $1.658$ & $4$ & $0.555$ & $3.365$  & $0.811$ & $14.834$\\ \hline 
    $sp(8)$ &  $1.276$ & $1.366$ & $5$ & $1.08$ & $5$ & $0.438$ & $3.96$  & $0.628$ & $46.6$\\ \hline 
    $sp(10)$ &  $0.854$ & $1.732$ & $6$ & $0.758$ & $6$ & $0.342$ & $4.56$  & $0.478$ & $108.017$\\ \hline  
    $sp(12)$ &  $0.614$ & $2.32$ & $7$ & $0.56$ & $7$ & $0.265$ & $5.16$  & $0.46$ & $263.91$\\ \hline
    \end{tabular}
}
\caption{The critical mass, $c_{\mbox{\scriptsize cr}}$, the fundamental representation Polyakov loop and the values of the coefficients (\ref{polyakov loop correlator coefficients})
 of its two-point correlator for the $Sp(2N)$ group, including the dimensionless string tensions $\hat\sigma$. }
\label{SP group values}
\end{table}

In the present framework of softly broken $Sp(2N)$ SYM on $\R^3 \times \S^1$ the transition remains first order, despite the fact that in the abelian semiclassical regime there is no change to the number of degrees of freedom immediately above and below the transition. 
At a technical level, the discontinuous nature can be traced to the form of the bion potential $V_{\rm bion}$, which contains cubic terms in the $\pmb b$ fields for all groups but $SU(2)$.

\subsubsection*{The $\theta$ dependence of $\mathbf{Sp(2N)}$}

In this section, we study the dependence of the phase transition on the vacuum angle $\theta$.  As is evident from the total potential (\ref{vtotal}), for a given $\theta$, we have $u=1,2,...c_2$ different branches. Thus, one needs to select the theory that corresponds to the true vacuum energy.  Setting $\pmb b=\pmb \sigma'=0$, we find that the correct theory is the one that corresponds to the minimum vacuum energy density:
\begin{eqnarray}
\rho_{\mbox{\scriptsize vacuum}}= - V^0_{\mbox{\scriptsize bion}} \; c_m \; \times\mbox{Min}_u\left\{\cos\left(\frac{\theta+2\pi u}{c_2}\right)\right\} \,.
\end{eqnarray}
In the following, and for all the subsequent groups, we perform our calculations for $\theta$ in the range $0\leq \theta<\pi$. Hence, we find that the branch $u=c_2$ corresponds to the vacuum branch. 

We take $\mathbf{Sp(6)}$ as a sample group. As in the case of zero vacuum angle, we find that the transition is first order for all $\theta$ in the chosen range. However, contrary to the case $\theta=0$, where all the broken minima happen at $\pmb b \neq 0$ and $\pmb \sigma'=0$,  the broken minima in the presence of a non-zero $\theta$ occur  at $\pmb b \neq 0$ and $\pmb \sigma'\neq 0$. 

In Table~\ref{Theta for sp6}  we see that the critical value of the mass parameter $c_{\mbox{\scriptsize cr}}$ is a decreasing function of $\theta$. In addition, we find that the discontinuity of the Polyakov loop increases with increasing $\theta$. These findings have the same qualitative behavior as in $SU(N)$ case found in \cite{Anber:2013sga}, and confirmed by lattice simulations in \cite{D'Elia:2012vv,D'Elia:2013eua}.

\begin{table}
\centerline{
\begin{tabular}{|c||c|c|c|c|c|c|c|c|c|c|c|}
\hline
 $\theta$ & $0$ & $\frac{\pi}{12}$ & $\frac{2\pi}{12}$ & $\frac{3\pi}{12}$ & $\frac{4\pi}{12}$ & $\frac{5\pi}{12}$ & $\frac{6\pi}{12}$ & $\frac{7\pi}{12}$ & $\frac{8\pi}{12}$ & $\frac{9\pi}{12}$ & $\frac{10\pi}{12}$ \\\hline
 $c_{\mbox{\scriptsize cr}} $ & $2.08$ & $2.079$ & $2.077$ & $2.072$ & $2.066$ & $2.057$ & $2.045$ & $2.03$ & $2.011$ & $1.987$ & $1.954$ \\\hline
 $\mbox{Tr}\langle\Delta\Omega(\theta) \rangle $ & $3.852$ & $3.856$ & $3.877$ & $3.903$ & $3.948$ & $4.002$ & $4.070$ & $4.156$ & $4.259$ & $4.38$ & $4.512$ \\\hline
      \end{tabular}
}
\caption{The $\theta$-dependence of the critical transition mass and the Polyakov loop discontinuity for the $Sp(6)$ group. }
\label{Theta for sp6}
\end{table}


\subsubsection{$\mathbf{Spin(2N)}$}

This is the double cover of the special orthogonal group $SO(2N)$. The rank of the group is $N$ and the  simple and positive roots are given by
\begin{eqnarray}
\nonumber
\{\pmb\alpha\}&=&\left\{\pmb e_i-\pmb e_{i+1}\,, 1\leq i\leq N-1, \mbox{and}\,\, \pmb \alpha_{N}=\pmb e_{N-1}+\pmb e_N\,\right\}\,,\\
\{\pmb\beta_+\} &=&\left\{\pmb e_i\pm\pmb e_{j}, 1\leq i< j\leq N\right\}\,.
\end{eqnarray}
The affine root is $\pmb\alpha_0=-\pmb e_1-\pmb e_2$, the dual Kac labels are $\{k_a^*\}=\{1,1,2,...,2,1,1\}$, and the dual Coxeter number is $c_2=2N-2$. The group has two spinor representations which are proper representations for the $Spin(2N)$ but not for the group $SO(2N)$. The weights of the two spinor representations are of the form 
\begin{eqnarray}
\{\pmb \nu\}_{1,2}=\left(\pm \frac{1}{2},\pm \frac{1}{2},,...,\pm \frac{1}{2},\right)\,.
\end{eqnarray}    
The two representations are distinguished by having an even or odd number of negative signs, and their dimension is $2^{N-1}$. The $Spin(2N)$ group has a $\mathbb Z_4$ or a $\mathbb Z_2 \times \mathbb Z_2$ center group depending on weather $N$ is odd or even, respectively. If $N$ is odd, then the action of the $Z_4$ center symmetry on the modulus $\pmb \phi$ is given by
\begin{eqnarray}
\phi_i \rightarrow \left\{ \begin{array}{cc}\pi +\phi_N&\mbox{for}\,\, i=1\,,\\\pi-\phi_{N+1-i}&\mbox{for}\,\, i>1\,.  \end{array} \right.
\end{eqnarray}
For even $N$, the action of the $\mathbb Z_2\times \mathbb Z_2$ center is
\begin{eqnarray}
\phi_i \rightarrow \left\{ \begin{array}{cc}\pi \pm\phi_N&\mbox{for}\,\, i=1\,,\\\pi-\phi_{N+1-i}&\mbox{for}\,\, i>1\,,\\ \mp \pi \pm \phi_1&\mbox{for}\,i=N\,.  \end{array} \right.
\end{eqnarray}%

By examining the full potential, we find that the transition is first order for all $N\geq 3$. The critical mass values $c_{\mbox{\scriptsize cr}}$, the discontinuity of the Polyakov loop, as well as the parameters of the Polyakov loop correlator that appear in (\ref{polyakov loop correlator coefficients}) are listed in table (\ref{Spin even group values}) (we notice that we do not normalize the discontinuity of the Polyakov loop here as the spinor  representation has an exponentially large dimension). 

Both for the even- and odd- dimensional $Spin(N)$ groups, we observe, from Tables \ref{Spin even group values} and \ref{Spin odd group values} that the dimensionless string tension measured by the Polyakov loop correlator  at $c_m=0$ has a constant value, equal to $2 \sqrt{2}$. We note that this is not the lightest mass of any of the dual photons/$\pmb b$-fields and the appearance of an $N$-independent string tension is solely for group theory reasons. We also note that the decrease of the dimensionless string tension from $c_m=0$ to $c_m= c_{\rm cr}$ is much smaller (a few percent) than for $Sp(2N)$ (a factor of unity).

\begin{table}
\centerline{
\begin{tabular}{|c|c|c|c|c|c|c|c|c|c|}
\hline
    & $c_{\mbox{\scriptsize cr}}$& $\frac{4\pi}{g^2}\times  |\mbox{Tr}\langle \Delta \Omega\rangle |$&   $A_0$ &  $\hat\sigma_0$& $A_{\rm cr}$&  $\hat\sigma_{\rm cr}$ & $C$ & $\lambda_C$ &$D_{\rm cr_+}$  \\ \hline
    $spin(6)$ &  $2.953$ & $2.277$ & $4$ & $2\sqrt{2}$ & $4$ & $1.447$ & $1.605$  & $1.35$ & $5.18$\\ \hline  
    $spin(8)$ &  $1.689$ & $4.33$ & $6$ & $2\sqrt{2}$ & $6$ & $2.149$ & $1.93$  & $1.03$ & $18.75$\\ \hline 
    $spin(10)$ &  $1.063$ & $6.36$ & $8$ & $2\sqrt{2}$ & $8$ & $2.42$ & $2.134$  & $0.713$ & $40.43$\\ \hline 
    $spin(12)$ &  $0.727$ & $8.46$ & $10$ & $2\sqrt{2}$ & $10$ & $2.558$ & $2.27$  & $0.503$ & $71.716$\\ \hline  
    $spin(14)$ &  $0.527$ & $10.66$ & $12$ & $2\sqrt{2}$ & $12$ & $2.635$ & $2.37$  & $0.36$ & $113.659$\\ \hline
     $spin(16)$ &  $0.4$ & $13$ & $14$ & $2\sqrt{2}$ & $14$ & $2.68$ & $2.45$  & $0.28$ & $169$ \\ \hline
      \end{tabular}
}
\caption{The critical mass, $c_{\mbox{\scriptsize cr}}$, the spinor representation Polyakov loop and its correlator  coefficients (\ref{polyakov loop correlator coefficients}) for the $Spin(2N)$ group. The string tensions $\hat{\sigma}$ and other coefficients measured by the two spinor representations are identical. }
\label{Spin even group values}
\end{table}

 
To examine the effect of the vacuum  angle $\theta$ on the nature of phase transition, we take $Spin(6)$ as a representative sample. We find that the transition is first order, and that the critical value of the mass parameter $c_{\mbox{\scriptsize cr}}$ decreases with increasing $\theta$. We also find that the discontinuity of the Polyakov loop increases with $\theta$. This is illustrated in Table~\ref{Theta for so6}  and Figure~\ref{fig:SO6 c and omega}.

\begin{table}
\centerline{
\begin{tabular}{|c||c|c|c|c|c|c|c|c|c|c|c|}
\hline
 $\theta$ & $0$ & $\frac{\pi}{12}$ & $\frac{2\pi}{12}$ & $\frac{3\pi}{12}$ & $\frac{4\pi}{12}$ & $\frac{5\pi}{12}$ & $\frac{6\pi}{12}$ & $\frac{7\pi}{12}$ & $\frac{8\pi}{12}$ & $\frac{9\pi}{12}$ & $\frac{10\pi}{12}$ \\\hline
 $c_{\mbox{\scriptsize cr}} $ & $2.953$ & $2.951$ & $2.946$ & $2.937$ & $2.925$ & $2.908$ & $2.886$ & $2.859$ & $2.825$ & $2.782$ & $2.729$ \\\hline
 $\mbox{Tr}\langle\Delta\Omega(\theta) \rangle $ & $2.277$ & $2.279$ & $2.287$ & $2.299$ & $2.318$ & $2.342$ & $2.371$ & $2.407$ & $2.448$ & $2.493$ & $2.540$ \\\hline
      \end{tabular}
}
\caption{The $\theta$-dependence of the critical transition mass and the Polyakov loop discontinuity for the $Spin(6)$ group. }
\label{Theta for so6}
\end{table}


	\subsubsection{$\mathbf{Spin(2N+1)}$}

This is the double cover of the special orthogonal group $SO(2N+1)$. The rank is $N$ and the   simple and positive roots are given by
\begin{eqnarray}
\nonumber
\{\pmb\alpha\}&=&\left\{\pmb e_i-\pmb e_{i+1}\,, 1\leq i\leq N-1, \mbox{and}\,\, \pmb \alpha_{N}=\pmb e_N\,\right\}\,,\\
\{\pmb\beta_+\} &=&\left\{\pmb e_i\pm\pmb e_{j}, 1\leq i< j\leq N\,,\mbox{and}\, \pmb e_i\,, 1\leq i \leq N\right\}\,.
\end{eqnarray}
The affine root  is $\pmb \alpha_0=-\pmb e_1-\pmb e_2$, the dual Kac labels are $\{k_a^*\}=\{1,1,2,...,2,1\}$, and the dual Coxeter number is $c_2=2N-1$. The $Spin(2N+1)$ group has a single spinor representation of dimension $2^N$ with weights:
\begin{eqnarray}
\{\pmb \nu\}=\left(\pm \frac{1}{2},\pm \frac{1}{2},,...,\pm \frac{1}{2},\right)\,.
\end{eqnarray}    
In addition, the group has a $\mathbb Z_2$ center which acts on the modulus $\pmb \phi$ as
\begin{eqnarray}
\phi_i\rightarrow \left\{\begin{array}{cc} 1-\phi_1& \mbox{for}\, i=1\\ \phi_i& \mbox{for}\, j>1\,.  \end{array} \right.
\end{eqnarray}
For both even and odd $Spin(N)$ gauge theories we observe a first order non universal transition despite the availability of universality classes associated with the center symmetry. To the best of our knowledge, no lattice simulations have been performed for $Spin(N)$ pure YM theories.

\begin{table}
\centerline{
\begin{tabular}{|c|c|c|c|c|c|c|c|c|c|}
\hline
    & $c_{\mbox{\scriptsize cr}}$& $\frac{4\pi}{g^2}\times  |\mbox{Tr}\langle \Delta \Omega\rangle |$&    $A_0$ &  $\hat\sigma_0$& $A_{\rm cr}$&  $\hat\sigma_{\rm cr}$ & $C$ & $\lambda_C$ &$D_{\rm cr_+}$   \\ \hline
    $spin(5)$ &  $3.8$ & $1.503$ & $3$ & $2\sqrt{2}$ & $3$ & $0.593$ & $2.756$  & $0.878$ & $2.26$\\ \hline  
    $spin(7)$ &  $2.21$ & $4.69$ & $5$ & $2\sqrt{2}$ & $5$ & $1.89$ & $3.57$  & $1.23$ & $22$\\ \hline 
    $spin(9)$ &  $1.324$ & $7.55$ & $7$ & $2\sqrt{2}$ & $7$ & $2.3$ & $4$  & $0.86$ & $57$\\ \hline 
    $spin(11)$ &  $0.871$ & $10.46$ & $9$ & $2\sqrt{2}$ & $9$ & $2.48$ & $4.42$  & $0.595$ & $109.4$\\ \hline  
    $spin(13)$ &  $0.615$ & $13.58$ & $11$ & $2\sqrt{2}$ & $11$ & $2.6$ & $4.65$  & $0.42$ & $182$\\ \hline
    $spin(15)$ &  $0.457$ & $16.7$ & $13$ & $2\sqrt{2}$ & $13$ & $2.66$ & $4.8$  & $0.32$ & $278$\\ \hline
      \end{tabular}
}
\caption{The critical mass, $c_{\mbox{\scriptsize cr}}$, the spinor representation Polyakov loop and its correlator  coefficients (\ref{polyakov loop correlator coefficients})  for the $Spin(2N+1)$ group. }
\label{Spin odd group values}
\end{table}

The dependence of $c_{\mbox{\scriptsize cr}}$ and the discontinuity of the Polyakov loop on $\theta$ for $Spin(7)$ are illustrated in Table~\ref{Theta for so7}. The qualitative behavior  of these quantities exactly matches the previously studied group, and we do not give any further comments.

\begin{table}
\centerline{
\begin{tabular}{|c||c|c|c|c|c|c|c|c|c|c|c|}
\hline
 $\theta$ & $0$ & $\frac{\pi}{12}$ & $\frac{2\pi}{12}$ & $\frac{3\pi}{12}$ & $\frac{4\pi}{12}$ & $\frac{5\pi}{12}$ & $\frac{6\pi}{12}$ & $\frac{7\pi}{12}$ & $\frac{8\pi}{12}$ & $\frac{9\pi}{12}$ & $\frac{10\pi}{12}$ \\\hline
 $c_{\mbox{\scriptsize cr}} $ & $2.211$ & $2.210$ & $2.206$ & $2.199$ & $2.189$ & $2.175$ & $2.159$ & $2.138$ & $2.113$ & $2.083$ & $2.047$ \\\hline
 $\mbox{Tr}\langle\Delta\Omega(\theta) \rangle $ & $4.696$ & $4.699$ & $4.707$ & $4.719$ & $4.735$ & $4.753$ & $4.778$ & $4.8$ & $4.824$ & $4.842$ & $4.847$ \\\hline
      \end{tabular}
}
\caption{The $\theta$-dependence of the critical transition mass and the Polyakov loop discontinuity for the $Spin(7)$ group. }
\label{Theta for so7}
\end{table}


\subsection{Exceptional groups with  center symmetry}

\label{results43}

The pattern for the exceptional groups with center symmetry, $E_6$ and $E_7$, is as in all cases with center symmetry. Upon increasing the soft gaugino mass with $\Lambda$ and $m_W$ fixed, we find a center-breaking first order phase transition. In both these cases, we verify that there exist 3 (for $E_6$) and 2 (for $E_7$) degenerate broken vacua. We have not calculated any Polyakov loop traces or correlators, but simply verified the symmetry breaking pattern and determined the critical values of $c_m$. In the next two subsections, we present our findings. 

\subsubsection{$\mathbf{E_6}$}

This is the rank $6$ special group with dimension $78$. A convenient way to express the roots of this group is to write all quantities in terms of the standard unit vectors in eight dimensional space $\mathbb R^8$.   The set of the simple and positive roots are given by
\begin{eqnarray}
\nonumber
\{\pmb\alpha\}&=&\left\{\pmb \alpha_i=\pmb e_{i+2}-\pmb e_{i+1}\,(1\geq i\geq 5),\, \pmb \alpha_6=\frac{1}{2}\left(\sum_{i=1}^4\pmb e_i-\sum_{i=5}^8 \pmb e_i\right)\right\}\,,\\
\nonumber
\{\pmb\beta\}&=&\left\{\frac{1}{2}\left(\pmb e_1+\sum_{i=2}^7(-1)^{n_i}\pmb e_i-\pmb e_8\right)\left(\mbox{three odd} \,n_i\right)\,, \pmb e_i-\pmb e_j\,(7\geq i >j\geq 2)\,, \pmb e_1-\pmb e_8   \right\}\,.\\
\end{eqnarray}
The affine root is $\pmb \alpha_0=-\pmb e_1+\pmb e_8$, and the dual Kac labels are $\{k_a^*\}=\{1,1,2,3,2,1,2\}$, and the dual Coxeter number is $c_2=12$. The roots  lay on the intersection of two hyper planes such that $\phi_1+\phi_8=\sum_{i=2}^7\phi_i=0$. These conditions are used to eliminate $\phi_7$ and $\phi_8$.
The group has a $\mathbb Z_3$ center which acts on the modulus $\pmb \phi$ as:
\begin{eqnarray}
\left(\begin{array}{c}\phi_1\\\phi_2\\\phi_3\\\phi_4\\\phi_5\\\phi_6\end{array}\right)\rightarrow \left(\begin{array}{c}\pi+\frac{1}{2}\phi_2-\frac{1}{2}\phi_3\\-\frac{\pi}{3}+\frac{1}{2}\phi_2+\frac{1}{2}\phi_3+\sum_{i=4}^6\phi_i\\  -\frac{\pi}{3}-\frac{1}{2}\phi_2-\frac{1}{2}\phi_3-\phi_6\\ -\frac{\pi}{3}-\frac{1}{2}\phi_2-\frac{1}{2}\phi_3-\phi_5\\ -\frac{\pi}{3}-\frac{1}{2}\phi_2-\frac{1}{2}\phi_3-\phi_4\\-\frac{\pi}{3}+\frac{1}{2}\phi_2+\frac{1}{2}\phi_3+\phi_1
\end{array}\right)\,.
\end{eqnarray}
It is easy to check that the supersymmetric minimum
\begin{eqnarray}
\nonumber
\pmb \phi_0&=&\pmb \phi_0^{(0)}+\frac{g^2}{4\pi}\pmb \phi_0^{(1)} = (2.879, -1.31, -0.78, -0.262, 0.262, 0.78, 1.31,  
-2.879) \\
&&+\frac{g^2}{4\pi}(0.155, 0.155, -0.155,-0.119, 0.119, 0.155, -0.155, -0.155)\,,
\end{eqnarray}
is invariant under the $\mathbb Z_3$ symmetry. 

As we increase $c_m$, the theory experiences a first order transition at $c_{\mbox{\scriptsize cr}}=0.360$. There are three broken minima which are related via the $\mathbb Z_3$ symmetry.


The $\theta$ dependence of $c_{\mbox{\scriptsize cr}}$ and the discontinuity of the Polyakov loop are illustrated in 
Table~\ref{Theta for E6}.

\begin{table}
\centerline{
\begin{tabular}{|c||c|c|c|c|c|c|c|c|c|c|c|}
\hline
 $\theta$ & $0$ & $\frac{\pi}{12}$ & $\frac{2\pi}{12}$ & $\frac{3\pi}{12}$ & $\frac{4\pi}{12}$ & $\frac{5\pi}{12}$ & $\frac{6\pi}{12}$ & $\frac{7\pi}{12}$ & $\frac{8\pi}{12}$ & $\frac{9\pi}{12}$ & $\frac{10\pi}{12}$ \\\hline
 $c_{\mbox{\scriptsize cr}} $ & $0.6153$ & $0.6151$ & $0.6147$ & $0.6140$ & $0.613$ & $0.6112$ & $0.610$ & $0.608$ & $0.606$ & $0.604$ & $0.601$ \\\hline
      \end{tabular}
}
\caption{The $\theta$-dependence of the critical transition mass and the Polyakov loop discontinuity for the $E_6$ group. }
\label{Theta for E6}
\end{table}


\subsubsection{$\mathbf{E_7}$}

This is the rank $7$ special group with dimension $133$. As in the $E_6$ case, it is convenient to express the roots of this group  in terms of the standard unit vectors in eight dimensional space.  The set of the simple and positive roots are given by
\begin{eqnarray}
\nonumber
\{\pmb\alpha\}&=&\left\{\pmb \alpha_i=\pmb e_{i+1}-\pmb e_{i+2}\,(1\leq i\leq 6),\, \pmb \alpha_7=\frac{1}{2}\left(-\sum_{i=1}^4\pmb e_i+\sum_{i=5}^8 \pmb e_i\right)\right\}\,,\\
\nonumber
\{\pmb\beta\}&=&\left\{\frac{1}{2}\left(-\pmb e_1+\sum_{i=2}^8(-1)^{n_i}\pmb e_i-\pmb e_8\right)\left(\mbox{three odd} \,n_i\right),\, \pmb e_i-\pmb e_j\,(2 \leq i \leq j\leq 8),\right.\\
&&\left. \quad\quad\pmb e_i-\pmb e_1 (2\leq i\leq 8)   \right\}\,.
\end{eqnarray}
The affine root is $\pmb \alpha_0=\pmb e_2-\pmb e_1$, and the dual Kac labels are $\{k_a^*\}=\{1,2,3,4,3,2,1,2\}$, and the dual Coxeter number is $c_2=18$. The roots lay on a  hyperplane in the eight dimensional space such that we impose the constraint $\sum_{i=1}^8\phi_i=0$ to eliminate $\phi_8$.  The $E_7$ group has a $\mathbb Z_2$ center symmetry which acts on $\pmb \phi$ as
\begin{eqnarray}
\phi_i=\left\{\begin{array}{cc} -\frac{3}{2}\pi-\phi_{9-i}&\mbox{for}\, i=1,8\,,\\\frac{\pi}{2}-\phi_{9-i}&\mbox{for}\, 2 \leq i\leq 7\,. \end{array}  \right.
\end{eqnarray}
It is easy to check that the supersymmetric minimum
\begin{eqnarray}
\nonumber
\pmb \phi_0&=&\pmb \phi_0^{(0)}+\frac{g^2}{4\pi}\pmb \phi_0^{(1)}\\
\nonumber 
& = &(-4.276, 1.658, 1.309, 0.960, 0.611, 0.262, -0.0873, -0.436) \\
&&+\frac{g^2}{4\pi}(-0.342, 0.111, 0.217, 0.120, -0.120, -0.217, -0.111, 0.342)\,,
\end{eqnarray}
is invariant under the $\mathbb Z_2$ symmetry. 

As we increase $c_m$, the group experiences a first order transition at $c_{\mbox{\scriptsize cr}}=0.360$. There are two broken minima which are related via the $\mathbb Z_2$ symmetry.

\subsection{Exceptional groups without center symmetry}

\label{results44}

Of all exceptional groups without a center, only 
the physics of the thermal pure YM $G_2$ theory was studied on the lattice. A first order transition without change of symmetry was found \cite{Pepe:2006er}, with properties rather similar to the ones found here as we discuss below. Further detailed studies of large-volume scaling confirmed the first order nature of the transition \cite{Cossu:2007dk}. The transition in the groups without center is thus discontinuous, rather than a smooth crossover, despite the absence of an order parameter for confinement.

In the softly broken $G_2$, $F_4$ and $E_8$ SYM$^*$ theories we find results very similar to the $G_2$ lattice studies described above. 
We now briefly describe our findings for each case.

\subsubsection{$\mathbf{G_2}$ revisited}
\label{g2revisit}

This group has rank $2$ and dimension $14$. This group does not have a center. The sets of the simple and positive roots are given by
\begin{eqnarray}
\nonumber
\{\pmb\alpha\}&=&\left\{\pmb \alpha_1=\left(\frac{1}{\sqrt{2}},-\sqrt{\frac{3}{2}}\right),\, \pmb \alpha_2=\left(0,\sqrt{\frac{2}{3}}\right)\right\}\,,\\
\nonumber
\{\pmb\beta_+\}&=&\left\{\pmb \alpha_1,\, \pmb \alpha_2,\,\pmb \alpha_1+\pmb \alpha_2,\,\pmb \alpha_1+2\pmb\alpha_2,\, \pmb\alpha_1+3\pmb\alpha_2,\,2\pmb\alpha_1+3\pmb \alpha_2 \right\}\,. 
\end{eqnarray}
The affine root is $\pmb\alpha_0=\left(-\sqrt{2},0\right)$, the dual Kac labels are $\{k_a^*\}=\{1,2,1\}$, and the dual Coxeter number is $c_2=4$. The Cartan matrices of the $7$-th dimensional fundamental representation (or equivalently the weights of the fundamental representation) are:
\begin{eqnarray}
\nonumber
H_1 = \mbox{diag}\left(\frac{1}{\sqrt{2}},-\frac{1}{\sqrt{2}},0,-\frac{1}{\sqrt{2}},\frac{1}{\sqrt{2}},0,0\right)\,, 
H_2 = \mbox{diag}\left(\frac{1}{\sqrt{6}},\frac{1}{\sqrt{6}},-\sqrt{\frac{2}{3}},-\frac{1}{\sqrt{6}},-\frac{1}{\sqrt{6}} ,\sqrt{\frac{2}{3}},0\right)\,.
\end{eqnarray}
The supersymmetric minimum for $G_2$ has $g^2$ dependence:
\begin{eqnarray}
\pmb \phi_0=\pmb\phi_0^{(0)}+\frac{g^2}{4\pi}\pmb\phi_0^{(1)}=(3.33, 0.64)+\frac{g^2}{4\pi}(0.025,-0.2389)\,.
\end{eqnarray}
The $g^2$ dependence gets contributions from both tree-level and one-loop corrections. The trace of the Polyakov loop in the fundamental representation at the supersymmetric minimum reads\footnote{This corrects the result of \cite{Poppitz:2012nz} which did not take into account the contribution of the one-loop determinants (which are now part of $\pmb \phi_0^{(1)}$) to the expectation value of the Polyakov loop; we notice that this does not affect the value of the discontinuity.}:
\begin{eqnarray}
\label{polyakovlow}
\langle\mbox{Tr}\Omega \rangle=\mbox{Tr}_f\left[e^{i\pmb \phi_0\cdot H} \right]\cong\mbox{Tr}_f\left[e^{i\pmb \phi_0^{(0)}\cdot H}\left(1+i\frac{g^2}{4\pi}\pmb \phi_0^{(1)}\cdot H\right) \right]=\frac{g^2}{4\pi}0.0746\,.
\end{eqnarray}
By studying the total potential, we find that the $G_2$ group experiences a first order phase transition at the critical value $c_{\mbox{\scriptsize cr}}=3.174$. The trace of the Polyakov loop in the fundamental representation at the broken minimum is
\begin{eqnarray}
\label{polyakovhigh}
\langle\mbox{Tr}\Omega \rangle=\mbox{Tr}_f\left[e^{i\left(\pmb \phi_0+\frac{g^2}{4\pi}\pmb b\right)\cdot H} \right]\cong\mbox{Tr}_f\left[e^{i\pmb \phi_0^{(0)}\cdot H}\left(1+i\frac{g^2}{4\pi}\left(\pmb \phi_0^{(1)}+\pmb b\right)\cdot H\right) \right] = \frac{g^2}{4\pi}3.437.
\end{eqnarray}
Polyakov loop correlator $\langle\Tr \Omega(x)\Tr\Omega(y)^\dagger\rangle$ is now, instead of (\ref{polyakov loop correlator coefficients}),
\begin{eqnarray}
\langle\Tr \Omega(x)\Tr\Omega^\dagger(y)\rangle=\left\{ \begin{array}{cc} 0.0056 \left(\frac{g^2 }{4\pi}\right)^2 + \frac{g^2 R}{4r}\times 4\;e^{-3.586 m_0 r}\,,& c_m=0\\ 0.0056 \left(\frac{g^2 }{4\pi}\right)^2 +
\frac{g^2 R}{4r}\times 4\;e^{-2.194 m_0 r}\,,& \;\; \;c_m=c_{\mbox{\scriptsize cr}_-} \\
11.3 \left(\frac{g^2}{4\pi^2}\right)^2  + \frac{g^2 R}{4r}\times 3.98 \;e^{-1.49 m_0 r}, &~\; \;\;c_m=c_{\mbox{\scriptsize cr}_+}\,.
\end{array}  \right.
\label{polyakov loop correlator coefficients G2}
\end{eqnarray}
Naturally, as expected in a theory without center symmetry, the above correlators show that there is no linear confinement of fundamental charges, but rather ``string breaking" (in the lattice strong-coupling expansion, this can be seen in the study of the fundamental Wilson loop for $G_2$: a  transition from area to perimeter law takes place for quark separations of the order of 8 lattice spacings, to   leading order in the strong-coupling expansion \cite{Holland:2003jy}). Within the regime of validity of our semiclassical abelian description, however, there is no range of $r$ where the decaying exponential in (\ref{polyakov loop correlator coefficients G2}) is dominant over the rest.\footnote{As opposed to the study of SQCD \cite{Poppitz:2013zqa} in a setup similar to the one of the present paper  where the quark mass provided  a parameter controlling the smallness of the constant term.}  

The lattice studies of thermal $G_2$ pure YM theory \cite{Pepe:2006er}  found that the trace of the Polyakov loop changes from a small (close to zero) value below $T_c$ to a large positive value above $T_c$, ``exactly" as  a look at our Eqns.~(\ref{polyakovlow}) and (\ref{polyakovhigh}) would indicate. A  detailed comparison between the peaks in the histograms of the distributions of $\Tr \Omega$ in Fig.~4 of Ref.~\cite{Pepe:2006er} and the expectation values of $\Tr \Omega$ from our Eqns.~(\ref{polyakovlow}) and (\ref{polyakovhigh}) is difficult to perform, because of the different regimes of the calculations (but nonetheless the similarity appears striking).


The theta dependence of $c_{\mbox{\scriptsize cr}}$ and the discontinuity of the Polyakov loop on $\theta$ is illustrated in Table~\ref{Theta for G2}.

\begin{table}
\centerline{
\begin{tabular}{|c||c|c|c|c|c|c|c|c|c|c|c|}
\hline
 $\theta$ & $0$ & $\frac{\pi}{12}$ & $\frac{2\pi}{12}$ & $\frac{3\pi}{12}$ & $\frac{4\pi}{12}$ & $\frac{5\pi}{12}$ & $\frac{6\pi}{12}$ & $\frac{7\pi}{12}$ & $\frac{8\pi}{12}$ & $\frac{9\pi}{12}$ & $\frac{10\pi}{12}$ \\\hline
 $c_{\mbox{\scriptsize cr}} $ & $3.174$ & $3.172$ & $3.166$ & $3.156$ & $3.142$ & $3.122$ & $3.098$ & $3.067$ & $3.028$ & $2.981$ & $2.921$ \\\hline
 $\mbox{Tr}\langle\Delta\Omega(\theta) \rangle $ & $3.362$ & $3.365$ & $3.376$ & $3.394$ & $3.420$ & $3.451$ & $3.494$ & $3.542$ & $3.596$ & $3.658$ & $3.716$ \\\hline
      \end{tabular}
}
\caption{The $\theta$-dependence of the critical transition mass and the Polyakov loop discontinuity for the $G_2$ group.}
\label{Theta for G2}
\end{table}


\subsubsection{$\mathbf{F_4}$}

This is one of the five special groups, it has rank $4$ and dimension $52$. It is the isometry group of a $16$ dimensional Riemannian manifold known as the octonionic projective plane.  This group does not have a center. The set of the simple roots are given by
\begin{eqnarray}
\{\pmb\alpha\}=\left\{\pmb \alpha_1=\pmb e_2-\pmb e_3,\, \pmb \alpha_2=\pmb e_3-\pmb e_4,\, \pmb \alpha_3=\pmb e_4,\,\pmb \alpha_4=\frac{1}{2}\left(\pmb e_1-\pmb e_2-\pmb e_3-\pmb e_4\right)\right\}\,.
\end{eqnarray}
The affine root is $\pmb\alpha_0=-\pmb e_1-\pmb e_2$, the dual Kac labels are $\{k_a^*\}=\{1,2,3,2,1\}$, and the dual Coxeter number is $c_2=9$.

Examining the full potential reveals a first order phase transition at $c_{\mbox{\scriptsize cr}}=0.922$ not associated with symmetry breaking.


The theta dependence of $c_{\mbox{\scriptsize cr}}$ and the discontinuity of the Polyakov loop on $\theta$ are illustrated in Table~\ref{Theta for F4}.

\begin{table}
\centerline{
\begin{tabular}{|c||c|c|c|c|c|c|c|c|c|c|c|}
\hline
 $\theta$ & $0$ & $\frac{\pi}{12}$ & $\frac{2\pi}{12}$ & $\frac{3\pi}{12}$ & $\frac{4\pi}{12}$ & $\frac{5\pi}{12}$ & $\frac{6\pi}{12}$ & $\frac{7\pi}{12}$ & $\frac{8\pi}{12}$ & $\frac{9\pi}{12}$ & $\frac{10\pi}{12}$ \\\hline
 $c_{\mbox{\scriptsize cr}} $ & $0.922$ & $0.921$ & $0.920$ & $0.918$ & $0.916$ & $0.913$ & $0.909$ & $0.904$ & $0.899$ & $0.893$ & $0.886$ \\\hline
      \end{tabular}
}
\caption{The $\theta$-dependence of the critical transition mass and the Polyakov loop discontinuity for the $F_4$ group. }
\label{Theta for F4}
\end{table}


\subsubsection{$\mathbf{E_8}$}

This is the last of the five special groups. It is of rank $8$ and dimension $248$. This group does not have a center. The roots are given in terms of the standard unit vectors in eight dimensional space.  The set of the simple and positive roots are given by
\begin{eqnarray}
\nonumber
\{\pmb\alpha\}&=&\left\{\pmb \alpha_i=\pmb e_{i+1}-\pmb e_{i+2}\,(1\leq i\leq 6),\, \pmb \alpha_7=\frac{1}{2}\left(\pmb e_1+\pmb e_8-\sum_{i=2}^7\pmb e_i\right),\,\pmb \alpha_8=\pmb e_7+\pmb e_8\right\}\,,\\
\nonumber
\{\pmb\beta\}&=&\left\{\frac{1}{2}\left(\pmb e_1+\sum_{i=2}^8(-1)^{n_i}\pmb e_i\right)\left(\mbox{sum} \,n_i\,\mbox{even}\right),\, \pmb e_i\pm\pmb e_j\,(1 \leq i \leq j\leq 8)  \right\}\,.
\end{eqnarray}
The affine root is $\pmb \alpha_0=-\pmb e_1-\pmb e_2$, and the dual Kac labels are $\{k_a^*\}=\{1,2,3,4,5,6,4,2,3\}$, and the dual Coxeter number is $c_2=30$. The smallest  representation of the theory is the adjoint representation of  dimension $248$. Hence, the weights of the fundamental representation  are the roots themselves. 

The supersymmetric minimum is found at
\begin{eqnarray}
\nonumber
\pmb \phi_0&=&\pmb \phi_0^{(0)}+\frac{g^2}{4\pi}\pmb \phi_0^{(1)}=(4.817,1.257,1.047,0.838,0.628,0.419,0.209,0)\\
&&+\frac{g^2}{4\pi}(0.669,-0.014,0.293,0.398,0.359,0.2087,-0.0329,-0.0719)\,.
\end{eqnarray}
The trace of the Polyakov loop  in the fundamental (for $E_8$, the same as the adjoint) representation at the supersymmetric minimum reads
\begin{eqnarray}
\langle\mbox{Tr}\Omega \rangle=\mbox{Tr}\left[e^{i\pmb \phi_0\cdot H} \right]\cong\mbox{Tr}\left[e^{i\pmb \phi_0^{(0)}\cdot H}\left(1+i\frac{g^2}{4\pi}\pmb \phi_0^{(1)}\cdot H\right) \right]=-1-\frac{g^2}{4\pi}\times 4.47\,.
\end{eqnarray}
As we increase $c_m$, the theory experiences a first order transition at $c_{\mbox{\scriptsize cr}}=0.1936$. The trace of the Polyakov loop  at the broken minimum is
\begin{eqnarray}
\langle\mbox{Tr}\Omega \rangle\cong -1+\frac{g^2}{4\pi}\times 80.06\, ,	
\end{eqnarray}
and the discontinuity at the transition is $|\mbox{Tr}\langle \Delta \Omega\rangle |\simeq 84.53\; \frac{g^2}{4\pi}$.

\section{Future directions}
\label{future}

The conjectured continuity between quantum SYM$^*$ on $\mathbb R^3 \times \mathbb S^1_L$, where $L$ is the circle circumference,  and thermal YM on $\mathbb R^3\times \mathbb S^1_\beta$ gives some hope that one can use reliable semi-classical technology to learn about strongly coupled systems. Up to now, there has been excellent qualitative agreement between the predictions of this continuity and full-scale lattice simulations. We are still far from a complete understanding of this continuity and ample evidence in support of, or even against, it has to be collected before such an understanding can emerge. 

 It is still plausible that the agreement we have seen until now  is a mere coincidence and there is no deep physical connection between the quantum SYM and thermal YM transitions. 
Indeed, finding a counter example to the conjectured continuity will teach us about what goes wrong with it and in turn will sharpen our understanding of the topological excitations relevant  to the confinement/deconfinement mechanism. Here, we describe possible future directions to further test the continuity:

\begin{enumerate} 

\item There has been significant progress towards formulating and studying ${\cal N}=1$ SYM on the lattice, see the recent work  \cite{Giedt:2008xm,Demmouche:2010sf,Bergner:2014saa}. It appears that the phase diagram on Fig.~\ref{fig:phase} should be amenable to a lattice study, especially since we are not interested in taking the chiral limit for the gaugino. Reaching the semiclassical limit in lattice simulations is, perhaps, difficult, but at least part of the phase diagram should be accessible. Thus, we might   learn whether unexpected phases appear along the  dotted line on Fig.~\ref{fig:phase}.

\item Recently, there has been an extensive effort to study QCD with adjoint fermions, QCD(adj), motivated in part by the observation that these theories obey volume-independence
\cite{Kovtun:2007py} and have a weak-coupling confining regime \cite{Myers:2007vc,Unsal:2007jx}. In particular, in a series of previous works \cite{Anber:2011gn,Anber:2012ig,Anber:2013doa}, we exploited a duality, found by two of us and M. \" Unsal in \cite{Anber:2011gn}, between QCD(adj) compactified on a small spatial circle of size $L$, and considered at temperatures  near deconfinement,  and $2$d ``affine"  $XY$-spin models, to study the deconfinement transition in QCD(adj). In particular, in \cite{Anber:2012ig}, we performed Monte Carlo simulations to the dual $XY$ model to find that the transition experienced by $SU(3)$ QCD(adj) is a first order transition, which is in accord to the $4$d lattice simulations of \cite{Karsch:1998qj}.
It is thus  tempting to use a modified version of the continuity conjecture studied in this paper  to infer the order of the transition  in QCD(adj). This can perhaps be done within massive QCD(adj) (it is in any case easier to simulate on lattice than massless QCD(adj)), which can provide a direct test for the proposed continuity.

\item  QCD at nonzero temperature and finite density (nonzero chemical potential) is an important topic for practical as well as theoretical reasons. However, the chemical potential introduces a sign problem which makes the Monte Carlo simulations inapplicable. In order to circumvent this, one instead uses an imaginary chemical potential \cite{Alford:1998sd, Roberge:1986mm,D'Elia:2002gd,Misumi:2014raa} which ensures the positivity of the measure and allows for a direct use of the Monte Carlo simulations. However, it is well known that the partition function with fermions at finite temperature and imaginary chemical potential is equivalent to the partition function with fermions that satisfy twisted boundary conditions: $\psi(\vec x, x_4+L)=e^{i\phi}\psi(\vec x, x_4)$. Hence, one can exploit the conjectured continuity to study the order of the phase transition of the finite density QCD with either fundamental or adjoint matter via a detailed study of the softly broken SYM with chiral matter in the appropriate representation and twisted boundary conditions.

\end{enumerate}

\acknowledgments

We thank Philip Argyres, Massimo D'Elia, Takuya Kanazawa, and Mithat \" Unsal for discussions.
 We acknowledge support by a Discovery Grant from NSERC.

\appendix

\section{Group theory notation and the one-loop perturbative GPY potential}
\label{gpypotential}

\subsection{Simple roots, positive roots, weights, lattices, and traces}
\label{groupsummary}

We begin by summarizing some group theory facts that we will use. It is not our purpose to give an introduction to the subject here; there exist large literature on the subject, see for example  Appendix A of \cite{Argyres:2012ka} and references therein. We only pragmatically  set the notation and attempt to give some physical intuition behind its meaning. 

 The generators of any of the simple Lie groups we consider can be split into the $r=$ rank($G$) Cartan generators, labelled by $H^i$, $i=1,...,r$ and the rest of the generators $E_{\pmb \beta}$ and $E_{- \pmb \beta} = E^\dagger_{\pmb \beta}$. The meaning of the index $\pmb \beta$ will be explained shortly (evidently, there are (dim($G$)$- r)/2$ values that $\pmb \beta$ can take). The Cartan generators are Hermitean  while the $E_{\pmb\beta}$ are not. In this paper, we will need the commutation relations between the Cartan generators and between the Cartan generators and the rest, i.e.
\begin{eqnarray}
\[H^i, H^j\] &=& 0 \nonumber \\
\[H^i, E_{\pm \pmb \beta} \] &=& \pm \beta^i E_{\pm \pmb \beta}~. \label{commutators}
\end{eqnarray}
As is clear from the second relation in (\ref{commutators}) the 
  (dim($G$)$- r)/2$ ``labels" $\pmb \beta$'s are, in fact,  $r$-components vectors, i.e. each of them can be thought as a collection of $r$ numbers, $
  \pmb \beta = ( \beta^1, ..., \beta^r)$, which determine the commutator of the $r$ Cartan generators with $E_{\pmb \beta}$. The  $r$ component vectors $\pmb \beta$ that label the generators $E_{\pmb \beta}$ are called the ``positive roots" and we shall denote this set of (dim($G$)$- r)/2$ vectors by $\{\pmb\beta_+\}$ in what follows. There exists a subset of $r$ vectors in $\{\pmb\beta_+\}$, called the ``simple roots" $\pmb \alpha_a$, $a=1,...r$. Any positive root $\pmb \beta$ can be expressed as an integer linear combination of the simple roots with non-negative coefficients. In Section \ref{results}, for each group we study, we give a useful explicit representation of the simple and positive roots.
  
 A physical interpretation of the positive roots $\pmb \beta$ (and negative roots $-\pmb \beta$) follows from the fact that they represent the eigenvalues of the Cartan generators $H^i$, $i=1,...,r$, in the adjoint representation. Recall that the adjoint representation is of dimension dim$(G$), hence it is the one is defined by the commutation relations of the algebra (\ref{commutators})---i.e. the action of the generator $H^i$ on the ``vector" $E_{\pmb \beta}$ is given by the commutator $\[H^i, E_{\pmb \beta}\]$, etc. Recall also that in SYM, the Cartan generators correspond to the unbroken $U(1)^r$ gauge bosons. 
 Thus the roots $\pmb \beta$ (and $- \pmb \beta$) label  the charges of the adjoint representation fields (heavy ``W-bosons": gauge bosons or fermions) under the unbroken $U(1)^r$ gauge symmetry.\footnote{
 For example, the sum over $\pmb \beta_+$ in the one-loop Casimir energy due the  gaugino field (the one-loop GPY potential for the holonomy $\pmb \phi = (\phi^1,..., \phi^r)$) calculated in the next Section and given in Eqn.~(\ref{gpyheavy}) is  interpreted as a sum over the loop contribution of each massive W-boson's Kaluza-Klein tower. Similarly, the quantum correction to the monopole vertex given in Eqn.~(\ref{quantum corrections to monopole}) is also interpreted as due to the Kaluza-Klein tower of  each massive W-boson supermultiplet. In each of these cases the contributions of W-bosons of charges $\pmb \beta$ and $- \pmb \beta$ are equal and  add up, leaving only a sum over positive roots.}

For our further purposes, we will need to introduce a few  other concepts. 
The $r$ simple roots $\pmb \alpha_i$, $i=1,...,r$, are the basis vectors of a lattice in $r$-dimensional space, called the ``root lattice" (recall that any  root $\pm\pmb \beta$ is a linear combination of the simple roots). The lattice dual to the root lattice is called the ``co-weight lattice" and is spanned by the co-weight vectors,  $\pmb w_b^*$, $b=1,...r$, defined by the orthogonality relations:
\begin{equation}
\label{coweightlattice}
\pmb w_b^* \cdot \pmb \alpha_a = \delta_{ba}. 
\end{equation}
The dot product is understood in the usual $r$-dimensional Euclidean sense.

Of particular importance to us will be the ``weight  lattice". This is, once again, a lattice in  $r$-dimensional space. Physically, the points on the vertices of this lattice represent all possible charges, under the unbroken $U(1)^r$ gauge fields, that fields in any (not just the adjoint) representation of the gauge group can have. The weight lattice is spanned by the $r$ ``fundamental weights" $\pmb w_a$, 
$a=1,...r$, which obey
\begin{equation}
\label{fundamentalweights}
{\pmb w_a \cdot 2 \pmb \alpha_b \over (\pmb \alpha_b \cdot \pmb \alpha_b) }  = \delta_{ab}~.
\end{equation}
The above relation can also be written, introducing the ``co-roots" $\pmb \alpha^*_a$, $a=1,...,r$,
\begin{equation}
\label{coroots}
\pmb \alpha_a^* = {2 \pmb \alpha_a \over (\pmb \alpha_b \cdot \pmb \alpha_b)} 
\end{equation}
as
\begin{equation}
\label{fundamentalweights2}
\pmb w_a \cdot \pmb \alpha_b^*  = \delta_{ab}~.
\end{equation}
In particular,  fundamental weights and any root vector obey the relation:
\begin{eqnarray}
\label{weightroots}
\frac{\pmb w_a \cdot \pmb\beta}{\pmb \beta^2}=\frac{n}{2}\,,\quad n \in \Z\,,
\end{eqnarray}
which is used in Sections \ref{holonomyperiod}, \ref{dualphotonperiod}.

The weight lattice is finer than the root lattice---recall that the root lattice is the lattice of charges under $U(1)^r$ of fields in the adjoint representation, while the weight lattice includes the charges of all representations.\footnote{This should be familiar already from $SU(2)$: adjoint fields have integer charges under the $T^3$ generator, while fundamentals have half-integer charges.}  In Section \ref{holonomyperiod}, we used the fact that the eigenvalue of the Cartan generators $H^i$ in any representation are integer linear combinations of the fundamental weights, $\pmb w_a$ ($a=1,...,r$), as well the property (\ref{fundamentalweights2}) to deduce the periodicity of the holonomy.

We can associate a lattice also with the co-roots $\pmb \alpha^*$ of (\ref{coroots}), called the co-root lattice; it is clear from (\ref{fundamentalweights2}) that the co-root lattice is dual to the weight lattice (the co-weight lattice is dual to the root lattice, as per (\ref{coweightlattice})). The importance of the co-root lattice for us lies in the fact that the  $U(1)^r$  magnetic charges of the   monopole-instantons (that play an important role in this paper)   lie on the vertices of the co-root lattice, i.e. they are integer linear combinations of the co-roots, see e.g. Eqn.~(\ref{magnchargealpha}). In Section \ref{dualphotonperiod}, we use the properties of the co-roots to argue that the periodicity of the dual photon fields is given by the  fundamental weights (in the case when all representations are allowed, i.e. the gauge group is taken to be the universal cover). 

The sum over roots is also used in our  calculations of various traces. For example,  a trace in the adjoint representation of a product of $U(1)^r$ fields, e.g.
$ \pmb\phi \cdot H \;
\pmb\psi \cdot H \; \pmb \eta \cdot H.. $, where $\pmb \phi = (\phi^1, ... \phi^r)$ and $ \pmb\phi \cdot H = \sum_{i=1}^r \phi^i H^i$  (and similar for $\pmb \psi = (\psi^1,... \psi^r)$, etc.), can be written as
\begin{eqnarray}
\mbox{Tr}_{adj}\left[\pmb\phi \cdot H \;
\pmb\psi \cdot H \; \pmb \eta \cdot H...\right] =\sum_{\pmb\beta} \phi^i \beta^i \psi^j\beta^j \eta^k \beta^k....\,.
\label{grand formula for trace adj}
\end{eqnarray}
When the adjoint representation is replaced by some other representation, the sum over roots (the eigenvalues of $H^i$ in the adjoint representations) must be replaced by the sum over its weights---the eigenvalues of $H^i$ in that representation. We use this in the calculation of the Polyakov loop observables  in the fundamental and spinor representations, see Appendix \ref{polyakovloop} and Section \ref{results}. 

One other root has a significance for us: this is the affine (or lowest) root, denoted by $\pmb \alpha_0$. It has the property that all other roots can be found by adding to it non-negative integer sums of the simple roots. The affine root enters the description of the fundamental domain (Weyl chamber) of the holonomy, see (\ref{weylchamber}). Given a choice of simple roots, the lowest root can be expressed as
\begin{eqnarray}
\pmb \alpha_0   =-\sum_{a=1}^rk_a\pmb \alpha_a ~,
\end{eqnarray}
where the integers $k_a$ are called the Kac labels. Throughout this paper we normalize the lowest root s.t. $\pmb \alpha_0 \cdot \pmb \alpha_0 = 2$.

The dual of the  affine root 
 $\pmb\alpha_0^* = {2\pmb\alpha_0 \over (\pmb \alpha_0 \cdot \pmb \alpha_0)}$ obeys $\pmb \alpha_0^* = \pmb \alpha_0$. It 
 is also called  the ``affine co-root". The affine co-root can be expressed as a linear combination of the co-roots $\pmb \alpha_a^*$, $a=1,...r$:
\begin{eqnarray}
\pmb \alpha_0^* =  -\sum_{a=1}^rk_a^*\pmb\alpha_a^*\,,
\end{eqnarray}
where the integers $k_a^*$ are the so-called dual Kac labels (for each group we study, we give the values of $k_a^*$ in Section \ref{results}). The physical significance of the affine co-root is that $\pmb \alpha_0^*$ labels the $U(1)^r$ magnetic charge of  the ``twisted" monopole-instanton which contributes to the superpotential on $\R^3 \times \S^1$, but not in the $\R^3$ limit. Finally, introducing $k_0^* \equiv 1$, the dual Coxeter number of the gauge group, $c_2$ is expressed as a sum over dual Kac labels
\begin{equation}
\label{dualcoxeter}
c_2 = \sum_{a=0}^r k_a^*.
\end{equation}
 The dual Coxeter number $c_2$ determines the beta function of SYM theory. 

\subsection{The one-loop GPY potential for  general gauge group with nonzero gaugino mass}

\label{gauginoGPY}
The four dimensional  action for the gaugino field is given by\footnote{In this Section, we use the conventional notation $\lambda_\alpha$ for the four-dimensional gaugino field. The same field was labelled $\zeta_\alpha$ in Section \ref{abelianization}, where $\lambda_\alpha$ was instead used to label a redefined field, see discussion after (\ref{wlinear}); this slight mismatch of notation has no bearing on the resulting holonomy potential.  }
\begin{eqnarray}
{\cal L}={2  \over g^2} \int_0^{2\pi R} dx^3\int d^3x\mbox{Tr}\left[i\bar \lambda_{\dot\alpha} \bar\sigma^{m \;\dot\alpha\alpha}D_m\lambda_{\alpha}+\frac{m}{2}\lambda^\alpha\lambda_\alpha+\frac{m}{2}\bar\lambda_{\dot\alpha}\bar\lambda^{\dot\alpha}\right]\,
\label{lambda1}
\end{eqnarray}
where $D_m\lambda_{\alpha}=\partial_m\lambda_{\alpha}+i\left[v_m,\lambda_\alpha \right]$. 
The gaugino field is taken in the fundamental representation with generators normalized as $\Tr T^a_F T^b_F = \delta_{ab}$. Note that (\ref{lambda1})  is consistent with the normalization of the dimensionally reduced action (\ref{the action as integral over WS}) for the Cartan components.
Next, we expand $\lambda_\alpha$ in Fourier modes along the $x^3$ direction, using the decomposition of the generators from the previous Section, as follows:
\begin{eqnarray}
\nonumber
\lambda_\alpha(x^\mu,x^3)=\sum_{p \in Z} \left[\pmb\lambda_\alpha^p(x^\mu)\cdot H+
\sum_{\pmb\beta+} \lambda_{\alpha \;\pmb\beta}^p(x^\mu) E_{\pmb\beta}+\sum_{\pmb\beta_+} \lambda_{\alpha\; \pmb\beta}^{*\; p}(x^\mu) E_{-\pmb\beta}\right] e^{i\frac{p x^3}{R}}\,, 
\end{eqnarray}
where $\{\pmb\beta\}$ denotes the set of positive roots.   Remembering now that $v_3=\frac{\pmb \phi \cdot H}{2\pi R}$, and   $\[H^i, E_{\pm \pmb \beta}\] = \pm \pmb \beta^i  E_{\pm \pmb \beta}$, we find
\begin{eqnarray}
\left[v_3, \lambda_\alpha \right]=\frac{1}{2\pi R}\sum_{p\in Z}\sum_{\pmb \beta_+}
\pmb\beta \cdot \pmb\phi \; \lambda_{\alpha\; \pmb\beta}^p  \; E_{\pmb \beta} e^{i\frac{p x^3}{R}}-\frac{1}{2\pi R}\sum_{p\in Z}\sum_{\pmb\beta_+}
\pmb \beta \cdot \pmb \phi \;  \lambda_{\alpha\; \pmb\beta}^{*\; p} \; E_{-\pmb \beta} e^{i\frac{p x^3}{R}}\,, 
\end{eqnarray}
and using $\mbox{Tr}\left[E_{\pmb \beta}E_{-\pmb \beta'} \right]=\delta_{\pmb\beta\pmb \beta'}$ and $\mbox{Tr}\left[E_{\pmb \beta}E_{\pmb \beta'} \right]=0$, we find that the contribution of the derivative along the compact direction is:
\begin{eqnarray}
\int_0^{2\pi R} dx^3 \bar \lambda_{\dot\alpha} \bar\sigma^ {3\;\dot\alpha\alpha}D_3\lambda_{\alpha}=\sum_{p\in Z}\sum_{\pmb \beta_+}\left[\frac{p}{R}+\frac{\pmb \beta\cdot \pmb \phi}{2\pi R} \right] \bar \lambda_{\dot\alpha \; \pmb\beta}^{*\;p} \bar\sigma^{3\; \dot\alpha\alpha} \lambda_{\alpha\;\pmb\beta}^p+\mbox{c.c.}\,.
\end{eqnarray}
Similarly, we find that the derivatives in the noncompact directions contribute 
\begin{eqnarray}
\int_0^{2\pi R} dx^3 \bar \lambda_{\dot\alpha} \bar\sigma^ {\mu\;\dot\alpha\alpha}D_\mu\lambda_{\alpha}=\sum_{p\in Z}\sum_{\pmb \beta_+} \bar \lambda_{\dot\alpha\; \pmb\beta}^{*\; p} \;\bar\sigma^{\mu \; \dot\alpha\alpha}\partial_\mu \lambda_{\alpha\; \pmb\beta}^p+\mbox{c.c}+\mbox{higher order corrections}\,, 
\end{eqnarray}
with a similar expression for the mass term. Since the Lagrangian is quadratic in $\lambda$, we can calculate the determinant easily.  Collecting  the above expression, the mass term, and also performing a Fourier transform in the $x^\mu$ directions, we obtain the determinant of the gaugino in the holonomy background as a sum over Kaluza-Klein modes and positive roots:
\begin{eqnarray}
\nonumber
\log\mbox{Det}_{ \mbox{gaugino}}&=&2 \sum_{p \in Z}\sum_{\pmb \beta_+}\int \frac{d^3 k}{(2\pi)^3}\log\left[m^2+k^2+\left(\frac{p}{R}+\frac{\pmb \beta\cdot \pmb \phi}{2\pi R}\right)^2 \right]\\
\nonumber
&=&-2 \;\mbox{lim}_{s\rightarrow 0}  \frac{d}{ds}\sum_{p \in Z}\sum_{\pmb \beta_+}\int \frac{d^3 k}{(2\pi)^3}\left[m^2+k^2+\left(\frac{p}{R}+\frac{\pmb \beta\cdot \pmb \phi}{2\pi R}\right)^2 \right]^{-s}\,.
\end{eqnarray}
Performing the $k$ integral, then performing the sum using the zeta function, as indicated in the second line above,  we find the massive gaugino contribution to the Gross-Pisarski-Yaffe (GPY) effective potential
\begin{eqnarray}
V_{GPY}^{(gaugino)}= \frac{m^2}{ \pi^3 R}\sum_{p=1}^\infty \sum_{\pmb \beta_+}\frac{K_2(2\pi p mR)}{p^2}\cos(p \pmb \beta\cdot \pmb \phi)\,.
\label{gpyheavy}
\end{eqnarray}
Further, we use $K_2(x)\big\vert_{x\rightarrow 0} \sim {2 \over x^2} - {1 \over2}$ to find:
\begin{eqnarray}
V_{GPY}^{(gaugino)}\big\vert_{m\rightarrow 0}\approx  \frac{4}{\pi^2 (2\pi R)^3}\sum_{p=1}^\infty \sum_{\pmb \beta_+} {\cos(p \pmb \beta\cdot \pmb \phi)\over p^4}  - \frac{m^2}{2 \pi^3 R}\sum_{p=1}^\infty \sum_{\pmb \beta_+} {\cos(p \pmb \beta\cdot \pmb \phi)\over p^2} ~.\end{eqnarray}
The bosonic contribution cancels the $m=0$ part of the gaugino contribution (the first term above, which was previously calculated for general gauge groups in \cite{Argyres:2012ka}), leaving us with the ${\cal O}(m^2)$ GPY potential:
\begin{eqnarray}
\label{gauginogpy}
V_{GPY}^{{\cal O}(m^2)}=   - \frac{m^2}{2 \pi^3 R}\sum_{p=1}^\infty \sum_{\pmb \beta_+} {\cos(p \pmb \beta\cdot \pmb \phi)\over p^2} ~.\end{eqnarray}

\section{Review of monopole-instanton solutions}

The Euclidean action of pure Yang-Mills is given by
\begin{eqnarray}
S=\frac{1}{2g^2}\int_{\mathbb R^3\times\mathbb S^1} \mbox{Tr}\left[v_{mn}^aX^av_{mn}^b X^b\right]=\frac{2\pi R}{g^2}\int d^3 x \mbox{Tr}\left[B_{\mu}^aX^aB_\mu^bX^b+E_\mu^aX^aE_\mu^bX^b\right]\,,
\label{main action}
\end{eqnarray}
where $B^a_\mu=\frac{1}{2}\epsilon_{\mu\nu\rho}v_{\nu\rho}^a$, and $E_\mu^a=D_\mu v_3^a$. The generators $X^a$ are taken in the fundamental representation, and we use the normalization $\mbox{Tr}[X^aX^b]=\delta_{ab}C(F)$. In fact, the normalization in (\ref{main action}) is chosen arbitrarily since one can absorb any prefactor in the normalization of $g$. However, as we show below, a conventional choice can be set by assigning an appropriate instanton number to the BPST instanton. 

The action (\ref{main action}) can be written as 
\begin{eqnarray}
S=\frac{2\pi R}{g^2}\int d^3 x \mbox{Tr}\left[\left(B_\mu^a\mp D_\mu v_3^a\right)X^a\left(B_\mu^b\mp D_\mu v_3^b\right)X^b\right]\pm2\mbox{Tr}\left[B_\mu^aX^a D_\mu v_3^bX^b\right]\,.
\label{intermediate stage}
\end{eqnarray}
The last term in (\ref{intermediate stage}) can be manipulated using integration by parts:
\begin{eqnarray}
\nonumber
\int d^3 x B_\mu^a D_\mu v_3^b&=&\int d^3 x B_\mu^a\left(\partial_\mu v_3^b+f_{bcd}v_\mu^cv_3^d\right)\\
&=&\int d^3 x \partial_\mu \left(B_\mu^a v_3^a\right)-v_3^bD_\mu B_\mu^a\,. 
\end{eqnarray}
Using the equation of motion $D_\mu B_\mu^a=0$, we find that the action takes the form
\begin{eqnarray}
\nonumber
S&=&\frac{2\pi R}{g^2}\int d^3 x \mbox{Tr}\left[\left(B_\mu^a\mp D_\mu v_3^a\right)X^a\left(B_\mu^b\mp D_\mu v_3^b\right)X^b\right]\\
&&\pm\frac{4\pi R}{g^2}\int_{S^2} d^2S_\mu \mbox{Tr}\left[B_\mu^a X^a v_3^bX^b\right]\,.
\end{eqnarray}
A self dual or anti-self dual BPS monopole-instanton satisfies $B_\mu^a=\pm D_\mu v_3^a$, with action
\begin{eqnarray}
\nonumber
S=\pm\frac{4\pi R}{g^2}\int_{S^2} d^2S_\mu \mbox{Tr}\left[B_\mu^a X^a v_3^bX^b\right]\,.
\label{action for general group}
\end{eqnarray}

We also define the instanton number as
\begin{eqnarray}
{\cal K}=\frac{{\cal I}}{16\pi^2}\int_{\mathbb R^3\times \mathbb S^1}\mbox{Tr}\left[v_{mn}^aX^a\tilde v_{mn}^b X^b \right]\,.
\end{eqnarray}
An (anti)self-dual solution satisfies $v_{mn}=\pm\tilde v_{mn}$, or equivalently $B_\mu^a=\pm D_\mu v_3^a$, and hence we have
\begin{eqnarray}
{\cal K}=\pm {\cal I} \frac{g^2}{8\pi ^2}S\,.
\label{instanton number and action}
\end{eqnarray} 
At this stage, we use the conventional choice that ${\cal K}=1$  corresponds to a single BPST instanton with action $S= 8\pi^2/ g^2$, which fixes ${\cal I}=1$.

\subsection{ The $\mathbf{SU(2)}$ monopole}

In the $SU(2)$ case, it is convenient to choose the generators $X^a=\frac{\tau^a}{2}$, where $\{\tau^a\}$ are the Pauli matrices. In this case we have $\mbox{Tr}[X^aX^b]=\delta_{ab}/2$, and the action in this case is given by
\begin{eqnarray}
S=\frac{1}{4g^2}\int_{\mathbb R^3\times \mathbb S^1}v_{mn}^a v_{mn}^a\,.
\end{eqnarray}
The explicit solution for the anti-self-dual BPS monopole, in the hedgehog gauge, is given by
\begin{eqnarray}
\nonumber
v_{\mu}=v_{\mu}^c\tau^c\,,\\
v_3=\Psi^c\tau^c\,,
\end{eqnarray}
and
\begin{eqnarray}
\nonumber
\Psi^c(x, v)&=&\frac{x_c}{|x|^2}\left(v|x|\coth(v|x|)+1\right)\,,\\
v_{\mu}^c(x, v)&=&\epsilon_{\mu\nu c}\frac{x_\nu}{|x|^2}\left(1-\frac{v|x|}{\sinh(v|x|)}\right)\,,
\label{explicit functions}
\end{eqnarray}
where $v$ is the vacuum expectation value of the field $v_3$.
The monopole action is given by 
\begin{eqnarray}
S=\frac{4\pi Lv}{g^2},
\end{eqnarray}
and the asymptotic magnetic field in the singular gauge reads
\begin{eqnarray}
B_\mu|_{|x|\rightarrow\infty}=\frac{1}{2}\epsilon_{\mu\nu\lambda}v_{\nu\lambda}|_{|x|\rightarrow\infty}=-\frac{1}{2}\frac{x_\mu}{|x^3|}\tau^3\,,
\end{eqnarray}
and using (\ref{instanton number and action}) we find the instanton number
\begin{eqnarray}
{\cal K}=\pm \frac{vL}{2\pi}\,.
\end{eqnarray}
%

\subsection{ The fundamental monopole solutions in a general gauge group}

The solution for any gauge group $G$ is given by the embedding $SU(2)\subset G$ associated to every simple root $\pmb \alpha$:
\begin{eqnarray}
t^1=\frac{1}{\sqrt{2\pmb \alpha^2}}\left(E_{\pmb \alpha}+E_{-\pmb \alpha}\right)\,,\quad t^2=\frac{1}{i\sqrt{2\pmb \alpha^2}}\left(E_{\pmb \alpha}-E_{-\pmb \alpha}\right)\,,\quad t^3=\frac{1}{2}\pmb \alpha^*\cdot   H\,, 
\end{eqnarray}  
which obey the $SU(2)$ algebra $[t^a,t^b]=i\epsilon_{abc}t^c$. For a general gauge group, we use the convention $\mbox{Tr}[X^aX^b]=\delta_{ab}$, which fixes the length of the long roots to $2$. Hence, the action reads
\begin{eqnarray}
S=\frac{1}{2g^2}\int_{\mathbb R^3 \times \mathbb S^1} v_{mn}^av_{mn}^b\,,
\end{eqnarray}
which explains the $1/2$ factor difference between the $SU(2)$ case and the convention we use for a general gauge group.

The explicit solution of the monopole is given by:
\begin{eqnarray}
\nonumber
v_{\mu}&=&v_{\mu}^c(x,v) t^c\,,\\
v_3&=&\Psi^c(x,v)t^c+\pmb Q\cdot  H\,,
\end{eqnarray}
where $v_\mu^c$ and $\Psi^c$ are the ones from  (\ref{explicit functions}), with  the vacuum expectation value $v$  given by the projection of $\pmb v_3$ along the Cartan generators:
\begin{eqnarray}
v=\pmb v_3\cdot \pmb \alpha=\frac{\pmb \alpha\cdot \pmb\phi}{2\pi R},
\end{eqnarray}
while $\pmb Q$ is determined below (to avoid confusion and a clutter of notation, note that $\pmb \phi$ here and below denotes the value of the field at infinity).  Notice that $v$ is automatically positive whenever  $\pmb \phi$ is in the Weyl chamber (\ref{weylchamber}) and $\pmb\alpha$ is a simple root. Thus, the asymptotic behavior of $\Psi^c\tau^c$ reads
\begin{eqnarray}
\Psi^c\tau^c|_{|x|\rightarrow\infty}=\frac{x^c}{|x|}\; t^c\; \frac{\pmb \alpha\cdot \pmb\phi}{2\pi R}\,,
\end{eqnarray}
or in the singular gauge
\begin{eqnarray}
\Psi^c\tau^c|_{|x|\rightarrow\infty}=t^3 \; \frac{\pmb \alpha\cdot \pmb\phi}{2\pi R}=\frac{1}{2}\; \frac{\pmb \alpha\cdot \pmb\phi}{2\pi R}\; \pmb \alpha^*\cdot  H\,.
\end{eqnarray}
The value of $\pmb Q$ is determined by requiring that $v_3\rightarrow \frac{\pmb \phi\cdot H}{2\pi R}$ as $|x|\rightarrow \infty$, from which we find
\begin{eqnarray}
\pmb Q=\pmb\phi-\frac{1}{2}\pmb \alpha^* v\,.
\end{eqnarray}

 We conclude by a summary of the monopole-instanton solutions relevant for the generation of a superpotential.   For the fundamental monopole-instanton solutions associated with the simple  (co)-roots $\pmb \alpha^*_i$, $i=1,...,r$, the long range magnetic field is given by
\begin{eqnarray}
\label{magnchargealpha1}
B_\mu^{\pmb \alpha_i }=-\frac{x_\mu}{|x|^3}\frac{\pmb\alpha^*_i \cdot   H}{2}\,, ~ i=1,...,r,
\end{eqnarray}
and, using (\ref{action for general group}) and (\ref{instanton number and action}),   the action and the instanton number by
\begin{eqnarray}
\label{actionalpha1}
S^{\pmb \alpha_i }=\frac{4\pi }{g^2}\pmb \alpha^*_i \cdot\pmb \phi\,,\quad {\cal K}^{\pmb \alpha_i}=\frac{1 }{2\pi}\pmb \alpha^*_i \cdot\pmb \phi\,, ~ i=1,...,r.
\end{eqnarray}

Finally, we recall that on $\R^3 \times \S^1$ there is an entire tower of ``Kaluza-Klein" monopole-instanton solutions. Most of them have large topological charges and action and are irrelevant for the generation of a superpotential in  SYM  $\R^3 \times \S^1$ since they have too many fermion zero modes. However, there one solution is singled out, described first in terms of D-branes in \cite{Lee:1997vp}, see also  \cite{Lee:1998bb,Kraan:1998pm}, associated with the lowest  co-root  $\pmb \alpha_0^*=\pmb \alpha_0$. Its construction is described elsewhere, see \cite{Davies:1999uw} for general gauge groups or \cite{Poppitz:2008hr} for $SU(N)$. We will need two facts: the $\pmb \alpha_0^*$-solution is also self-dual, has long-range magnetic field
\begin{eqnarray}
\label{magnchargealpha2}
B_\mu^{\pmb \alpha_0 }=-\frac{x_\mu}{|x|^3}\frac{\pmb \alpha _0^* \cdot   H}{2}\,,
\end{eqnarray}
and   action and  instanton number
\begin{eqnarray}
\label{actionalpha2}
S^{\pmb \alpha_0 }=\frac{4\pi }{g^2}(2 \pi + \pmb \alpha^*_0 \cdot\pmb \phi)\,,\quad {\cal K}^{\pmb \alpha_0 }=  \frac{2\pi + \pmb \alpha^*_0 \cdot\pmb \phi }{2\pi}\,.
\end{eqnarray}

\section{Quantum corrections to the monopole-instanton vertex}

\label{monopoledets}
\subsection{The index theorem}
The index theorem in monopole-instanton backgrounds on $\R^3 \times \S^1$ was first studied in the mathematical literature \cite{Nye:2000eg} and then derived using physicist's tools in  \cite{Poppitz:2008hr}. Here we give a quick derivation along the lines of the latter reference, rewritten using the Lie-algebraic notation of this paper. The index theorem and the index function play a role in calculating the one-loop monopole-instanton determinants.

As in \cite{Poppitz:2008hr}, one begins with a  mass-dependent ``index" in the representation ${\cal R}$:
\begin{eqnarray}
I^{\pmb \alpha}_{{\cal R}}(M^2)&=&\mbox{tr}\left[\frac{M^2}{\Delta_-+M^2}\right]-\mbox{tr}\left[\frac{M^2}{\Delta_++M^2}\right]\,,
\end{eqnarray}
where $\Delta_-=D^\dagger D=-D_m D^m-\frac{1}{2}\sigma_{mn}v^{mn}$, $\Delta_+=DD^\dagger=-D_m D^m$, where all operators are assumed to be taken in the $\pmb \alpha$-monopole-instanton background. Expanding $\Delta_-$ as a power series in $\sigma_{mn}v^{mn}$, and taking into account the axial anomaly we find
\begin{eqnarray}
I_{{\cal R}}^{\pmb \alpha}(M^2)=I_{1{\cal R}}^{\pmb \alpha}+I_{2{\cal R}}^{\pmb \alpha}(M^2),
\end{eqnarray}
where
\begin{eqnarray}
I_{1{\cal R}}^{\pmb \alpha}=-\frac{1}{32\pi^2}\int_{\mathbb R^3 \times \mathbb S^1}\mbox{Tr}_{{\cal R}}\left[v_{mn}^a X^a\tilde v_{mn}^b X^b \right],
\end{eqnarray}
and
\begin{eqnarray}\nonumber
I_{2{\cal R}}^{\pmb \alpha}(M^2)=2\int_0^L dy \int_{S^2_\infty}d^2 \sigma_\mu \mbox{Tr}_{{\cal R}}\left\langle x;y\left| iD_3 \frac{1}{-\partial_\mu^2+M^2-D_3^2}B^\mu \frac{1}{-\partial_\mu^2+M^2-D_3^2}\right|x;y \right\rangle,
\end{eqnarray}
and $-iD_3\rightarrow \frac{p}{R}+v_3|_{|x|\rightarrow \infty}$ for every $p \in \mathbb Z$. Hence we find for $I_{2{\cal R}}(M^2)$:
\begin{eqnarray}
I_{2{\cal R}}^{\pmb \alpha}(M^2)=\frac{1}{2}\sum_{p\in Z}\mbox{tr}_{{\cal R}}\left[\frac{\left(\frac{ p}{R}+\frac{\pmb \phi\cdot \pmb H}{2\pi R}\right)\pmb \alpha^*\cdot\pmb H}{\sqrt{M^2+\left(\frac{ p}{R}+\frac{\pmb \phi\cdot \pmb H}{2\pi R}\right)^2}} \right]\,.
\end{eqnarray}
Also, one can use integration by parts to write $I_{1{\cal R}}$ as:
\begin{eqnarray}
I_{1{\cal R}}^{\pmb \alpha}=-\frac{1}{32\pi^2}(2\pi R)\int_{S^2_{\infty}} dS_{\mu}\mbox{Tr}_{{\cal R}}\left[v_3B_\mu \right]=\mbox{Tr}_{{\cal R}}\left[\frac{\pmb \phi \cdot \pmb H}{2\pi}\pmb \alpha^*\cdot \pmb H\right]\,.
\end{eqnarray}
In the limit $M \rightarrow 0$, we use the identity $\sum_{p\in Z}\mbox{sign}\left(a+\frac{ P}{R}\right)=1-2aR+2\lfloor aR\rfloor$, to find the index $I_{\cal R}(0)$:
\begin{eqnarray}
\nonumber
I_{{\cal R}}^{\pmb \alpha}(0)&=&\mbox{Tr}_{{\cal R}}\left[ \left(\frac{1}{2}-\frac{\pmb \phi \cdot \pmb H}{2\pi}+ \lfloor\frac{\pmb \phi \cdot \pmb H}{2\pi} \rfloor \right)\pmb \alpha^*\cdot \pmb H\right]+\mbox{Tr}_{{\cal R}}\left[\frac{\pmb \phi \cdot \pmb H}{2\pi}\pmb \alpha^*\cdot \pmb H\right]\\
&=&\mbox{Tr}_{{\cal R}}\left[  \lfloor\frac{\pmb \phi \cdot \pmb H}{2\pi} \rfloor \pmb \alpha^*\cdot \pmb H\right]\,,
\label{index theorem}
\end{eqnarray}
where we have used $\mbox{Tr}_{{\cal R}}[H]=0$. 
In the adjoint representation, we can   write the trace as sum over all roots $\pmb \beta$: 
\begin{eqnarray}
I_{adj}^{\pmb \alpha}(0)=\sum_{\pmb \beta} \lfloor\frac{\pmb \phi \cdot \pmb \beta}{2\pi} \rfloor \pmb \alpha^*\cdot \pmb \beta=\sum_{\pmb \beta_+} \lfloor\frac{\pmb \phi \cdot \pmb \beta}{2\pi} \rfloor \pmb \alpha^*\cdot \pmb \beta-\sum_{\pmb \beta_+} \lfloor\frac{-\pmb \phi \cdot \pmb \beta}{2\pi} \rfloor \pmb \alpha^*\cdot \pmb \beta\,,
\label{index theorem adj}
\end{eqnarray}
where we have split the sum over the positive and negative roots, and used $\pmb\beta_-=-\pmb \beta_+$. Using $\lfloor -a \rfloor=-1-\lfloor a\rfloor$, and recalling the fact that $\pmb \phi\cdot \pmb \beta$ lies in the first Weyl chamber (\ref{weylchamber}) and hence $\lfloor \frac{\pmb \phi\cdot \pmb \beta}{2\pi}\rfloor=0$, we obtain
\begin{eqnarray}
I_{adj}^{\pmb \alpha_i}(0)=\sum_{\pmb \beta_+}\pmb \alpha_i^* \cdot \pmb \beta\,~, i = 1,...r.
\end{eqnarray}
For the $\pmb\alpha_0$ solution, the result, for $\pmb \phi$ in  (\ref{weylchamber}), is:
\begin{eqnarray}
I_{adj}^{\pmb \alpha}(0)=2 c_2 +  \sum_{\pmb \beta_+}\pmb \alpha_0^* \cdot \pmb \beta\,.
\end{eqnarray}
\subsection{The ``index function" and the loop-corrected monopole-instanton vertex}
\label{determinant}

As in \cite{Poppitz:2012sw}, we shall use the index theorem to find the quantum corrections to the monopole background. To this end, we use the mass-dependent ``index" to calculate the determinant. The relation (\ref{abc1}) between the integral of the mass-dependent ``index" and the ratio of one-loop regularized determinants in the monopole-instanton background plays a crucial role in this calculation. 

In this Section, we will need the expressions for $I_{1{adj}}^{\pmb \alpha}(M^2)$ and $I_{2{ adj}}^{\pmb \alpha}(M^2)$. Using the identity $\mbox{Tr}_{adj}[H^iH^j]=\delta^{ij}T(\mbox{adj})$, we find
\begin{eqnarray}
\nonumber
I_{1adj}^{\pmb \alpha}(M^2)&=&T(\mbox{adj})\frac{\pmb \alpha^* \cdot \pmb \phi}{2\pi}\,,\\
I_{2adj}^{\pmb \alpha}(M^2)&=&\frac{1}{2}\sum_{p\in Z}\sum_{\pmb\beta}\left[\frac{\left(\frac{ p}{R}+\frac{\pmb \phi\cdot \pmb \beta}{2\pi R}\right)\pmb \alpha^*\cdot\pmb \beta}{\sqrt{M^2+\left(\frac{ p}{R}+\frac{\pmb \phi\cdot \pmb \beta}{2\pi R}\right)^2}} \right]\,,
\end{eqnarray}
where $T(\mbox{adj})=2c_2$, and $c_2=\sum_{a=0}^rk_a^*$ is the dual Coxeter number.

 We start with
\begin{eqnarray}
\nonumber
\int_{\mu^2}^{\Lambda^2_{PV}}\frac{dM^2}{M^2}I^{\pmb \alpha}(M^2)&=&\mbox{tr}\log\left[\Delta_-+M^2 \right]_{\mu^2}^{\Lambda^2_{PV}}-\mbox{tr}\log\left[ \Delta_++M^2\right]_{\mu^2}^{\Lambda^2_{PV}}\\
&=&\mbox{tr}\log\left[\frac{\Delta_-+\Lambda^2_{PV}}{\Delta_-+\mu^2}\right]-\mbox{tr}\log\left[\frac{\Delta_++\Lambda^2_{PV}}{\Delta_++\mu^2}\right]\,.
\end{eqnarray}
Using the identity $\mbox{tr}\log A=\log \mbox{Det} A$, we find
\begin{eqnarray}
\int_{\mu^2}^{\Lambda^2_{PV}}\frac{dM^2}{M^2}I^{\pmb \alpha}(M^2)=\log \mbox{Det}\left[\frac{\left(\Delta_++\mu^2\right)\left(\Delta_-+\Lambda^2_{PV}\right)}{\left(\Delta_-+\mu^2\right)\left(\Delta_++\Lambda^2_{PV}\right)} \right]\,.
\label{abc1}
\end{eqnarray}

The usefulness of the expression (\ref{abc1}) is that it determines the one-loop corrections around the supersymmetric monopole-instanton background, see Appendix A of \cite{Poppitz:2012sw}. The contribution of the nonzero modes of the fluctuations of the fermions and gauge fields in the $\pmb \alpha$-monopole-instanton background is given by 
the ratio $R_{\pmb \alpha^*}$ of determinants  defined as%
\begin{eqnarray}
R_{\pmb \alpha^*}=\mbox{lim}_{\mu\rightarrow 0}\left( \frac{\mu^{I^{\pmb \alpha}_{{\cal R}}(0)}\left(\Delta_++\mu^2\right)\left(\Delta_-+\Lambda^2_{PV}\right)}{\left(\Delta_-+\mu^2\right)\left(\Delta_++\Lambda^2_{PV}\right)} \right)^{3/4}\,,
\end{eqnarray}
where $I^{\pmb \alpha}_{\cal R}(0)$ is the number of zero modes of the operator $\Delta_-$. We find
\begin{eqnarray}
\left(e^{-S^{\pmb \alpha}_0}R\right)_{\pmb \alpha^*}=e^{-S^{\pmb \alpha}_0}\mbox{lim}_{\mu \rightarrow 0}\left(\mu^{I^{\pmb \alpha}_{{\cal R}}(0)}e^{I^{\pmb \alpha}_{1 \mbox{\scriptsize adj}} \log \frac{\Lambda_{PV}^2}{\mu^2}}\times e^{\int_{\mu^2}^{\Lambda^2_{PV}} dM^2 \frac{I^{\pmb \alpha}_{2\mbox{\scriptsize adj}}(M^2)}{M^2}} \right)^{3/4}\,,
\end{eqnarray}
where 
\begin{eqnarray}
S^{\pmb \alpha}_0=\frac{4\pi}{g^2\left(\Lambda_{PV}\right)}\pmb \alpha^*\cdot \pmb\phi
\end{eqnarray}
 is the bare monopole action defined at the scale $\Lambda_{PV}$.
Using the SYM running coupling ($\Lambda$ is the strong-coupling scale (\ref{lambdascale}) to one loop)
\begin{eqnarray}
\frac{4\pi}{g^2(\mu)}=\frac{3}{8\pi}T(\mbox{adj})\log\frac{\mu^2}{\Lambda^2}\,,
\label{QCD beta function}
\end{eqnarray}
we find
\begin{eqnarray}
\nonumber
-S_0^{\pmb \alpha}+\frac{3}{4}I^{\pmb \alpha}_{1\mbox{\scriptsize adj}}\log \frac{\Lambda^2_{PV}}{\mu^2}&=&\underbrace{-\frac{4\pi}{g^2\left(\Lambda_{PV}\right)}\pmb \alpha^*\cdot \pmb \phi+\frac{3}{8\pi}T(\mbox{adj})\pmb\alpha^*\cdot \pmb \phi \log (2 \pi R \Lambda_{PV})^2}_{-\frac{4\pi}{g^2(1/2\pi R)}\pmb\alpha^*\cdot \pmb \phi}\\
\nonumber
&&-\frac{3}{8}\frac{T(\mbox{adj})}{\pi}\pmb \alpha^*\cdot \pmb \phi \log (\mu 2 \pi R)^2\\
&=&-\frac{4\pi}{g^2(1/2\pi R)}\pmb\alpha^*\cdot \pmb \phi-\frac{3}{4}\frac{T(\mbox{adj})}{\pi}\pmb \alpha^*\cdot \pmb \phi \log (2 \pi R\mu) \,.
\end{eqnarray}

  Next, we calculate $\int_{\mu^2}^{\Lambda_{PV}^2} \frac{dM^2}{M^2}I^{\pmb \alpha}_{2 \mbox{\scriptsize adj}}(M^2)$. Given the integral 
\begin{eqnarray}
\int_\mu^\infty \frac{dM}{M\sqrt{A^2+M^2}}=\frac{1}{|A|}\sinh^{-1}\left[\frac{|A|}{\mu} \right]=\frac{1}{|A|}\log\left(\frac{2|A|}{\mu}\right)+{\cal O}(\mu^2)\,,
\end{eqnarray}
we find
\begin{eqnarray}
\nonumber
\int_{\mu^2}^{\Lambda_{PV}^2} \frac{dM^2}{M^2}I^{\pmb \alpha}_{2 \mbox{\scriptsize adj}}(M^2)&=&\sum_{p\in Z}\sum_{\pmb\beta}\left[ \mbox{sign}\left(\frac{\pmb \phi\cdot \pmb \beta}{2\pi R}+\frac{ p}{R}\right)\pmb\alpha^*\cdot \pmb \beta \log\frac{2\left|\frac{ p}{R}+\frac{\pmb \phi\cdot \pmb \beta}{2\pi R}\right|}{\mu}\right]\\
\nonumber
&=&\sum_{p\in Z}\sum_{\pmb\beta}\left[ \mbox{sign}\left(\frac{\pmb \phi\cdot \pmb \beta}{2\pi R}+\frac{ p}{R}\right)\pmb\alpha^*\cdot \pmb \beta \log\frac{2}{\mu R}\right]\\
\nonumber
&&+\sum_{p\in Z}\sum_{\pmb\beta}\left[ \mbox{sign}\left(\frac{\pmb \phi\cdot \pmb \beta}{2\pi R}+\frac{ p}{R}\right)\pmb\alpha^*\cdot \pmb \beta\log\left|p+\frac{\pmb \phi\cdot \pmb \beta}{2\pi }\right|\right]\,.\\
\end{eqnarray}
 Using $\sum_{p\in Z}\mbox{sign}\left(a+\frac{ P}{R}\right)=1-2aR+2\lfloor aR\rfloor$, and the index theorem in the adjoint representation, we find
\begin{eqnarray}
\sum_{p\in Z}\sum_{\pmb\beta}\left[ \mbox{sign}\left(\frac{\pmb \phi\cdot \pmb \beta}{2\pi R}+\frac{ p}{R}\right)\pmb\alpha^*\cdot \pmb \beta  \log\frac{2}{\mu R}\right]
=\left(2I_{\mbox{\scriptsize adj}}(0)-2I_{1\mbox{\scriptsize adj}}\right)\log \frac{2}{\mu R}\,.
\end{eqnarray}

Collecting everything we find:
\begin{eqnarray}
\nonumber
\log(e^{-S_0^{\pmb \alpha}}R^{\pmb \alpha})&=&-\frac{4\pi}{g^2(1/2\pi R)}\pmb\alpha^*\cdot\pmb \phi-\frac{3}{4\pi}T(\mbox{adj})\pmb \alpha^*\cdot\pmb\phi\log(4\pi)+\frac{3}{2}I_{\mbox{\scriptsize adj}}(0)\log \frac{2}{ R}\\
\nonumber
&&+\frac{3}{4}\sum_{p\in Z}\sum_{\pmb\beta}\left[\mbox{sign}\left(p+\frac{\pmb \phi\cdot \pmb \beta}{2\pi }\right)\pmb\alpha^*\cdot \pmb \beta \log \left|p+\frac{\pmb \phi\cdot \pmb \beta}{2\pi } \right| \right]\\
\nonumber
&=&-\frac{4\pi}{g^2\left(\frac{2}{R}\right)}\pmb\alpha^*\cdot\pmb \phi+\frac{3}{2}I_{\mbox{\scriptsize adj}}(0)\log \frac{2}{ R}\\
&&+\frac{3}{4}\sum_{p\in Z}\sum_{\pmb\beta}\left[\mbox{sign}\left(p+\frac{\pmb \phi\cdot \pmb \beta}{2\pi }\right) \pmb\alpha^*\cdot \pmb \beta \log \left|p+\frac{\pmb \phi\cdot \pmb \beta}{2\pi } \right| \right]\,,
\label{semi final expression for logR}
\end{eqnarray}
where we have used (\ref{QCD beta function}) to simplify the first two terms in the first line. 

Further, to calculate the last term in (\ref{semi final expression for logR}) we observe that
\begin{eqnarray}
\sum_{p\in Z}\mbox{sign}(p+a)\log|p+a|=-\frac{d}{ds}\left[\xi(s,a)-\xi(s,1-a) \right]_{s=0}\,,
\end{eqnarray}
where $0<a<1$, and $\xi(s,a)=\sum_{p\geq0}\frac{1}{|p+a|^s}$ is the incomplete gamma function. Similarly we have
\begin{eqnarray}
\sum_{p\in Z}\mbox{sign}(p-a)\log|p-a|=-\frac{d}{ds}\left[\xi(s,1-a)-\xi(s,a) \right]_{s=0}\,.
\end{eqnarray}
Then, using $\xi'(0,x)=\log \Gamma(x)-\frac{1}{2}\log(2\pi)$, $I_{\mbox{\scriptsize adj}}(0)=2$, and taking into account that $\frac{\pmb \phi\cdot \pmb \beta}{2\pi }$ can be positive or negative,  we finally have
\begin{eqnarray}
\left(e^{-S_0}R\right)_{\pmb \alpha^*}=\left(\frac{2}{R}\right)^3e^{-\pmb\alpha^*\cdot\left[\frac{4\pi}{g^2\left(\frac{2}{R}\right)}  \pmb \phi +\frac{3}{2}\sum_{\pmb\beta_+}\left(\pmb \beta\log \frac{\Gamma\left(\frac{\pmb \beta\cdot \pmb \phi}{2\pi}\right)}{\Gamma\left(1-\frac{\pmb \beta\cdot \pmb \phi}{2\pi}\right)} \right)\right]}\,.
\label{quantum corrections to monopole}
\end{eqnarray}

\section{The Polyakov loop and its correlator}

\label{polyakovloop}

The trace of the Polyakov loop is given by (notice that, as opposed to the main text, for brevity, in this Appendix we use $\Omega$ to denote the trace of the Polyakov loop):
\begin{eqnarray}
\Omega(x)=\mbox{tr}\left[e^{i \pmb H\cdot \pmb \phi(x)}\right]\,.
\end{eqnarray}
Writing $\phi$ as $\pmb \phi=\pmb\phi_0^{(0)}+\frac{g^2}{4\pi}\pmb\phi_0^{(1)}+\frac{g^2}{4\pi}\pmb b$, and expanding in powers of $g$ we obtain
\begin{eqnarray}
\langle \Omega \rangle \cong\mbox{tr}\left[e^{i \pmb H\cdot \pmb \phi_0^{(0)}}\left(1+i\frac{g^2}{4\pi}\pmb\phi_0^{(1)}+i\frac{g^2}{4\pi}\pmb b\right)\cdot \pmb H\right]+{\cal O}(g^4)\,.
\end{eqnarray}
In addition, we can write $\pmb b =\pmb v+ \tilde{\pmb b}$, where $\pmb v$ is the vacuum expectation value of $\pmb b$. Then,  we define the discontinuity of the Polyakov loop as
\begin{eqnarray}
\langle \Delta\Omega \rangle \equiv i\frac{g^2}{4\pi}\mbox{tr}\left[e^{i \pmb H\cdot \pmb \phi_0^{(0)}}\pmb v\cdot \pmb H\right]+{\cal O}(g^4)\,.
\end{eqnarray}

We are also interested in  calculating the Polyakov correlator given by:
\begin{eqnarray}
\langle  \Omega(x)\Omega^\dagger(y) \rangle\,,
\end{eqnarray}
which reads upon expanding in powers of $g^2$
\begin{eqnarray}
\nonumber
\langle  \Omega(x)\Omega^\dagger(y) \rangle&\cong&\left\langle \mbox{tr}\left[e^{i \pmb H\cdot \pmb \phi_0^{(0)}} \left(1+i\frac{g^2}{4\pi}\pmb\phi_0^{(1)}+i\frac{g^2}{4\pi}\pmb b(x)\right)\cdot \pmb H\right] \right.\\
\nonumber
&&\left.\times\mbox{tr}\left[e^{-i \pmb H\cdot \pmb \phi_0^{(0)}}
\left(1-i\frac{g^2}{4\pi}\pmb\phi_0^{(1)}-i\frac{g^2}{4\pi}\pmb b(y)\right)\cdot \pmb H\right]   \right\rangle\\
\nonumber
&=& \left|\mbox{tr}\left[e^{i \pmb H\cdot \pmb \phi_0^{(0)}} \left(1+i\frac{g^2}{4\pi}\pmb\phi_0^{(1)}\right)\cdot \pmb H\right]\right|^2\\
\nonumber
&&+i\frac{g^2}{4\pi}\mbox{tr}\left[e^{-i \pmb H\cdot \pmb \phi_0^{(0)}}\pmb H\cdot  \langle \pmb b(x)\rangle \left(1+i\frac{g^2}{4\pi}\pmb\phi_0^{(1)}\right)\cdot \pmb H\right]+\mbox{h.c.}\\
&&+\left(\frac{g^2}{4\pi}\right)^2\mbox{tr}\left[e^{i \pmb H\cdot \pmb \phi_0^{(0)}}H_i\right]\mbox{tr}\left[e^{-i \pmb H\cdot \pmb \phi_0^{(0)}}H_j\right]\langle b_i(x)b_j(y) \rangle\,.
\end{eqnarray} 
Using  $\pmb b =\pmb v+ \tilde{\pmb b}$, we obtain:
\begin{eqnarray}
\nonumber
\langle  \Omega(x)\Omega^\dagger(y) \rangle&\cong&
\left|\mbox{tr}\left[e^{i \pmb H\cdot \pmb \phi_0^{(0)}} \left(1+i\frac{g^2}{4\pi}\pmb\phi_0^{(1)}\right)\cdot \pmb H\right]\right|^2\\
\nonumber
&&+i\frac{g^2}{4\pi}\mbox{tr}\left[e^{-i \pmb H\cdot \pmb \phi_0^{(0)}}  \pmb H\cdot \pmb v \left(1+i\frac{g^2}{4\pi}\pmb\phi_0^{(1)}\right)\cdot \pmb H\right]+\mbox{h.c.}\\
\nonumber
&&+\left(\frac{g^2}{4\pi}\right)^2\mbox{tr}\left[e^{i \pmb H\cdot \pmb \phi_0^{(0)}}H_i\right]\mbox{tr}\left[e^{-i \pmb H\cdot \pmb \phi_0^{(0)}}H_j\right]\left\{v_iv_j+\langle {\tilde b}_i(x){\tilde b}_j(y) \rangle\right\}\,.\\
\end{eqnarray} 

Now, we turn to the calculations of the correlator $\langle \tilde b_i(x)\tilde b_j(y)\rangle$. We start from (\ref{bosonic potential}) and make the change of variables
\begin{eqnarray}
\pmb \xi=\frac{g}{4\pi\sqrt{\pi R}}O \tilde{\pmb b}\,,
\end{eqnarray}
where $O$ is an orthogonal matrix, and expand $V_T$ to the second power of $\pmb b$ to obtain:
\begin{eqnarray}
L_{\pmb b}=\frac{1}{2}\partial_\mu \pmb \xi \cdot \partial_\mu \pmb \xi+\frac{1}{2}\pmb \xi O{\cal M}^2 O^T\pmb \xi\,.
\end{eqnarray}
The square mass matrix ${\cal M}^2$ is given by
\begin{eqnarray}
{\cal M}_{ij}^2=\frac{16\pi^3 R}{g^2}\frac{\partial^2 V_T}{\partial b_i \partial b_j}|_{\pmb b=\pmb v}=m_0^2\left[\frac{\partial^2 (V_{T}/V_{bion}^0)}{\partial b_i \partial b_j}\right]|_{\pmb b=\pmb v}\,,
\end{eqnarray}
where $m_0^2=\frac{16 \pi^3 R}{g^2}V_{bion}^0$.
The matrix  ${\Lambda^2=O\cal M}^2 O^T$ is a diagonal matrix  with elements $\Lambda^2=\mbox{diag}(\lambda_{11}^2,.... \lambda_{rr}^2)$. Thus, the $\langle\xi_i(0)\xi_j(x)\rangle$ correlator is given by
\begin{eqnarray}
\langle\xi_i(0)\xi_j(x)\rangle=\delta_{ij}\frac{e^{-\lambda_{ii} r}}{4 \pi r}\,.
\end{eqnarray}
Hence, we obtain
\begin{eqnarray}
\langle \tilde b_i(0) \tilde b_j(x) \rangle=\frac{16\pi^3 R}{g^2}\sum_{l=1}^r (O^T)_{il} O_{lj}\frac{e^{-\lambda_{ll}r}}{4\pi r}\,.
\end{eqnarray}
Finally, the Polyakov correlator reads
\begin{eqnarray}
\nonumber
\langle  \Omega(x)\Omega^\dagger(y) \rangle&\cong&
\left|\mbox{tr}\left[e^{i \pmb H\cdot \pmb \phi_0^{(0)}} \left(1+i\frac{g^2}{4\pi}\pmb\phi_0^{(1)}\right)\cdot \pmb H\right]\right|^2\\
\nonumber
&&+i\frac{g^2}{4\pi}\mbox{tr}\left[e^{-i \pmb H\cdot \pmb \phi_0^{(0)}} \pmb H\cdot \pmb v \left(1+i\frac{g^2}{4\pi}\pmb\phi_0^{(1)}\right)\cdot \pmb H\right]+\mbox{h.c.}\\
\nonumber
&&+\frac {g^2 R}{4r}\mbox{tr}\left[e^{i \pmb H\cdot \pmb \phi_0^{(0)}}H_i\right]\mbox{tr}\left[e^{-i \pmb H\cdot \pmb \phi_0^{(0)}}H_j\right] (O^T)_{il} O_{lj} e^{-\lambda_{ll}r}\\
&&+\left(\frac{g^2}{4\pi}\right)^2\mbox{tr}\left[e^{i \pmb H\cdot \pmb \phi_0^{(0)}}H_i\right]\mbox{tr}\left[e^{-i \pmb H\cdot \pmb \phi_0^{(0)}}H_j\right]v_iv_j\,.
\end{eqnarray} 
For groups with center and at the supersymmetric vacuum one has
\begin{eqnarray}
\nonumber
\langle  \Omega(x)\Omega^\dagger(y) \rangle\cong
\frac {g^2 R}{4r}\sum_{i,j,l=1}^r\mbox{tr}\left[e^{i \pmb H\cdot \pmb \phi_0^{(0)}}H_i\right]\mbox{tr}\left[e^{-i \pmb H\cdot \pmb \phi_0^{(0)}}H_j\right] (O^T)_{il} O_{lj} e^{-\lambda_{ll}r}\,,
\end{eqnarray} 
from which we define the string tension, as described in Section \ref{results41}.

\bibliography{Deconfinement_General_Group_arxiv}

\providecommand{\href}[2]{#2}\begingroup\raggedright\begin{thebibliography}{10}

\bibitem{Poppitz:2012sw}
E.~Poppitz, T.~Schaefer, and M.~Unsal, {\it {Continuity, Deconfinement, and
  (Super) Yang-Mills Theory}},  {\em JHEP} {\bf 1210} (2012) 115,
  [\href{http://xxx.lanl.gov/abs/1205.0290}{{\tt arXiv:1205.0290}}].

\bibitem{Poppitz:2012nz}
E.~Poppitz, T.~Schaefer, and M.~Unsal, {\it {Universal mechanism of
  (semi-classical) deconfinement and theta-dependence for all simple groups}},
  {\em JHEP} {\bf 1303} (2013) 087,
  [\href{http://xxx.lanl.gov/abs/1212.1238}{{\tt arXiv:1212.1238}}].

\bibitem{Polyakov:1978vu}
A.~M. Polyakov, {\it {Thermal Properties of Gauge Fields and Quark
  Liberation}},  {\em Phys.Lett.} {\bf B72} (1978) 477--480.

\bibitem{Susskind:1979up}
L.~Susskind, {\it {Lattice Models of Quark Confinement at High Temperature}},
  {\em Phys.Rev.} {\bf D20} (1979) 2610--2618.

\bibitem{Gross:1980br}
D.~J. Gross, R.~D. Pisarski, and L.~G. Yaffe, {\it {QCD and Instantons at
  Finite Temperature}},  {\em Rev.Mod.Phys.} {\bf 53} (1981) 43.

\bibitem{Svetitsky:1982gs}
B.~Svetitsky and L.~G. Yaffe, {\it {Critical Behavior at Finite Temperature
  Confinement Transitions}},  {\em Nucl.Phys.} {\bf B210} (1982) 423.

\bibitem{Bhattacharya:1990hk}
T.~Bhattacharya, A.~Gocksch, C.~Korthals~Altes, and R.~D. Pisarski, {\it
  {Interface tension in an SU(N) gauge theory at high temperature}},  {\em
  Phys.Rev.Lett.} {\bf 66} (1991) 998--1000.

\bibitem{Smilga:1993vb}
A.~V. Smilga, {\it {Are Z(N) bubbles really there?}},  {\em Annals Phys.} {\bf
  234} (1994) 1--59.

\bibitem{Aharony:1998qu}
O.~Aharony and E.~Witten, {\it {Anti-de Sitter space and the center of the
  gauge group}},  {\em JHEP} {\bf 9811} (1998) 018,
  [\href{http://xxx.lanl.gov/abs/hep-th/9807205}{{\tt hep-th/9807205}}].

\bibitem{KorthalsAltes:1999xb}
C.~Korthals-Altes, A.~Kovner, and M.~A. Stephanov, {\it {Spatial 't Hooft loop,
  hot QCD and Z(N) domain walls}},  {\em Phys.Lett.} {\bf B469} (1999)
  205--212, [\href{http://xxx.lanl.gov/abs/hep-ph/9909516}{{\tt
  hep-ph/9909516}}].

\bibitem{Armoni:2008yp}
A.~Armoni, S.~P. Kumar, and J.~M. Ridgway, {\it {Z(N) Domain walls in hot N=4
  SYM at weak and strong coupling}},  {\em JHEP} {\bf 0901} (2009) 076,
  [\href{http://xxx.lanl.gov/abs/0812.0773}{{\tt arXiv:0812.0773}}].

\bibitem{Aharony:2003sx}
O.~Aharony, J.~Marsano, S.~Minwalla, K.~Papadodimas, and M.~Van~Raamsdonk, {\it
  {The Hagedorn - deconfinement phase transition in weakly coupled large N
  gauge theories}},  {\em Adv.Theor.Math.Phys.} {\bf 8} (2004) 603--696,
  [\href{http://xxx.lanl.gov/abs/hep-th/0310285}{{\tt hep-th/0310285}}].

\bibitem{Pisarski:2001pe}
R.~D. Pisarski, {\it {Tests of the Polyakov loops model}},  {\em Nucl.Phys.}
  {\bf A702} (2002) 151--158,
  [\href{http://xxx.lanl.gov/abs/hep-ph/0112037}{{\tt hep-ph/0112037}}].

\bibitem{Fukushima:2003fw}
K.~Fukushima, {\it {Chiral effective model with the Polyakov loop}},  {\em
  Phys.Lett.} {\bf B591} (2004) 277--284,
  [\href{http://xxx.lanl.gov/abs/hep-ph/0310121}{{\tt hep-ph/0310121}}].

\bibitem{Ratti:2005jh}
C.~Ratti, M.~A. Thaler, and W.~Weise, {\it {Phases of QCD: Lattice
  thermodynamics and a field theoretical model}},  {\em Phys.Rev.} {\bf D73}
  (2006) 014019, [\href{http://xxx.lanl.gov/abs/hep-ph/0506234}{{\tt
  hep-ph/0506234}}].

\bibitem{Braun:2010cy}
J.~Braun, A.~Eichhorn, H.~Gies, and J.~M. Pawlowski, {\it {On the Nature of the
  Phase Transition in SU(N), Sp(2) and E(7) Yang-Mills theory}},  {\em
  Eur.Phys.J.} {\bf C70} (2010) 689--702,
  [\href{http://xxx.lanl.gov/abs/1007.2619}{{\tt arXiv:1007.2619}}].

\bibitem{Diakonov:2012dx}
D.~Diakonov, C.~Gattringer, and H.-P. Schadler, {\it {Free energy for
  parameterized Polyakov loops in SU(2) and SU(3) lattice gauge theory}},  {\em
  JHEP} {\bf 1208} (2012) 128, [\href{http://xxx.lanl.gov/abs/1205.4768}{{\tt
  arXiv:1205.4768}}].

\bibitem{Greensite:2012dy}
J.~Greensite, {\it {The potential of the effective Polyakov line action from
  the underlying lattice gauge theory}},  {\em Phys.Rev.} {\bf D86} (2012)
  114507, [\href{http://xxx.lanl.gov/abs/1209.5697}{{\tt arXiv:1209.5697}}].

\bibitem{Dumitru:2012fw}
A.~Dumitru, Y.~Guo, Y.~Hidaka, C.~P.~K. Altes, and R.~D. Pisarski, {\it
  {Effective Matrix Model for Deconfinement in Pure Gauge Theories}},  {\em
  Phys.Rev.} {\bf D86} (2012) 105017,
  [\href{http://xxx.lanl.gov/abs/1205.0137}{{\tt arXiv:1205.0137}}].

\bibitem{Haas:2013qwp}
L.~M. Haas, R.~Stiele, J.~Braun, J.~M. Pawlowski, and J.~Schaffner-Bielich,
  {\it {Improved Polyakov-loop potential for effective models from functional
  calculations}},  \href{http://xxx.lanl.gov/abs/1302.1993}{{\tt
  arXiv:1302.1993}}.

\bibitem{Smith:2013msa}
D.~Smith, A.~Dumitru, R.~Pisarski, and L.~von Smekal, {\it {Effective potential
  for SU(2) Polyakov loops and Wilson loop eigenvalues}},  {\em Phys.Rev.} {\bf
  D88} (2013), no.~5 054020, [\href{http://xxx.lanl.gov/abs/1307.6339}{{\tt
  arXiv:1307.6339}}].

\bibitem{Simic:2010sv}
D.~Simic and M.~Unsal, {\it {Deconfinement in Yang-Mills theory through
  toroidal compactification with deformation}},  {\em Phys.Rev.} {\bf D85}
  (2012) 105027, [\href{http://xxx.lanl.gov/abs/1010.5515}{{\tt
  arXiv:1010.5515}}].

\bibitem{Anber:2011gn}
M.~M. Anber, E.~Poppitz, and M.~Unsal, {\it {2d affine XY-spin model/4d gauge
  theory duality and deconfinement}},  {\em JHEP} {\bf 1204} (2012) 040,
  [\href{http://xxx.lanl.gov/abs/1112.6389}{{\tt arXiv:1112.6389}}].

\bibitem{Anber:2012ig}
M.~M. Anber, S.~Collier, and E.~Poppitz, {\it {The $SU(3)/\Z_3$ QCD(adj)
  deconfinement transition via the gauge theory/'affine' XY-model duality}},
  {\em JHEP} {\bf 1301} (2013) 126,
  [\href{http://xxx.lanl.gov/abs/1211.2824}{{\tt arXiv:1211.2824}}].

\bibitem{Anber:2013doa}
M.~M. Anber, S.~Collier, E.~Poppitz, S.~Strimas-Mackey, and B.~Teeple, {\it
  {Deconfinement in $\mathcal{N}=1$ super Yang-Mills theory on $\mathbb{R}^3
  \times \mathbb{S}^1$ via dual-Coulomb gas and "affine" XY-model}},  {\em
  JHEP} {\bf 1311} (2013) 142, [\href{http://xxx.lanl.gov/abs/1310.3522}{{\tt
  arXiv:1310.3522}}].

\bibitem{Unsal:2010qh}
M.~Unsal and L.~G. Yaffe, {\it {Large-$N$ volume independence in conformal and
  confining gauge theories}},  {\em JHEP} {\bf 1008} (2010) 030,
  [\href{http://xxx.lanl.gov/abs/1006.2101}{{\tt arXiv:1006.2101}}].

\bibitem{Poppitz:2008hr}
E.~Poppitz and M.~Unsal, {\it {Index theorem for topological excitations on
  $\R^3 \times \S^1$ and Chern-Simons theory}},  {\em JHEP} {\bf 0903} (2009)
  027, [\href{http://xxx.lanl.gov/abs/0812.2085}{{\tt arXiv:0812.2085}}].

\bibitem{Seiberg:1996nz}
N.~Seiberg and E.~Witten, {\it {Gauge dynamics and compactification to
  three-dimensions}},  \href{http://xxx.lanl.gov/abs/hep-th/9607163}{{\tt
  hep-th/9607163}}.

\bibitem{Aharony:1997bx}
O.~Aharony, A.~Hanany, K.~A. Intriligator, N.~Seiberg, and M.~Strassler, {\it
  {Aspects of $N=2$ supersymmetric gauge theories in three-dimensions}},  {\em
  Nucl.Phys.} {\bf B499} (1997) 67--99,
  [\href{http://xxx.lanl.gov/abs/hep-th/9703110}{{\tt hep-th/9703110}}].

\bibitem{Davies:1999uw}
N.~M. Davies, T.~J. Hollowood, V.~V. Khoze, and M.~P. Mattis, {\it {Gluino
  condensate and magnetic monopoles in supersymmetric gluodynamics}},  {\em
  Nucl.Phys.} {\bf B559} (1999) 123--142,
  [\href{http://xxx.lanl.gov/abs/hep-th/9905015}{{\tt hep-th/9905015}}].

\bibitem{Lucini:2012gg}
B.~Lucini and M.~Panero, {\it {$SU(N)$ gauge theories at large-$N$}},  {\em
  Phys.Rept.} {\bf 526} (2013) 93--163,
  [\href{http://xxx.lanl.gov/abs/1210.4997}{{\tt arXiv:1210.4997}}].

\bibitem{Unsal:2012zj}
M.~Unsal, {\it {Theta dependence, sign problems and topological interference}},
   {\em Phys.Rev.} {\bf D86} (2012) 105012,
  [\href{http://xxx.lanl.gov/abs/1201.6426}{{\tt arXiv:1201.6426}}].

\bibitem{Parnachev:2008fy}
A.~Parnachev and A.~R. Zhitnitsky, {\it {Phase Transitions, theta Behavior and
  Instantons in QCD and its Holographic Model}},  {\em Phys.Rev.} {\bf D78}
  (2008) 125002, [\href{http://xxx.lanl.gov/abs/0806.1736}{{\tt
  arXiv:0806.1736}}].

\bibitem{Thomas:2011ee}
E.~Thomas and A.~R. Zhitnitsky, {\it {Topological Susceptibility and Contact
  Term in QCD. A Toy Model}},  {\em Phys.Rev.} {\bf D85} (2012) 044039,
  [\href{http://xxx.lanl.gov/abs/1109.2608}{{\tt arXiv:1109.2608}}].

\bibitem{Anber:2013sga}
M.~M. Anber, {\it {Theta dependence of the deconfining phase transition in pure
  $SU(N_c)$ Yang-Mills theories}},
  \href{http://xxx.lanl.gov/abs/1302.2641}{{\tt arXiv:1302.2641}}.

\bibitem{D'Elia:2012vv}
M.~D'Elia and F.~Negro, {\it {$\theta$-dependence of the deconfinement
  temperature in Yang-Mills theories}},  {\em Phys.Rev.Lett.} {\bf 109} (2012)
  072001, [\href{http://xxx.lanl.gov/abs/1205.0538}{{\tt arXiv:1205.0538}}].

\bibitem{D'Elia:2013eua}
M.~D'Elia and F.~Negro, {\it {On the phase diagram of Yang-Mills theories in
  the presence of a theta therm}},  {\em Phys.Rev.} {\bf D88} (2013) 034503,
  [\href{http://xxx.lanl.gov/abs/1306.2919}{{\tt arXiv:1306.2919}}].

\bibitem{Pepe:2006er}
M.~Pepe and U.-J. Wiese, {\it {Exceptional Deconfinement in $G_2$ Gauge
  Theory}},  {\em Nucl.Phys.} {\bf B768} (2007) 21--37,
  [\href{http://xxx.lanl.gov/abs/hep-lat/0610076}{{\tt hep-lat/0610076}}].

\bibitem{Cossu:2007dk}
G.~Cossu, M.~D'Elia, A.~Di~Giacomo, B.~Lucini, and C.~Pica, {\it {$G_2$ gauge
  theory at finite temperature}},  {\em JHEP} {\bf 0710} (2007) 100,
  [\href{http://xxx.lanl.gov/abs/0709.0669}{{\tt arXiv:0709.0669}}].

\bibitem{Unsal:2007jx}
M.~Unsal, {\it {Magnetic bion condensation: A New mechanism of confinement and
  mass gap in four dimensions}},  {\em Phys.Rev.} {\bf D80} (2009) 065001,
  [\href{http://xxx.lanl.gov/abs/0709.3269}{{\tt arXiv:0709.3269}}].

\bibitem{Poppitz:2011wy}
E.~Poppitz and M.~Unsal, {\it {Seiberg-Witten and 'Polyakov-like' magnetic bion
  confinements are continuously connected}},  {\em JHEP} {\bf 1107} (2011) 082,
  [\href{http://xxx.lanl.gov/abs/1105.3969}{{\tt arXiv:1105.3969}}].

\bibitem{Argyres:2012ka}
P.~C. Argyres and M.~Unsal, {\it {The semi-classical expansion and resurgence
  in gauge theories: new perturbative, instanton, bion, and renormalon
  effects}},  {\em JHEP} {\bf 1208} (2012) 063,
  [\href{http://xxx.lanl.gov/abs/1206.1890}{{\tt arXiv:1206.1890}}].

\bibitem{Wess:1992cp}
J.~Wess and J.~Bagger, {\it {Supersymmetry and supergravity, Princeton
  University Press, 1992}}, .

\bibitem{Lee:1997vp}
K.-M. Lee and P.~Yi, {\it {Monopoles and instantons on partially compactified
  $D$-branes}},  {\em Phys.Rev.} {\bf D56} (1997) 3711--3717,
  [\href{http://xxx.lanl.gov/abs/hep-th/9702107}{{\tt hep-th/9702107}}].

\bibitem{Lee:1998bb}
K.-M. Lee and C.-h. Lu, {\it {SU(2) calorons and magnetic monopoles}},  {\em
  Phys.Rev.} {\bf D58} (1998) 025011,
  [\href{http://xxx.lanl.gov/abs/hep-th/9802108}{{\tt hep-th/9802108}}].

\bibitem{Kraan:1998pm}
T.~C. Kraan and P.~van Baal, {\it {Periodic instantons with nontrivial
  holonomy}},  {\em Nucl.Phys.} {\bf B533} (1998) 627--659,
  [\href{http://xxx.lanl.gov/abs/hep-th/9805168}{{\tt hep-th/9805168}}].

\bibitem{Holland:2003kg}
K.~Holland, M.~Pepe, and U.~Wiese, {\it {The Deconfinement phase transition of
  Sp(2) and Sp(3) Yang-Mills theories in (2+1)-dimensions and
  (3+1)-dimensions}},  {\em Nucl.Phys.} {\bf B694} (2004) 35--58,
  [\href{http://xxx.lanl.gov/abs/hep-lat/0312022}{{\tt hep-lat/0312022}}].

\bibitem{HoyosBadajoz:2007ds}
C.~Hoyos-Badajoz, B.~Lucini, and A.~Naqvi, {\it {Confinement, screening and the
  center on S**3 x S**1}},  {\em JHEP} {\bf 0804} (2008) 075,
  [\href{http://xxx.lanl.gov/abs/0711.0659}{{\tt arXiv:0711.0659}}].

\bibitem{Lovelace:1982hz}
C.~Lovelace, {\it {Universality at large-N}},  {\em Nucl.Phys.} {\bf B201}
  (1982) 333.

\bibitem{Unsal:2006pj}
M.~Unsal and L.~G. Yaffe, {\it {(In)validity of large N orientifold
  equivalence}},  {\em Phys.Rev.} {\bf D74} (2006) 105019,
  [\href{http://xxx.lanl.gov/abs/hep-th/0608180}{{\tt hep-th/0608180}}].

\bibitem{Bershadsky:1998cb}
M.~Bershadsky and A.~Johansen, {\it {Large N limit of orbifold field
  theories}},  {\em Nucl.Phys.} {\bf B536} (1998) 141--148,
  [\href{http://xxx.lanl.gov/abs/hep-th/9803249}{{\tt hep-th/9803249}}].

\bibitem{DelDebbio:2006df}
L.~Del~Debbio, G.~M. Manca, H.~Panagopoulos, A.~Skouroupathis, and E.~Vicari,
  {\it {Theta-dependence of the spectrum of SU(N) gauge theories}},  {\em JHEP}
  {\bf 0606} (2006) 005, [\href{http://xxx.lanl.gov/abs/hep-th/0603041}{{\tt
  hep-th/0603041}}].

\bibitem{Polyakov:1976fu}
A.~M. Polyakov, {\it {Quark Confinement and Topology of Gauge Groups}},  {\em
  Nucl.Phys.} {\bf B120} (1977) 429--458.

\bibitem{Lecheminant:2006hj}
P.~Lecheminant, {\it {Nature of the deconfining phase transition in the
  2+1-dimensional $SU(N)$ Georgi-Glashow model}},  {\em Phys.Lett.} {\bf B648}
  (2007) 323--328, [\href{http://xxx.lanl.gov/abs/hep-th/0610046}{{\tt
  hep-th/0610046}}].

\bibitem{Diakonov:2007nv}
D.~Diakonov and V.~Petrov, {\it {Confining ensemble of dyons}},  {\em
  Phys.Rev.} {\bf D76} (2007) 056001,
  [\href{http://xxx.lanl.gov/abs/0704.3181}{{\tt arXiv:0704.3181}}].

\bibitem{Shuryak:2014gja}
E.~Shuryak, {\it {On Chiral Symmetry Breaking, Topology and Confinement}},
  \href{http://xxx.lanl.gov/abs/1401.2032}{{\tt arXiv:1401.2032}}.

\bibitem{Poppitz:2013zqa}
E.~Poppitz and T.~Sulejmanpasic, {\it {(S)QCD on $\R^3 \times \S^1$: Screening
  of Polyakov loop by fundamental quarks and the demise of semi-classics}},
  \href{http://xxx.lanl.gov/abs/1307.1317}{{\tt arXiv:1307.1317}}.

\bibitem{Shuryak:2013tka}
E.~Shuryak and T.~Sulejmanpasic, {\it {Holonomy potential and confinement from
  a simple model of the gauge topology}},
  \href{http://xxx.lanl.gov/abs/1305.0796}{{\tt arXiv:1305.0796}}.

\bibitem{Schafer:1996wv}
T.~Schaefer and E.~V. Shuryak, {\it {Instantons in QCD}},  {\em Rev.Mod.Phys.}
  {\bf 70} (1998) 323--426, [\href{http://xxx.lanl.gov/abs/hep-ph/9610451}{{\tt
  hep-ph/9610451}}].

\bibitem{Faccioli:2013ja}
P.~Faccioli and E.~Shuryak, {\it {QCD topology at finite temperature:
  Statistical mechanics of self-dual dyons}},  {\em Phys.Rev.} {\bf D87}
  (2013), no.~7 074009, [\href{http://xxx.lanl.gov/abs/1301.2523}{{\tt
  arXiv:1301.2523}}].

\bibitem{Argyres:2012vv}
P.~Argyres and M.~Unsal, {\it {A semiclassical realization of infrared
  renormalons}},  {\em Phys.Rev.Lett.} {\bf 109} (2012) 121601,
  [\href{http://xxx.lanl.gov/abs/1204.1661}{{\tt arXiv:1204.1661}}].

\bibitem{Dunne:2012ae}
G.~V. Dunne and M.~Unsal, {\it {Resurgence and Trans-series in Quantum Field
  Theory: The $CP^{N-1}$ Model}},  {\em JHEP} {\bf 1211} (2012) 170,
  [\href{http://xxx.lanl.gov/abs/1210.2423}{{\tt arXiv:1210.2423}}].

\bibitem{Cherman:2014ofa}
A.~Cherman, D.~Dorigoni, and M.~Unsal, {\it {Decoding perturbation theory using
  resurgence: Stokes phenomena, new saddle points and Lefschetz thimbles}},
  \href{http://xxx.lanl.gov/abs/1403.1277}{{\tt arXiv:1403.1277}}.

\bibitem{Giedt:2008xm}
J.~Giedt, R.~Brower, S.~Catterall, G.~T. Fleming, and P.~Vranas, {\it {Lattice
  super-Yang-Mills using domain wall fermions in the chiral limit}},  {\em
  Phys.Rev.} {\bf D79} (2009) 025015,
  [\href{http://xxx.lanl.gov/abs/0810.5746}{{\tt arXiv:0810.5746}}].

\bibitem{Demmouche:2010sf}
K.~Demmouche, F.~Farchioni, A.~Ferling, I.~Montvay, G.~Munster, et~al., {\it
  {Simulation of 4d N=1 supersymmetric Yang-Mills theory with Symanzik improved
  gauge action and stout smearing}},  {\em Eur.Phys.J.} {\bf C69} (2010)
  147--157, [\href{http://xxx.lanl.gov/abs/1003.2073}{{\tt arXiv:1003.2073}}].

\bibitem{Bergner:2014saa}
G.~Bergner, P.~Giudice, G.~MŸnster, S.~Piemonte, and D.~Sandbrink, {\it {Phase
  structure of the N=1 supersymmetric Yang-Mills theory at finite
  temperature}},  \href{http://xxx.lanl.gov/abs/1405.3180}{{\tt
  arXiv:1405.3180}}.

\bibitem{deBoer:1997kr}
J.~de~Boer, K.~Hori, and Y.~Oz, {\it {Dynamics of N=2 supersymmetric gauge
  theories in three-dimensions}},  {\em Nucl.Phys.} {\bf B500} (1997) 163--191,
  [\href{http://xxx.lanl.gov/abs/hep-th/9703100}{{\tt hep-th/9703100}}].

\bibitem{Intriligator:2013lca}
K.~Intriligator and N.~Seiberg, {\it {Aspects of 3d N=2 Chern-Simons-Matter
  Theories}},  \href{http://xxx.lanl.gov/abs/1305.1633}{{\tt arXiv:1305.1633}}.

\bibitem{Aharony:2013hda}
O.~Aharony, N.~Seiberg, and Y.~Tachikawa, {\it {Reading between the lines of
  four-dimensional gauge theories}},  {\em JHEP} {\bf 1308} (2013) 115,
  [\href{http://xxx.lanl.gov/abs/1305.0318}{{\tt arXiv:1305.0318}}].

\bibitem{Smilga:2004zr}
A.~V. Smilga and A.~Vainshtein, {\it {Background field calculations and
  nonrenormalization theorems in 4-D supersymmetric gauge theories and their
  low-dimensional descendants}},  {\em Nucl.Phys.} {\bf B704} (2005) 445--474,
  [\href{http://xxx.lanl.gov/abs/hep-th/0405142}{{\tt hep-th/0405142}}].

\bibitem{Davies:2000nw}
N.~M. Davies, T.~J. Hollowood, and V.~V. Khoze, {\it {Monopoles, affine
  algebras and the gluino condensate}},  {\em J.Math.Phys.} {\bf 44} (2003)
  3640--3656, [\href{http://xxx.lanl.gov/abs/hep-th/0006011}{{\tt
  hep-th/0006011}}].

\bibitem{Unsal:2008ch}
M.~Unsal and L.~G. Yaffe, {\it {Center-stabilized Yang-Mills theory:
  Confinement and large-$N$ volume independence}},  {\em Phys.Rev.} {\bf D78}
  (2008) 065035, [\href{http://xxx.lanl.gov/abs/0803.0344}{{\tt
  arXiv:0803.0344}}].

\bibitem{Douglas:1995nw}
M.~R. Douglas and S.~H. Shenker, {\it {Dynamics of SU(N) supersymmetric gauge
  theory}},  {\em Nucl.Phys.} {\bf B447} (1995) 271--296,
  [\href{http://xxx.lanl.gov/abs/hep-th/9503163}{{\tt hep-th/9503163}}].

\bibitem{Hisano:1997ua}
J.~Hisano and M.~A. Shifman, {\it {Exact results for soft supersymmetry
  breaking parameters in supersymmetric gauge theories}},  {\em Phys.Rev.} {\bf
  D56} (1997) 5475--5482, [\href{http://xxx.lanl.gov/abs/hep-ph/9705417}{{\tt
  hep-ph/9705417}}].

\bibitem{Holland:2003mc}
K.~Holland, M.~Pepe, and U.~Wiese, {\it {The Deconfinement phase transition in
  Yang-Mills theory with general Lie group G}},  {\em Nucl.Phys.Proc.Suppl.}
  {\bf 129} (2004) 712--714,
  [\href{http://xxx.lanl.gov/abs/hep-lat/0309062}{{\tt hep-lat/0309062}}].

\bibitem{Armoni:2007kd}
A.~Armoni, M.~Shifman, and M.~Unsal, {\it {Planar Limit of Orientifold Field
  Theories and Emergent Center Symmetry}},  {\em Phys.Rev.} {\bf D77} (2008)
  045012, [\href{http://xxx.lanl.gov/abs/0712.0672}{{\tt arXiv:0712.0672}}].

\bibitem{Holland:2003jy}
K.~Holland, P.~Minkowski, M.~Pepe, and U.~Wiese, {\it {Exceptional confinement
  in G(2) gauge theory}},  {\em Nucl.Phys.} {\bf B668} (2003) 207--236,
  [\href{http://xxx.lanl.gov/abs/hep-lat/0302023}{{\tt hep-lat/0302023}}].

\bibitem{Kovtun:2007py}
P.~Kovtun, M.~Unsal, and L.~G. Yaffe, {\it {Volume independence in large N(c)
  QCD-like gauge theories}},  {\em JHEP} {\bf 0706} (2007) 019,
  [\href{http://xxx.lanl.gov/abs/hep-th/0702021}{{\tt hep-th/0702021}}].

\bibitem{Myers:2007vc}
J.~C. Myers and M.~C. Ogilvie, {\it {New phases of SU(3) and SU(4) at finite
  temperature}},  {\em Phys.Rev.} {\bf D77} (2008) 125030,
  [\href{http://xxx.lanl.gov/abs/0707.1869}{{\tt arXiv:0707.1869}}].

\bibitem{Karsch:1998qj}
F.~Karsch and M.~Lutgemeier, {\it {Deconfinement and chiral symmetry
  restoration in an SU(3) gauge theory with adjoint fermions}},  {\em
  Nucl.Phys.} {\bf B550} (1999) 449--464,
  [\href{http://xxx.lanl.gov/abs/hep-lat/9812023}{{\tt hep-lat/9812023}}].

\bibitem{Alford:1998sd}
M.~G. Alford, A.~Kapustin, and F.~Wilczek, {\it {Imaginary chemical potential
  and finite fermion density on the lattice}},  {\em Phys.Rev.} {\bf D59}
  (1999) 054502, [\href{http://xxx.lanl.gov/abs/hep-lat/9807039}{{\tt
  hep-lat/9807039}}].

\bibitem{Roberge:1986mm}
A.~Roberge and N.~Weiss, {\it {Gauge Theories With Imaginary Chemical Potential
  and the Phases of {QCD}}},  {\em Nucl.Phys.} {\bf B275} (1986) 734.

\bibitem{D'Elia:2002gd}
M.~D'Elia and M.-P. Lombardo, {\it {Finite density QCD via imaginary chemical
  potential}},  {\em Phys.Rev.} {\bf D67} (2003) 014505,
  [\href{http://xxx.lanl.gov/abs/hep-lat/0209146}{{\tt hep-lat/0209146}}].

\bibitem{Misumi:2014raa}
T.~Misumi and T.~Kanazawa, {\it {Adjoint QCD on $\mathbb{R}^3\times S^1$ with
  twisted fermionic boundary conditions}},
  \href{http://xxx.lanl.gov/abs/1405.3113}{{\tt arXiv:1405.3113}}.

\bibitem{Nye:2000eg}
T.~M. Nye and M.~A. Singer, {\it {An $L^2$ index theorem for Dirac operators on
  $\S^1 \times \R^3$}},  {\em J.Funct.Anal.} (2000)
  [\href{http://xxx.lanl.gov/abs/math/0009144}{{\tt math/0009144}}].

\end{thebibliography}\endgroup
\bibliographystyle{JHEP}

\end{document}